\documentclass[iop]{emulateapj}
\usepackage{lscape}
\usepackage{multirow}
\usepackage{natbib}

\newcommand{\ee}[1]{\mbox{${} \times 10^{#1}$}}
\newcommand{\lbol}{\mbox{$L_{bol}$}} 
\newcommand{\lint}{\mbox{$L_{int}$}} 
\newcommand{\tbol}{\mbox{$T_{bol}$}} 
\newcommand{\lbolsmm}{\mbox{$L_{bol}/L_{smm}$}} 

\newcommand{\tmb}{\mbox{$T_{\rm mb}$}}
\newcommand{\etamb}{$\eta_{\rm mb}$}

\newcommand{\comeron}{Comer\'{o}n}

\newcommand{\degree}{\mbox{$^{\circ}$}}
\newcommand{\am}{\mbox{\arcmin}}
\newcommand{\as}{\mbox{\arcsec}}
\newcommand{\cms}{\mbox{cm s$^{-1}$}}
\newcommand{\kms}{\mbox{km s$^{-1}$}}

\newcommand{\um}{$\mu$m}
\newcommand{\lsun}{\mbox{L$_\odot$}}
\newcommand{\msun}{\mbox{M$_\odot$}}

\newcommand{\mflow}{\mbox{$M_{\rm flow}$}} 
\newcommand{\pflow}{\mbox{$P_{\rm flow}$}} 
\newcommand{\eflow}{\mbox{$E_{\rm flow}$}} 
\newcommand{\lflow}{\mbox{$L_{\rm flow}$}} 
\newcommand{\fflow}{\mbox{$F_{\rm flow}$}} 

\newcommand{\hh}{\mbox{{\rm H}$_2$}}

\newcommand{\ammonia}{\mbox{{\rm NH}$_3$}}
\newcommand{\co}{$^{12}$CO}
\newcommand{\coo}{$^{13}$CO}

\newcommand{\nthp}{N$_2$H$^+$}

\newcommand{\cojone}{$^{12}$CO (1--0)}
\newcommand{\cojtwo}{$^{12}$CO (2--1)}
\newcommand{\cojthree}{$^{12}$CO (3--2)}

\newcommand{\cojten}{$^{12}$CO (10--9)}
\newcommand{\coojtwo}{$^{13}$CO (2--1)}
\newcommand{\coojthree}{$^{13}$CO (3--2)}
\newcommand{\cooojtwo}{C$^{18}$O (2--1)}

\slugcomment{\footnotesize For Resubmission to ApJ on \today}

\begin{document}
\title {Molecular Outflows Driven by Low-Mass Protostars. I. Correcting for Underestimates When Measuring Outflow Masses and Dynamical Properties}

\author{
Michael M.~Dunham\altaffilmark{1,2,3}, 
H\'ector G.~Arce\altaffilmark{2}, 
Diego Mardones\altaffilmark{4}, 
Jeong-Eun Lee\altaffilmark{5}, 
Brenda C.~Matthews\altaffilmark{6},
Amelia M.~Stutz\altaffilmark{7}, 
\& Jonathan P.~Williams\altaffilmark{8}
}

\altaffiltext{1}{Harvard-Smithsonian Center for Astrophysics, 60 Garden Street, MS 78, Cambridge, MA 02138, USA}

\altaffiltext{2}{Department of Astronomy, Yale University, P.O. Box 208101, New Haven, CT 06520, USA}

\altaffiltext{3}{mdunham@cfa.harvard.edu}

\altaffiltext{4}{Departamento de Astronom\'{i}a, Universidad de Chile, Casilla 36-D, Santiago, Chile}


\altaffiltext{5}{Department of Astronomy and Space Science, Kyung Hee University, Yongin, Gyeonggi 446-701, Republic of Korea}

\altaffiltext{6}{National Research Council of Canada, Herzberg Astronomy \& Astrophysics, 5071 W. Saanich Road, Victoria, BC V9E 2E7, Canada}

\altaffiltext{7}{Max-Planck-Institut f\"{u}r Astronomie, K\"{o}nigstuhl 17, D-69117, Heidelberg, Germany}

\altaffiltext{8}{Institute for Astronomy, University of Hawaii, Honolulu, Hawaii 96822, USA}

\begin{abstract}
We present a survey of 28 molecular outflows driven by low-mass protostars, 
all of which are sufficiently isolated spatially and/or kinematically to 
fully separate into individual outflows.  Using a combination of new and 
archival data from several single-dish telescopes, 17 outflows are mapped 
in \cojtwo\ and 17 are mapped in \cojthree, with 6 mapped in both transitions.  
For each outflow, we calculate and tabulate the mass (\mflow), momentum 
(\pflow), kinetic energy (\eflow), mechanical luminosity (\lflow), and force 
(\fflow) assuming optically thin emission in LTE at an excitation temperature, 
$T_{\rm ex}$, of 50 K.  We show that all of the calculated properties are 
underestimated when calculated under these assumptions.  Taken together, the 
effects of opacity, outflow emission at low velocities confused with ambient 
cloud emission, and emission below the sensitivities of the observations 
increase outflow masses and dynamical properties by an order of magnitude, on 
average, and factors of 50--90 in the most extreme cases.  Different (and 
non-uniform) excitation temperatures, inclination effects, and dissociation of 
molecular gas will all work to further increase outflow properties.  
Molecular outflows are thus almost certainly more massive and energetic than 
commonly reported.  Additionally, outflow properties are lower, on average, by 
almost 
an order of magnitude when calculated from the \cojthree\ maps compared to 
the \cojtwo\ maps, even after accounting for different opacities, 
map sensitivities, and possible excitation temperature variations.  
It has recently been argued in the literature that the \cojthree\ line 
is subthermally excited in outflows, and our results support this finding.
\end{abstract}

\keywords{ISM: jets and outflows - ISM: clouds - stars: formation - stars: low-mass - submillimeter: ISM}


\section{Introduction}\label{sec_intro}

Bipolar molecular outflows from protostars, first detected more than 30 
years ago \citep{snell1980:outflows}, are ubiquitous in the star formation 
process \citep[e.g.,][]{hatchell2007:outflows,hatchell2009:outflows}.  They 
are associated with both low and high-mass star formation 
\citep{wu2004:outflows}, and have even recently been detected in the substellar 
regime \citep{phanbao2008:bdoutflows,phanbao2011:bdoutflows}.  Since they are 
driven by accretion \citep[e.g.,][]{cabrit1992:outflows,bontemps1996:outflows}, 
molecular outflows can be used to measure the time-averaged accretion 
histories of their driving sources 
\citep{dunham2006:iram04191,dunham2010:l6737,lee2010:l1251a}.  They also 
carry away excess angular momentum, remove circumstellar material and shape the 
stellar initial mass function (IMF), and inject momentum and energy into 
the surrounding medium  
\citep[e.g.,][]{lada1985:outflows,bachiller1996:outflows,arce2007:ppv,banerjee2007:outflows,nakamura2007:outflows,hatchell2007:outflows,cunningham2009:outflows,nakamura2011:serpsouth,plunkett2013:ngc1333}, 
although the efficiency with which they accomplish each of these remains 
under debate.

Developing a complete understanding of the roles molecular outflows play in 
each of the above processes requires accurate measurements of the morphologies, 
masses, and energetics of outflows located in a diverse range of environments 
and driven by sources over all stages of protostellar evolution.  Numerous 
outflow surveys have been presented over the last three decades that have 
greatly improved our knowledge and understanding of the importance of outflows 
in the star formation process.  These studies have revealed correlations 
between outflow strengths and the properties of their driving sources 
\citep{cabrit1992:outflows,bontemps1996:outflows,wu2004:outflows,hatchell2007:outflows,curtis2010:outflows} and have directly measured the turbulent energy 
injected by outflows into their parent clusters 
\citep[e.g.,][]{arce2010:perseus,nakamura2011:serpsouth,ginsburg2011:outflows,plunkett2013:ngc1333}.  
However, since mapping the large extents of molecular outflows (which often 
have projected angular extents on the sky in excess of several arcminutes) to 
the sensitivies required to detect weak, high-velocity emission is necessarily 
expensive in terms of observing time, most of these studies suffer from one or 
more of the following limitations:  (1) Observations that only cover the 
central regions and do not map the full extent of the outflows; 
(2) Difficulty separating overlapping outflows along the line of sight in 
clustered regions; 
and (3) Compiling outflow masses and dynamical properties from previously published 
studies that adopt different methods and make different assumptions, leading 
to a heterogeneous dataset.

The simplest method to calculate the masses and dynamical properties of 
molecular outflows 
is to assume that the emission from outflowing gas in low-J rotational 
transitions of \co\ is optically thin, in local thermodynamic equilibrium 
at a single excitation 
temperature, and confined to velocities larger than those dominated by ambient 
cloud emission.  However, both \citet{downes2007:outflows} and 
\citet{offner2011:outflows} used synthetic observations of simulated outflows 
to show that the effects of line opacity, excitation temperature variations, 
low-velocity outflow emission confused with ambient cloud emission, 
inclination, and dissociation of molecular gas can increase outflow 
masses and dynamical properties by one or more orders of magnitude compared to the values 
obtained under the simple assumptions listed above.  The extent to which 
outflow surveys account for, and the methods they use to correct for, these 
effects vary widely from one to the next.  Specific examples will be discussed 
in the following sections of this paper, but most suffer from one or more of 
the limitations discussed above.  Indeed, a complete quantification of the 
magnitude of the corrections for all these effects with a large, 
statistically significant sample of well-separated outflows mapped in their 
entirety and analyzed with uniform methodology is currently lacking from 
the literature.  In light of the results of \citet{downes2007:outflows} and 
\citet{offner2011:outflows}, such a study is clearly needed.

With these motivations, we have undertaken a survey of 28 molecular 
outflows driven by low-mass protostars, all of which are sufficiently isolated 
spatially and/or kinematically to fully separate into individual outflows.  
Using a combination of new and archival data from several single-dish 
telescopes, 17 outflows are mapped in \cojtwo\ and 17 are mapped in \cojthree, 
with 6 mapped in both transitions.  Additional \coo\ observations are obtained 
for selected outflows.  In this paper we present an overview of the data 
collection and analysis.  We then calculate the masses and dynamical properties 
of all the outflows in a standard way assuming isothermal, optically thin 
emission in LTE.  We follow this with a detailed investigation of the 
correction factors 
to these quantities that are necessary for the various effects listed above, 
derived directly from our data.  
In a forthcoming paper we will explore the 
effects of these corrections on our current understanding of the evolution of 
protostellar outflows and the link between the accretion and outflow 
processes (M.~M.~Dunham et al.~2014, in preparation).

The organization of this paper is as follows.  We present an overview of 
the data collection and analysis in \S \ref{sec_data}, including the 
philosophy behind our target selection in \S \ref{sec_data_targets}, the 
observation strategy for the \co\ maps (\S \ref{sec_data_co}) and the 
selected \coo\ observations (\S \ref{sec_data_coo}), and the data reduction 
methods (\S \ref{sec_data_reduction}).  Our basic results are given in 
\S \ref{sec_results}, with \S \ref{sec_geometry} focusing on 
outflow geometrical properties and \S 
\ref{sec_mass_dynamical} giving details on our calculataion of the masses 
and dynamical properties under the simple assumptions listed above.  
We discuss the necessary corrections 
that must be applied to the outflow masses and dynamical properties in 
\S \ref{sec_correct} for the effects of opacity (\S \ref{sec_correct_opacity}), 
different (and non-uniform) excitation temperatures 
(\S \ref{sec_correct_temperature}), low-velocity outflow emission confused 
with ambient cloud emission (\S \ref{sec_correct_lowvel}), and emission below 
the sensitivities of the observations (\S \ref{sec_correct_velres}).  In 
\S \ref{sec_correct_other} we discuss other possible corrections that we 
are not able to derive from our data, including those due to inclination, 
dissociation of molecular gas, and calculation methods.  We provide a 
final overview and synthesis of the net effect of these corrections in 
\S \ref{sec_discussion_corrections}, and compare results from the two 
transitions of \co\ in \S 
\ref{sec_discussion_transitions}.  Finally, we summarize our results and 
outline necessary future work in \S \ref{sec_summary}.

\section{Description of the Data}\label{sec_data}

\subsection{Target Selection}\label{sec_data_targets}

\begin{deluxetable*}{lcccc}
\tabletypesize{\scriptsize}
\tablewidth{0pt}
\tablecaption{\label{tab_targets}List of Targets}  
\tablehead{
                 & \colhead{Map Center} & \colhead{Map Center} &                                                  &                         \\
                 & \colhead{R.A.}       & \colhead{Decl.}      & \colhead{Distance (Reference)\tablenotemark{a}}  & \colhead{Rest Velocity} \\
\colhead{Source} & \colhead{J2000}      & \colhead{J2000}      & \colhead{(pc)}                                   & \colhead{(\kms)}
}
\startdata
IRAS 03235$+$3004 & 03 26 37.6           & $+$30 15 24.2        & 250 (1)                                         & $+$5.1 \\
IRAS 03271$+$3013 & 03 30 15.5           & $+$30 23 43.0        & 250 (1)                                         & $+$5.9 \\
IRAS 03282$+$3035 & 03 31 21.0           & $+$30 45 27.8        & 250 (1)                                         & $+$7.1 \\
HH211             & 03 43 56.8           & $+$32 00 50.3        & 250 (1)                                         & $+$9.1 \\
IRAS 04166$+$2706 & 04 19 43.6           & $+$27 13 38.0        & 140 (2)                                         & $+$6.7 \\
IRAM 04191$+$1522 & 04 21 56.9           & $+$15 29 45.9        & 140 (2)                                         & $+$6.7 \\
HH25/26           & 05 46 04.9           & $-$00 14 52.0        & 430 (3)                                         & $+$10.1 \\
BHR86             & 13 07 37.2           & $-$77 00 09.0        & 178 (4)                                         & $+$3.7 \\
IRAS 15398$-$3359 & 15 43 01.3           & $-$34 09 15.0        & 150 (5)                                         & $+$5.1 \\
Lupus 3 MMS       & 16 09 18.1           & $-$39 04 53.4        & 200 (5)                                         & $+$4.8 \\
L1709-SMM1/5      & 16 31 35.6           & $-$24 01 29.3        & 125 (6)                                         & $+$2.5 \\
CB68              & 16 57 20.0           & $-$16 09 22.2        & 130 (7)                                         & $+$5.2 \\
L483              & 18 17 30.0           & $-$04 39 40.0        & 200 (8)                                         & $+$5.4 \\
Aqu-MM2/3/5       & 18 29 15.0           & $-$01 40 30.0        & 260 (9)                                         & $+$9.0 \\
SerpS-MM13        & 18 30 01.5           & $-$02 10 23.3        & 260 (9)                                         & $+$8.0 \\
CrA-IRAS32        & 19 02 58.7           & $-$37 07 35.9        & 130 (10)                                        & $+$5.6 \\
L673-7            & 19 21 34.8           & $+$11 21 23.0        & 240 (11)                                        & $+$7.1 \\
B335              & 19 37 00.9           & $+$07 34 09.8        & 150 (12)                                        & $+$8.3 \\
L1152             & 20 35 46.6           & $+$67 53 03.9        & 325 (13)                                        & $+$2.5 \\
L1157             & 20 39 06.2           & $+$68 02 15.0        & 300 (13)                                        & $+$2.6 \\
L1228             & 20 57 19.9           & $+$77 36 00.0        & 200 (14)                                        & $-$8.0 \\
L1014             & 21 24 07.6           & $+$49 59 08.9        & 258 (11)                                        & $+$4.2 \\
L1165             & 22 06 50.7           & $+$59 02 47.0        & 300 (15)                                        & $-$1.6 \\
L1251A-IRS3       & 22 30 31.9           & $+$75 14 08.8        & 300 (16)                                        & $-$3.9
\enddata
\tablenotetext{a}{Distance References:  (1) Enoch et al.~(2006); (2) Kenyon et al.~(1994); (3) Antoniucci et al.~(2008); (4) Whittet et al.~(1997); (5) \comeron\ (2008); (6) de Geus et al.~(1989); (7) Hatchell et al.~(2012); (8) Parker (1988); (9) Maury et al.~(2011); (10) Neuh{\"a}user \& Forbrich (2008); (11) Maheswar et al.~(2011); (12) Stutz et al.~(2008); (13) Kirk et al.~(2009); (14) Kun (1998); (15) Dobashi et al.~(1994); (16) Kun \& Prusti (1993).}
\end{deluxetable*}

The observations presented in this paper are a 
combination of new observations obtained from several single-dish 
(sub)millimeter telescopes and existing observations taken from telescope 
archives or provided by the authors of previously published data.  Given the 
motivations for this study described above in \S \ref{sec_intro}, we 
select targets based on the following three criteria:  
(1) a molecular outflow is either already known to exist or strongly suspected 
based on previous observations, (2) the outflow is sufficiently isolated 
spatially and/or kinematically from nearby outflows to prevent any issues with 
confusion when deriving properties, and (3) the full sample must span large 
ranges in both the bolometric luminosity and evolutionary status of the 
driving sources.

In total, we present maps of 28 outflows, 17 of which were mapped in \cojtwo\ 
and 17 in \cojthree\ (6 were mapped in both transitions).  These outflows are 
listed in Table \ref{tab_targets}, which lists the name of the driving source, 
the Right Ascension and Declination of the center of the map, the distance 
to the source (and reference for this distance), and rest velocity of the 
source.  All positions in the rest of the paper that are given in arcseconds of 
offset are relative to the positions listed in Table \ref{tab_targets}.  
A brief summary of the literature on each source is given in Appendix 
\ref{sec_appendix_sources}.  Further properties of the driving sources, 
including updated measurements of their bolometric luminosities and 
evolutionary status, will be given in a forthcoming paper aimed at compiling 
accurate, up-to-date measurements of source properties and evaluating the 
evolution of outflow activity from protostars.

\subsection{\co\ Observations}\label{sec_data_co}

\begin{deluxetable*}{lcccccccc}
\tabletypesize{\scriptsize}
\tablewidth{0pt}
\tablecaption{\label{tab_observations}Observation Log}  
\tablehead{
                  &                     &                       & \colhead{Observation} & \colhead{Map Size}    & \colhead{$\delta$v} & \colhead{1$\sigma$ rms} & \colhead{JCMT}    & \colhead{JCMT}     \\
\colhead{Source}  & \colhead{Telescope} & \colhead{Transition}  & \colhead{Date}        &\colhead{(arcmin$^2$)} & \colhead{(\kms)}    & \colhead{(K)}           & \colhead{Program} & \colhead{Map Type} 
}
\startdata
L1709-SMM1/5      & APEX                & \cojtwo\              & 2012 Apr              & 25                    & 0.1                 & 0.75                    & \nodata           & \nodata            \\
CB68              & APEX                & \cojtwo\              & 2012 Apr              & 25                    & 0.1                 & 0.72                    & \nodata           & \nodata            \\
Aqu-MM2/3/5       & APEX                & \cojtwo\              & 2012 Apr              & 81                    & 0.1                 & 0.50                    & \nodata           & \nodata            \\
SerpS-MM13        & APEX                & \cojtwo\              & 2012 Apr              & 40                    & 0.1                 & 0.55                    & \nodata           & \nodata            \\
CrA-IRAS32        & APEX                & \cojtwo\              & 2012 Apr              & 25                    & 0.1                 & 0.79                    & \nodata           & \nodata            \\
L673-7            & APEX                & \cojtwo\              & 2011 Oct, Nov         & 37                    & 0.1                 & 0.20                    & \nodata           & \nodata            \\
L673-7            & APEX                & \cojthree\            & 2012 Apr,             & 37                    & 0.1                 & 0.50                    & \nodata           & \nodata            \\
                  &                     &                       & May, Jun              &                       &                     &                         & \nodata           & \nodata            \\
L673-7            & APEX                & \coojthree\           & 2012 Jun,             & 37                    & 0.1                 & 0.40                    & \nodata           & \nodata            \\
                  &                     &                       & Jul, Oct              &                       &                     &                         & \nodata           & \nodata            \\
BHR86             & ASTE                & \cojthree\            & 2011 Jun              & 60                    & 0.5                 & 0.08                    & \nodata           & \nodata            \\
Lupus 3 MMS       & ASTE                & \cojthree\            & 2011 Jun              & 25                    & 0.5                 & 0.11                    & \nodata           & \nodata            \\
L483              & ASTE                & \cojthree\            & 2011 Jun              & 10                    & 0.5                 & 0.05                    & \nodata           & \nodata            \\
IRAS 03235$+$3004 & CSO                 & \cojtwo\              & 2012 Oct              & 25                    & 0.1                 & 0.32                    & \nodata           & \nodata            \\
IRAS 03282$+$3035 & CSO                 & \cojtwo\              & 2012 Oct              & 27                    & 0.1                 & 0.28                    & \nodata           & \nodata            \\
HH211             & CSO                 & \cojtwo\              & 2012 Oct              & 12                    & 0.1                 & 0.25                    & \nodata           & \nodata            \\
L1152             & CSO                 & \cojtwo\              & 2012 Oct              & 25                    & 0.1                 & 0.23                    & \nodata           & \nodata            \\
L1157             & CSO                 & \cojtwo\              & 2012 Oct              & 26                    & 0.1                 & 0.32                    & \nodata           & \nodata            \\
L1165             & CSO                 & \cojtwo\              & 2012 Oct              & 31                    & 0.1                 & 0.34                    & \nodata           & \nodata            \\
L1157             & CSO                 & \cojthree\            & 2012 Oct              & 16                    & 0.1                 & 1.6                     & \nodata           & \nodata            \\
IRAS 03235$+$3004 & JCMT                & \cojthree\            & 2007 Oct              & 3.5                   & 0.1                 & 0.29                    & M07BU08           & Jiggle             \\
IRAS 03271$+$3013 & JCMT                & \cojthree\            & 2007 Oct              & 3.5                   & 0.1                 & 0.35                    & M07BU08           & Jiggle             \\
IRAS 03282$+$3035 & JCMT                & \cojthree\            & 2007 Oct              & 3.5                   & 0.1                 & 0.53                    & M07BU08           & Jiggle             \\
HH211             & JCMT                & \cojthree\            & 2007 Dec              & 25                    & 0.1                 & 1.8                     & M06BGT02          & Raster             \\
IRAS 04166$+$2706 & JCMT                & \cojthree\            & 2007 Nov, 2009 Jan    & 180                   & 0.5                 & 0.48                    & GBS, M08BU26      & Raster             \\
IRAM 04191$+$1522 & JCMT                & \cojthree\            & 2008 Feb              & 24                    & 0.5                 & 0.23                    & M08AC08           & Raster             \\
HH25/26           & JCMT                & \cojthree\            & 2009 Jan              & 21                    & 0.5                 & 0.20                    & M08BU26           & Raster             \\
IRAS 15398$-$3359 & JCMT                & \cojthree\            & 2008 Jun              & 3.5                   & 0.1                 & 0.47                    & M08AN05           & Jiggle             \\
L1228             & JCMT                & \cojthree\            & 2008 Aug, Oct, Nov    & 105                   & 1.0                 & 0.33                    & M08BU11           & Raster             \\
L1014             & JCMT                & \cojthree\            & 2008 Jun              & 4                     & 0.5                 & 0.05                    & M08AC08           & Jiggle             \\
L1165             & JCMT                & \cojthree\            & 2008 Jun              & 49                    & 0.5                 & 0.24                    & M08AC03           & Raster             \\
L1251A-IRS3       & SRAO                & \cojtwo\              & 2009 Mar, Apr         & 160                   & 0.2                 & 0.17                    & \nodata           & \nodata            \\
B335              & SMT                 & \cojtwo\              & 2007 Apr              & 192                   & 0.33                & 0.14                    & \nodata           & \nodata 
\enddata
\end{deluxetable*}

In this section we summarize the observational details for the new and archival 
\co\ data used in this study.  All brightness temperatures given in this paper 
are in units of \tmb.  Assumed or measured values of \etamb\ for each 
telescope are listed and generally include a 10\% -- 20\% calibration 
uncertainty.  Table \ref{tab_observations} lists, for each 
map, the telescope used to obtain the map, the \co\ transition mapped, the 
observation date, the map size, the spectral resolution, and the 1$\sigma$ rms 
at this spectral resolution.  Also listed are additional details for the JCMT 
observations (see \S \ref{sec_jcmt} below).  Entries in Table 
\ref{tab_observations} are organized by telescope rather than by source.  
Finally, one \coo\ map is also listed and is described in more detail in 
\S \ref{sec_data_coo}.  

\subsubsection{Atacama Pathfinder Experiment}

A \cojtwo\ map of L673-7 was obtained at the Atacama Pathfinder 
Experiment (APEX) in 2011 October and November through APEX program 
C-088.F-1752B-2011.  Additional \cojtwo\ maps of Oph IRS 63, CB68, 
Aqu-MM2/3/5, SerpS-MM13, and CrA-IRAS32 were obtained at APEX in 2012 April
 through APEX program C-089.F-9757B-2012.  All data were obtained with 
the 230 GHz APEX-1 band of the Swedish Heterodyne Facility Instrument 
\citep[SHeFi;][]{belitsky2006:apex,risacher2006:apex} 
and the XFFTS fast fourier transform spectrometer, providing 2.5 GHz 
(3252 \kms) total bandwidth and 76 kHz (0.1 \kms) spectral resolution.  The 
beam FWHM is 27\as\ at 230 GHz, and the main-beam efficiency, \etamb, is 
0.82 \citep{vassilev2008:apexeta}.  All sources were mapped using the 
position-switched on-the-fly (otf) 
observing mode, with every second map observed 
at a position angle of 90\degree\ relative to the first.  

A \cojthree\ otf map of L673-7 was also obtained at APEX in 2012 April, May, 
and June through APEX program C-089.F-9758B-2012 with the 345 
GHz APEX-2 band of SHeFI and the XFFTS backend, providing 2.5 GHz 
(2167 \kms) total bandwidth and 76 kHz (0.07 \kms) spectral resolution.  The 
beam FWHM is 18\as\ at 345 GHz and \etamb\ is 0.73 \citep{gusten2006:apex}.  
The final map was smoothed to Nyquist sampled ($\sim 9$\as) pixels.  

\subsubsection{Atacama Submillimeter Telescope Experiment}

Maps of \cojthree\ of BHR86, Lupus 3 MMS, and L483 were obtained at 
the Atacama Submillimeter Telescope Experiment (ASTE) \citep{ezawa2004:aste} 
in 2011 June through the program CN2011B-070 with the 
CATS345 receiver and MAC digital spectro-correlator configured to provide 
512 MHz (445 \kms) bandwidth and 0.5 MHz (0.43 \kms) spectral resolution.  The 
beam FWHM is 21.5\as\ at 345 GHz and \etamb\ $= 0.6 \pm 0.1$ 
\citep[e.g., ][]{nakamura2011:serpsouth,miura2012:m33,watanabe2012:345ghz}.  
The maps were obtained using the position-switched otf observing mode, again 
with successive scans observed at perpendicular position angles.  
The final maps were smoothed to 11\as\ (approximately Nyquist sampled) 
pixels.  

\subsubsection{Caltech Submillimeter Observatory}

Maps of \cojtwo\ of IRAS 03235$+$3004, IRAS 03282$+$3035, HH211, L1152, 
L1157, and L1165 were obtained at the Caltech Submillimeter Observatory 
(CSO) in 2012 October with the 230 GHz sidecab receiver and a fast fourier 
transform spectrometer (FFTS) backend, providing 500 MHz (650 \kms) total 
bandwidth and 61 kHz (0.08 \kms) sectral resolution.  The beam FWHM is 
32.5\as\ at 230 GHz, and \etamb\ $= 0.73 \pm 0.02$ based on observations of 
Jupiter.  Position-switched otf maps were obtained for each source, with 
successive scans observed at perpendicular position angles.  The final maps 
were smoothed to Nyquist sampled ($\sim 16$\as) pixels.  

A \cojthree\ map of L1157 was also obtained at the CSO in 2012 October with 
the 345 GHz Barney receiver and FFTS backend, again providing a native spectral 
resolution of 61 kHz (0.05 \kms at 345 GHz).  The beam FWHM is 22\as, and 
\etamb\ $= 0.74 \pm 0.03$ based on observations of Jupiter.  
The final map was smoothed to Nyquist sampled ($\sim 11$\as) pixels.  

\subsubsection{James Clerk Maxwell Telescope}\label{sec_jcmt}

Maps of \cojthree\ of IRAM 04191$+$1522 and 
L1014 were obtained at the James Clerk Maxwell Telescope (JCMT) in 2008 
February and June with the Heterodyne Array Receiver Program B (HARP-B) 
band receiver \citep{buckle2009:harp} and Auto-Correlation Spectral Imaging 
System \citep[ACSIS;][]{dent2000:acsis,buckle2009:harp} backend through the 
JCMT 
observing program M08AC08.  Additionally, \cojthree\ maps of 
nine sources (IRAS 03235$+$3004, IRAS 03271$+$3013, IRAS 03282$+$3035, HH211, 
IRAS 04166$+$2706, HH25/26, IRAS 15398$-$3359, L1228, and L1165) 
that were obtained in other programs with HARP-B and the 
ACSIS backend were taken from the JCMT data archive\footnote{Available at 
http://www.jach.hawaii.edu/JCMT/archive/}.  HARP-B is a 16 element heterodyne 
receiver array arranged in a $4 \times 4$ grid with 30\as\ spacing between 
elements.  The beam FWHM is 14\as\ at 345 GHz \citep{buckle2009:harp}, and 
\etamb\ is taken to be 0.60 $\pm$ 0.02 \citep[mean and standard deviation of 
each individual receiver in the array;][]{buckle2009:harp}.  The data were 
taken in either the position-switched jiggle or raster map observing modes 
in a variety of backend configurations, and the final maps were smoothed to 
7\as\ (approximately Nyquist sampled) pixels.  The last two columns of Table 
\ref{tab_observations} list the JCMT program and observing mode in which the 
data were obtained.  

\begin{deluxetable*}{lcccc}
\tabletypesize{\scriptsize}
\tablewidth{0pt}
\tablecaption{\label{tab_cso_13co}Summary of CSO \coo\ Observations}  
\tablehead{
                  & \colhead{Observation} & \colhead{R.A. Offset\tablenotemark{a}}  & \colhead{Dec. Offset\tablenotemark{a}}  & \colhead{1$\sigma$ rms}  \\
\colhead{Source}  & \colhead{Date}        & \colhead{(arcseconds)} & \colhead{(arcseconds)} & \colhead{(K)}               
}
\startdata
\multicolumn{5}{c}{\coojtwo} \\
\hline
IRAS 03235$+$3004 & 2012 Oct & $-$110 & $+$40  & 0.04 \\
IRAS 03235$+$3004 & 2012 Oct & $+$30  & $-$10  & 0.04 \\
IRAS 03282$+$3035 & 2012 Oct & $-$20  & $+$10  & 0.04 \\
IRAS 03282$+$3035 & 2012 Oct & $+$20  & $-$5   & 0.04 \\
L673-7            & 2012 Sep & $-$90  & $-$60  & 0.06 \\
L673-7            & 2012 Sep & $-$60  & $-$60  & 0.06 \\
L673-7            & 2012 Sep & $-$30  & $-$30  & 0.05 \\
L673-7            & 2012 Sep & $+$30  & $+$30  & 0.06 \\
L673-7            & 2012 Sep & $+$60  & $+$30  & 0.06 \\
L673-7            & 2012 Sep & $+$60  & $+$60  & 0.06 \\
L1165             & 2012 Sep & $-$30  & $-$60  & 0.15 \\
L1165             & 2012 Sep & $+$00  & $-$30  & 0.06 \\
L1165             & 2012 Sep & $+$00  & $+$00  & 0.06 \\
L1165             & 2012 Sep & $+$30  & $+$30  & 0.05 \\
L1165             & 2012 Sep & $+$90  & $+$90  & 0.06 \\
L1251A-IRS3       & 2012 Oct & $-$10  & $+$100 & 0.05 \\
L1251A-IRS3       & 2012 Oct & $-$15  & $-$170 & 0.05 \\
\hline
\multicolumn{5}{c}{\coojthree} \\
\hline
IRAS 03235$+$3004 & 2012 Oct & $-$110 & $+$40  & 0.13 \\
IRAS 03235$+$3004 & 2012 Oct & $+$30  & $-$10  & 0.11 \\
IRAS 03282$+$3035 & 2012 Oct & $-$20  & $+$10  & 0.13 \\
IRAS 03282$+$3035 & 2012 Oct & $+$20  & $-$5   & 0.13 \\
L1165             & 2012 Oct & $+$00  & $-$30  & 0.11 
\enddata
\tablenotetext{a}{Offset in arcseconds from the positions listed in Table \ref{tab_targets}.}
\end{deluxetable*}

The data for HH211 were part of a large map of the IC348 cluster in Perseus 
and were previous published by 
\citet{curtis2010:data,curtis2010:outflows} and \citet{curtis2011:kinematics}; 
we refer 
the reader to those studies for a full description of the data collection and 
observation strategy.  We extracted a small region centered on HH211, and it 
is the area of this map that we list in Table \ref{tab_observations}.

For IRAS 04166$+$2706, we combined raster maps from two different programs:  
the JCMT Gould Belt Survey (GBS) 
\footnote{See http://www.jach.hawaii.edu/JCMT/surveys/gb/} and M08BU26.  
The observations from the GBS cover a larger area than those from M08BU26.  
The average 1$\sigma$ rms over the full, combined map is 0.48 K per 
spectral channel, and this is the value we list in Table \ref{tab_observations}.  
The rms decreases to $\sim$ 0.2 K in the region where the two programs overlap, 
which is also the region where the majority of the outflow emission is found.

\subsubsection{Seoul National Radio Astronomy Observatory}

A \cojtwo\ map of L1251A-IRS3 was obtained at the Seoul National 
Radio Astronomy Observatory (SRAO) in 2009 March and April.  These data were 
previously published by \citet{lee2010:l1251a}, in which full details of the 
instrumentation, observation strategy, and data reduction can be found.  
The beam FWHM is 48\as\ at 230 GHz, and \etamb\ is 0.57.  
The final map is presented on a 24\as\ spatial grid.  

\subsubsection{Submillimeter Telescope}

Observations of \cojtwo\ of B335 were obtained at the 
Submillimeter Telescope (HHT) in 2007 April with the 1.3 mm ALMA sideband 
separating receiver, providing a beam FWHM of 32\as\ at 230 GHz.  
These data were previously published by \citet{stutz2008:b335}, in which full 
details of the instrumentation, observation strategy and data reduction can be 
found.  The final map is presented on a 10\as\ spatial grid.  

\subsection{\coo\ Observations}\label{sec_data_coo}

\subsubsection{Atacama Pathfinder Experiment}

A \coojthree\ map of L673-7 covering the full extent of the outflow was 
obtained at APEX in 2012 June, July, and October through APEX program 
C-089.F-9758B-2012 with the 345 GHz APEX-2 band of SHeFI and the XFFTS 
backend, providing 2.5 GHz (2273 \kms) total bandwidth and 76 kHz (0.07 \kms) 
spectral resolution.  The beam FWHM is 19\as\ at 330 GHz and \etamb\ is 0.73 
\citep{gusten2006:apex}.  The final maps were smoothed to Nyquist sampled 
($\sim 10$\as) pixels, and the 1$\sigma$ rms per channel at this spectral 
resolution is 0.4 K, as listed in Table \ref{tab_observations}.

\subsubsection{Caltech Submillimeter Observatory}

Pointed \coojtwo\ observations toward bright positions in several of the 
outflows in this study were obtained at the CSO in 2012 September and 
October with the 230 GHz sidecab receiver and a fast fourier 
transform spectrometer (FFTS) backend, providing 500 MHz (682 \kms) total 
bandwidth and 61 kHz (0.08 \kms) sectral resolution.  
Additional \coojthree\ observations toward bright outflow positions were 
obtained at the CSO in 2012 October with the 345 GHz Barney receiver and FFTS 
backend, again providing a native spectral resolution of 61 kHz (0.06 \kms at 
345 GHz).  The beam FWHM is 34\as\ (23\as) at 220 (330) GHz, \etamb\ at 220 
GHz was measured to be $0.69 \pm 0.04$ ($0.77 \pm 0.03$) in 2012 September 
(October) based on observations of Jupiter, and \etamb\ at 330 GHz was 
measured to be $0.71 \pm 0.01$ based on observations of Jupiter.  
All maps were smoothed to spectral resolutions of 0.1 \kms.  For each pointed 
observation, Table \ref{tab_cso_13co} lists the source, observation date, 
position of the observation (measured in arcseconds of offset from the 
positions listed in Table \ref{tab_targets}), and 1$\sigma$ rms per 0.1 \kms\ 
channel.

\begin{figure*}
\epsscale{1.0}
\plotone{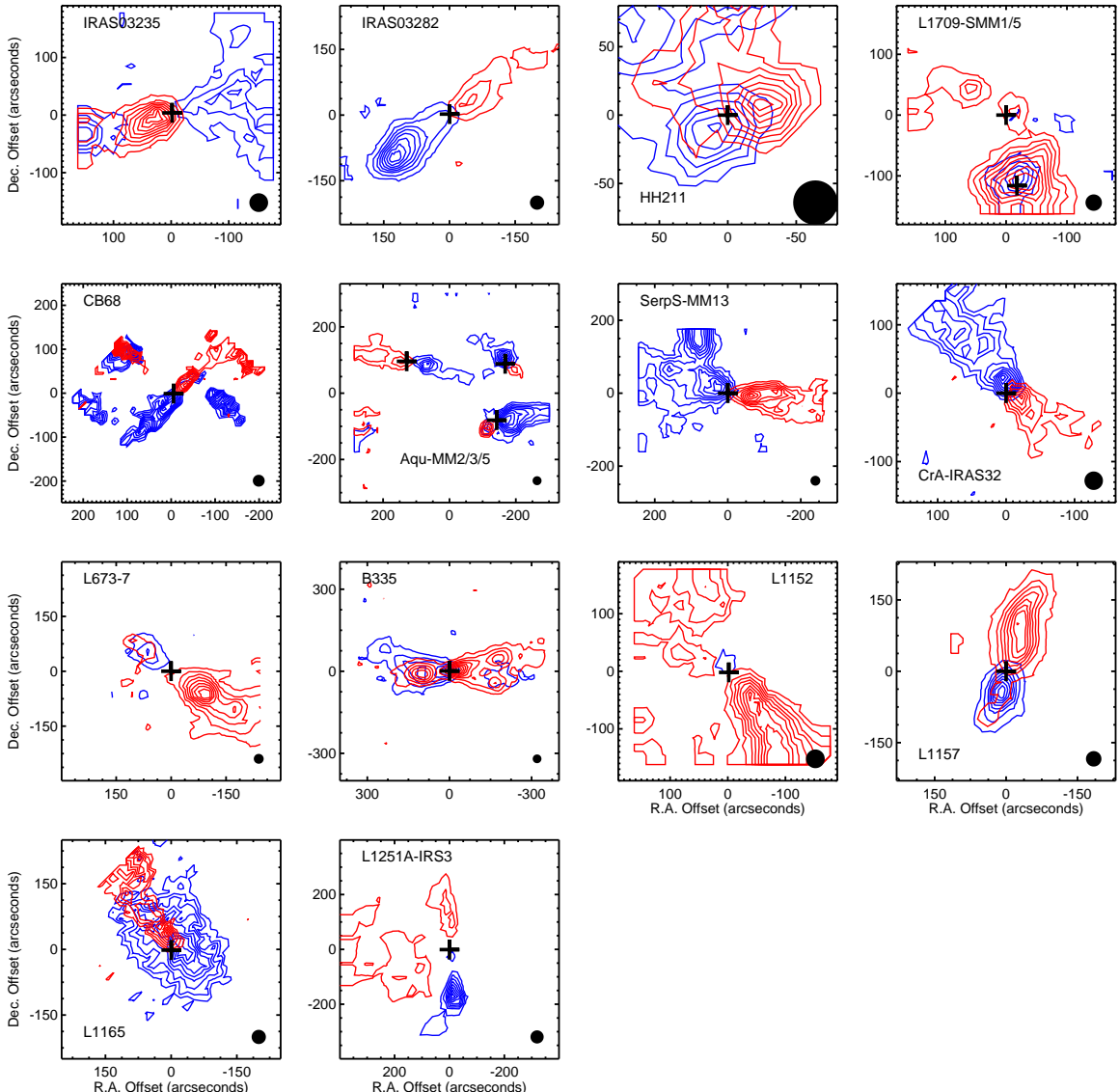}
\caption{\label{fig_redblue21}Integrated intensity contours showing 
blueshifted and redshifted emission for the 17 outflows mapped in \cojtwo.  
Each panel is labeled with the name(s) of the source(s).  The minimum and 
maximum velocities over which the emission is integrated are symmetrical about 
the rest velocity, and are chosen based on visual inspection of the velocity 
channels (see \S \ref{sec_mass_dynamical}).  In each panel we plot eight 
contour levels linearly spaced between three times the rms noise in each image 
and the maximum; the minimum contour and contour spacing are listed in Table 
\ref{tab_contours}.  The driving sources are marked with crosses and the beam 
sizes are shown as black ellipses in the lower right of each panel.}
\end{figure*}

\begin{figure*}
\epsscale{1.0}
\plotone{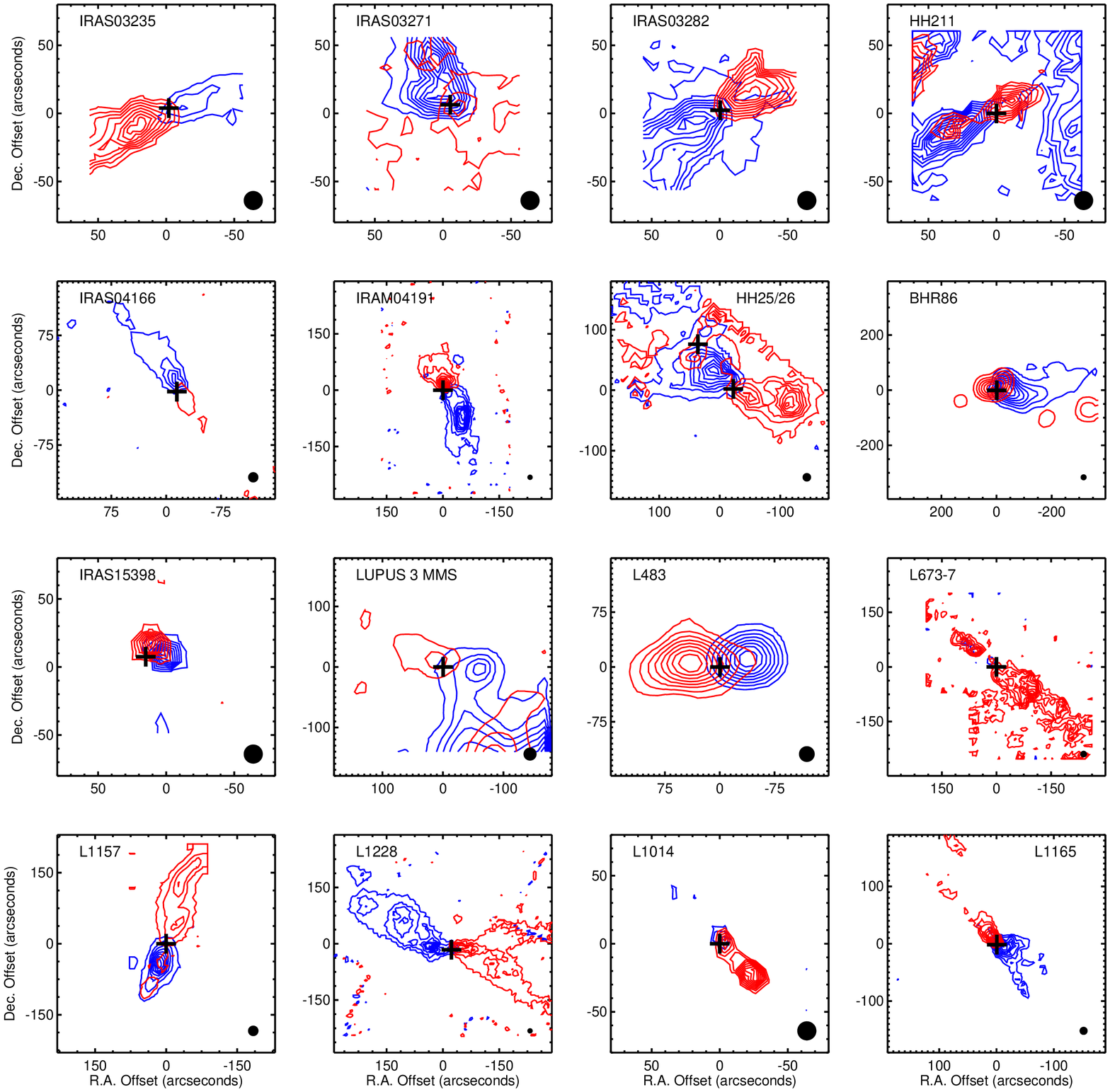}
\caption{\label{fig_redblue32}Integrated intensity contours showing 
blueshifted and redshifted emission for the 17 outflows mapped in \cojthree.  
Each panel is labeled with the name(s) of the source(s).  The minimum and 
maximum velocities over which the emission is integrated are symmetrical about 
the rest velocity, and are chosen based on visual inspection of the velocity 
channels (see \S \ref{sec_mass_dynamical}).  In each panel we plot eight 
contour levels linearly spaced between three times the rms noise in each image 
and the maximum; the minimum contour and contour spacing are listed in Table 
\ref{tab_contours}.  The driving sources are marked with crosses and the beam 
sizes are shown as black ellipses in the lower right of each panel.}
\end{figure*}

\subsection{Data Reduction}\label{sec_data_reduction}

Low-order polynomial baselines were subtracted from all of the raw data using 
the default software package for each telescope:  Continuum and Line Analysis 
Single-dish Software 
(CLASS\footnote{Available at: http://www.iram.fr/IRAMFR/GILDAS/}) for APEX, 
CSO, SRAO, and SMT, NEWSTAR\footnote{Available at: 
http://alma.mtk.nao.ac.jp/aste/guide/reduction/ \\index.html} for ASTE, and 
Starlink\footnote{Available at: 
http://www.jach.hawaii.edu/JCMT/spectral\_line/ \\data\_reduction/acsisdr/basics.html} for the JCMT.  These packages were then used to combine together all 
observations of a particular source and write out FITS datacubes on grids of 
Nyquist sampled spatial pixels.  Further analysis was performed using custom 
IDL procedures.

\section{Results}\label{sec_results}

\begin{deluxetable}{lcc}
\tabletypesize{\scriptsize}
\tablewidth{0pt}
\tablecaption{\label{tab_contours}Integrated Intensity Contour Levels for Figures \ref{fig_redblue21} and \ref{fig_redblue32}} 
\tablehead{
\colhead{} & \colhead{Minimum Contour Level} & \colhead{Contour Step} \\
\colhead{Source} & \colhead{(K \kms)} & \colhead{(K \kms)} 
}
\startdata
\multicolumn{3}{c}{\cojtwo\ (Figure \ref{fig_redblue21})} \\
\hline
IRAS 03235$+$3004           & 0.60 & 0.51 \\
IRAS 03282$+$3035           & 1.77 & 3.60 \\
HH211                       & 0.54 & 0.47 \\
L1709-SMM1                  & 1.12 & 1.27 \\
L1709-SMM5                  & 1.12 & 1.27 \\
CB68                        & 0.69 & 0.13 \\
Aqu-MM2                     & 2.81 & 2.16 \\
Aqu-MM3                     & 2.81 & 2.16 \\
Aqu-MM5                     & 2.81 & 2.16 \\
SerpS-MM13                  & 2.72 & 2.68 \\
CrA-IRAS32                  & 1.40 & 0.60 \\
L673-7                      & 0.80 & 0.94 \\
B335                        & 0.65 & 0.95 \\
L1152                       & 0.42 & 0.23 \\
L1157                       & 1.97 & 5.76 \\
L1165                       & 0.42 & 0.14 \\
L1251A-IRS3                 & 0.12 & 0.57 \\
\hline
\multicolumn{3}{c}{\cojthree\ (Figure \ref{fig_redblue32})} \\
\hline
IRAS 03235$+$3004          & 0.35 & 0.36 \\
IRAS 03271$+$3013          & 0.77 & 1.03 \\
IRAS 03282$+$3035          & 2.90 & 2.46 \\
HH211                      & 1.53 & 0.56 \\
IRAS 04166$+$2706          & 0.48 & 0.59 \\
IRAM 04191$+$1522          & 1.70 & 2.42 \\
HH25                       & 3.29 & 12.57 \\
HH26                       & 3.29 & 12.57 \\
BHR86                      & 0.80 & 1.02 \\
IRAS 15398$-$3359          & 0.90 & 1.09 \\
Lupus 3 MMS                & 0.57 & 0.90 \\
L483                       & 0.27 & 0.17 \\
L673-7                     & 0.87 & 0.32 \\
L1157                      & 3.83 & 10.21 \\
L1228                      & 4.35 & 7.86 \\
L1014                      & 0.20 & 0.05 \\
L1165                      & 0.78 & 0.44
\enddata
\end{deluxetable}

Figures \ref{fig_redblue21} and \ref{fig_redblue32} present integrated 
redshifted and blueshifted emission for each of the outflows mapped in 
\cojtwo\ and \cojthree, respectively, with the contour levels listed in Table 
\ref{tab_contours}.  The minimum and maximum velocities 
over which the emission is integrated are symmetrical about the rest velocity 
for the redshifted and blueshifted emission, and are chosen based on visual 
inspection of the velocity channels (see \S \ref{sec_mass_dynamical}).  As 
is evident from these Figures, we detect outflows from all of our targets.  
While most of these outflows have been mapped by previous authors (see 
Appendix \ref{sec_appendix_sources}), our results presented here represent 
the first complete molecular outflow maps published in the literature for 
Lupus 3 MMS, CrA-IRAS32, and L1152, and the first detection of the L1014 
molecular outflow with a single-dish facility.

\subsection{Outflow Geometrical Properties}\label{sec_geometry}

\begin{deluxetable}{lcc}
\tabletypesize{\scriptsize}
\tablewidth{0pt}
\tablecaption{\label{tab_outflow_geometry}Outflow Geometrical Properties}  
\tablehead{
\colhead{} & \colhead{$R_{\rm lobe}$} & \colhead{P.A.} \\
\colhead{Source} & \colhead{(AU)} & \colhead{(degrees)}
}
\startdata
\multicolumn{3}{c}{\cojtwo} \\
\hline
IRAS 03235$+$3004           & $\geq$4.5\ee{4} & 107     \\
IRAS 03282$+$3035           & 6.1\ee{4} & 126     \\
HH211                       & 1.5\ee{4} & 125     \\
L1709-SMM1                  & 1.6\ee{4} & 70      \\
L1709-SMM5\tablenotemark{a} & \nodata   & \nodata \\
CB68                        & 2.3\ee{4} & 135     \\
Aqu-MM2                     & 2.3\ee{4} & 45      \\
Aqu-MM3                     & 3.0\ee{4} & 107     \\
Aqu-MM5                     & 5.5\ee{4} & 77      \\
SerpS-MM13                  & $\geq$7.9\ee{4} & 78      \\
CrA-IRAS32                  & 2.1\ee{4} & 45      \\
L673-7                      & 4.5\ee{4} & 54      \\
B335                        & 4.9\ee{4} & 95      \\
L1152                       & $\geq$6.9\ee{4} & 35      \\
L1157                       & 5.5\ee{4} & 160     \\
L1165                       & 6.0\ee{4} & 45      \\
L1251A-IRS3                 & 7.6\ee{4} & 5       \\
\hline
\multicolumn{3}{c}{\cojthree} \\
\hline
IRAS 03235$+$3004          & $\geq$1.4\ee{4} & 115      \\
IRAS 03271$+$3013          & $\geq$1.4\ee{4} & 45       \\
IRAS 03282$+$3035          & $\geq$1.4\ee{4} & 115      \\
HH211                      & 1.4\ee{4} & 126      \\
IRAS 04166$+$2706          & 1.5\ee{4} & 40       \\
IRAM 04191$+$1522          & 2.2\ee{4} & 22       \\
HH25                       & 2.2\ee{4} & 155      \\
HH26                       & 6.2\ee{4} & 75       \\
BHR86                      & 4.0\ee{4} & 90       \\
IRAS 15398$-$3359          & 3.0\ee{3} & 58       \\
Lupus 3 MMS                & 2.1\ee{4} & 81       \\
L483                       & 2.2\ee{4} & 93       \\
L673-7                     & 5.5\ee{4} & 53       \\
L1157                      & 5.2\ee{4} & 160      \\
L1228                      & 6.1\ee{4} & 60       \\
L1014                      & 1.5\ee{4} & 45       \\
L1165                      & 3.8\ee{4} & 45       
\enddata
\tablenotetext{a}{Outflow geometrical properties are not possible to determine due to the pole-on geometry of this outflow.}
\end{deluxetable}

We measure and tabulate two geometrical properties of each outflow in Table 
\ref{tab_outflow_geometry}: the average length of each outflow and the 
outflow position 
angle.  The lobe length is measured by hand with a ruler using the integrated 
intensity contour maps presented in Figures \ref{fig_redblue21} and 
\ref{fig_redblue32}, and the value reported in Table 
\ref{tab_outflow_geometry} is the mean of the red and blue lobes.  The 
position angle is measured by hand with a protractor as the angle east of 
north, also using Figures \ref{fig_redblue21} and \ref{fig_redblue32}.  We 
estimate a typical measurement uncertainty of $\sim$2500 AU\footnote{This 
uncertainty is based on assuming a typical uncertainty of 10\as\ (approximately 
one-half to one-third of the 20--30\as\ beam sizes of most observations 
presented here), at a typical distance of 250 pc.  This leads to fractional 
uncertainties in lobe length (and quantities that depend on lobe length, as 
discussed below in \S \ref{sec_mass_dynamical}) of less than 20\% for all but 
one source, thus these uncertainties are negligible compared to the other 
effects explored in \S \ref{sec_correct}.} for the lobe 
length and $\sim$5\degree\ for the position angle, and we list lower limits 
for the lobe lengths for outflows that clearly extend beyond the edges of our 
maps.  We do not report values for either quantity for L1709-SMM2 due to the 
apparent pole-on geometry of this outflow, as inferred from the integrated 
intensity map.

With the relatively low spatial resolution of our single dish data, several 
outflows are either unresolved in width (direction perpendicular to the 
outflow axis) or only marginally resolved.  As a consequence we do not report 
opening angles for the outflows since many such measurements would be biased 
to larger angles.  Measurements of opening angles are better suited to 
interferometer studies of outflows, where the spatial resolution is high enough 
in most cases to resolve the outflows both along and perpendicular to their 
axes \citep[e.g.,][]{arce2006:outflows}.  

\subsection{Outflow Masses and Dynamical Properties}\label{sec_mass_dynamical}

We calculate the masses of the outflows, \mflow, and their dynamic 
properties (momentum, \pflow, kinetic energy, \eflow, luminosity, 
\lflow, and force, \fflow). 
%
For each outflow, we first calculate the column density of H$_2$, 
$N_{\mathrm{H_2}}$, within each velocity channel in each pixel.  Since some maps 
contain multiple outflows and most maps have increased noise near their edges, 
we only consider spatial pixels within regions drawn to encompass the 
outflow lobes.  
Assuming optically thin, LTE emission, 
$N_{\mathrm{H_2}} = f(J,T_{\rm ex},X_{\rm CO}) (\int \tmb \, \mathrm{d}$$ v)$, 
where $f(J,T_{\rm ex},X_{\rm CO})$ is a function of the quantum number of the 
lower state, $J$, the excitation temperature of the 
outflowing gas, $T_{\rm ex}$, and the CO abundance relative to \hh, 
$X_{\rm CO}$ (see Appendix \ref{sec_appendix_equations}).  
The assumed excitation temperature is 50 K and is 
discussed in \S \ref{sec_correct_temperature} below.  The integral 
$\int \tmb \, \mathrm{d}$$ v$ is over the velocity channel and is given by the 
main-beam temperature in that channel multipled by the channel width.  
We assume a standard CO abundance relative to \hh\ of $X_{\rm CO} = 10^{-4}$, 
which is generally uncertain to within about a factor of three 
\citep[e.g.,][]{frerking1982:cox,lacy1994:cox,hatchell2007:outflows}.

\begin{deluxetable*}{lccccccccc}
\tabletypesize{\scriptsize}
\tablewidth{0pt}
\tablecaption{\label{tab_outflow_dynamics}Uncorrected Outflow Dynamical Properties}  
\tablehead{
\colhead{} & \colhead{$v_{\rm min}$\tablenotemark{a}} & \colhead{$v_{\rm max}$\tablenotemark{a}} & \colhead{\mflow} & \colhead{\pflow} & \colhead{\eflow}& \colhead{$\tau_{\rm d}$} & \colhead{\lflow} & \colhead{\fflow} & \\
\colhead{Source} & \colhead{(\kms)} & \colhead{(\kms)} & \colhead{(\msun)} & \colhead{(\msun\ \kms)} & \colhead{(ergs)} & \colhead{(yr)} & \colhead{(\lsun)} & \colhead{(\msun\ \kms\ yr$^{-1}$)}
}
\startdata
\multicolumn{9}{c}{\cojtwo} \\
\hline
IRAS 03235$+$3004\tablenotemark{b}           & 2.0 & 4.5  & $\geq$1.1\ee{-2} & $\geq$2.7\ee{-2} & $\geq$7.2\ee{41} & 4.7\ee{4} & 1.3\ee{-4} & 5.7\ee{-7} \\
IRAS 03282$+$3035           & 6.0 & 19.0 & 4.7\ee{-2} & 4.3\ee{-1} & 4.3\ee{43} & 1.5\ee{4} & 2.3\ee{-2} & 2.8\ee{-5} \\
HH211                       & 2.9 & 6.0  & 2.8\ee{-3} & 1.0\ee{-2} & 4.0\ee{41} & 1.2\ee{4} & 2.8\ee{-4} & 8.7\ee{-7} \\
L1709-SMM1                  & 1.5 & 2.3  & 9.1\ee{-4} & 1.6\ee{-3} & 2.7\ee{40} & 3.3\ee{4} & 6.8\ee{-6} & 4.7\ee{-8} \\
L1709-SMM5\tablenotemark{b,c} & 2.0 & 5.1  & $\geq$7.8\ee{-3} & $\geq$2.3\ee{-2} & $\geq$6.9\ee{41} & \nodata   & \nodata    & \nodata    \\
CB68                        & 1.0 & 1.6  & 6.4\ee{-4} & 7.6\ee{-4} & 9.2\ee{39} & 6.8\ee{4} & 1.1\ee{-6} & 1.1\ee{-8} \\
Aqu-MM2                     & 3.0 & 9.6  & 1.4\ee{-2} & 6.9\ee{-2} & 3.7\ee{42} & 1.1\ee{4} & 2.7\ee{-3} & 6.0\ee{-6} \\
Aqu-MM3                     & 3.0 & 7.1  & 3.3\ee{-2} & 1.5\ee{-1} & 6.7\ee{42} & 2.0\ee{4} & 2.8\ee{-3} & 7.3\ee{-6} \\
Aqu-MM5                     & 3.0 & 7.4  & 6.5\ee{-3} & 2.6\ee{-2} & 1.1\ee{42} & 3.5\ee{4} & 2.5\ee{-4} & 7.3\ee{-7} \\
SerpS-MM13\tablenotemark{b} & 5.5 & 13.0 & $\geq$7.4\ee{-2} & $\geq$5.3\ee{-1} & $\geq$4.0\ee{43} & 2.9\ee{4} & 1.1\ee{-2} & 1.8\ee{-5} \\
CrA-IRAS32                  & 2.0 & 3.8  & 1.6\ee{-3} & 3.8\ee{-3} & 9.5\ee{40} & 2.6\ee{4} & 3.0\ee{-5} & 1.5\ee{-7} \\
L673-7                      & 3.0 & 7.5  & 2.0\ee{-2} & 7.8\ee{-2} & 3.2\ee{42} & 2.8\ee{4} & 9.3\ee{-4} & 2.7\ee{-6} \\
B335                        & 1.0 & 5.5  & 1.5\ee{-2} & 2.9\ee{-2} & 6.6\ee{41} & 4.2\ee{4} & 1.3\ee{-4} & 6.9\ee{-7} \\
L1152\tablenotemark{b}      & 2.0 & 3.5  & $\geq$1.4\ee{-2} & $\geq$3.3\ee{-2} & $\geq$7.6\ee{41} & 9.4\ee{4} & 6.7\ee{-5} & 3.5\ee{-7} \\
L1157                       & 2.0 & 22.0 & 1.4\ee{-1} & 9.4\ee{-1} & 8.7\ee{43} & 1.2\ee{4} & 6.1\ee{-2} & 7.9\ee{-5} \\
L1165                       & 2.0 & 3.0  & 8.9\ee{-3} & 2.1\ee{-2} & 5.1\ee{41} & 9.5\ee{4} & 4.4\ee{-5} & 2.2\ee{-7} \\
L1251A-IRS3                 & 2.3 & 5.3  & 2.6\ee{-2} & 8.7\ee{-2} & 3.0\ee{42} & 6.8\ee{4} & 3.6\ee{-4} & 1.3\ee{-6} \\
\hline
\multicolumn{9}{c}{\cojthree} \\
\hline
IRAS 03235$+$3004\tablenotemark{b}           & 2.6 & 4.3  & $\geq$4.4\ee{-4} & $\geq$1.4\ee{-3} & $\geq$4.6\ee{40} & 1.5\ee{4} & 2.5\ee{-5} & 9.1\ee{-8} \\
IRAS 03271$+$3013\tablenotemark{b}           & 1.8 & 4.9  & $\geq$1.8\ee{-3} & $\geq$4.9\ee{-3} & $\geq$1.4\ee{41} & 1.4\ee{4} & 8.6\ee{-5} & 3.6\ee{-7} \\
IRAS 03282$+$3035\tablenotemark{b}           & 3.0 & 9.9  & $\geq$9.3\ee{-3} & $\geq$4.4\ee{-2} & $\geq$2.2\ee{42} & 6.7\ee{3} & 2.7\ee{-3} & 6.5\ee{-6} \\
HH211                       & 2.0 & 2.7  & 9.4\ee{-4} & 2.1\ee{-3} & 4.9\ee{40} & 2.5\ee{4} & 1.7\ee{-5} & 8.7\ee{-8} \\
IRAS 04166$+$2706           & 2.0 & 2.5  & 3.1\ee{-4} & 7.1\ee{-4} & 1.7\ee{40} & 2.0\ee{4} & 4.8\ee{-6} & 2.5\ee{-8} \\
IRAM 04191$+$1522           & 2.0 & 7.7  & 5.4\ee{-3} & 1.8\ee{-2} & 6.8\ee{41} & 1.4\ee{4} & 4.2\ee{-4} & 1.3\ee{-6} \\
HH25                        & 4.0 & 10.5 & 1.8\ee{-2} & 8.5\ee{-2} & 4.5\ee{42} & 9.9\ee{3} & 3.7\ee{-3} & 8.5\ee{-6} \\
HH26                        & 4.0 & 24.5 & 2.7\ee{-1} & 2.0\ee{0}  & 1.8\ee{44} & 1.2\ee{4} & 1.3\ee{-1} & 1.6\ee{-4} \\
BHR86                       & 2.0 & 5.6  & 1.3\ee{-2} & 3.7\ee{-2} & 1.1\ee{42} & 3.4\ee{4} & 2.7\ee{-4} & 1.1\ee{-6} \\
IRAS 15398$-$3359           & 2.0 & 4.9  & 1.9\ee{-4} & 5.9\ee{-4} & 1.9\ee{40} & 2.9\ee{3} & 5.3\ee{-5} & 2.0\ee{-7} \\
Lupus 3 MMS                 & 2.0 & 4.0  & 1.7\ee{-3} & 4.4\ee{-3} & 1.1\ee{41} & 2.5\ee{4} & 3.8\ee{-5} & 1.8\ee{-7} \\
L483                        & 5.3 & 8.9  & 1.2\ee{-3} & 8.0\ee{-3} & 5.3\ee{41} & 1.2\ee{4} & 3.8\ee{-4} & 6.8\ee{-7} \\
L673-7                      & 2.0 & 3.6  & 4.0\ee{-3} & 9.4\ee{-3} & 2.3\ee{41} & 7.2\ee{4} & 2.6\ee{-5} & 1.3\ee{-7} \\
L1157                       & 1.4 & 8.3  & 4.0\ee{-2} & 1.3\ee{-1} & 5.0\ee{42} & 3.0\ee{4} & 1.4\ee{-3} & 4.3\ee{-6} \\
L1228                       & 2.0 & 12.0 & 6.7\ee{-2} & 2.7\ee{-1} & 1.3\ee{43} & 2.4\ee{4} & 4.4\ee{-3} & 1.1\ee{-5} \\
L1014                       & 1.3 & 3.0  & 9.3\ee{-5} & 1.7\ee{-4} & 3.1\ee{39} & 2.4\ee{4} & 1.1\ee{-6} & 7.2\ee{-9} \\
L1165                       & 1.6 & 4.0  & 2.4\ee{-3} & 5.9\ee{-3} & 1.5\ee{41} & 4.5\ee{4} & 2.8\ee{-5} & 1.3\ee{-7} 
\enddata
\tablenotetext{a}{$v_{\rm min}$ and $v_{\rm max}$ are measured relative to the ambient cloud velocity of each source.  They are the same for both blueshfited and redshifted emission since we adopt symmetrical velocity intervals (see text in \S \ref{sec_mass_dynamical} for details.)}
\tablenotetext{b}{The calculated values of \mflow, \pflow, and \eflow\ are lower limits only since the outflows extend beyond the mapped areas.}
\tablenotetext{c}{Properties that require measurement of outflow lobe length ($\tau_{\rm d}$ and thus \lflow\ and \fflow) cannot be calculated due to the pole-on geometry of this outflow.}
\end{deluxetable*}

The mass within each velocity channel in each pixel is then calculated as 
$M_{\rm v, pixel} = \mu_{\rm H_2} m_{\rm H} N_{\mathrm{H_2}} A_{\rm pixel}$, where 
$m_{\rm H}$ is the mass of a hydrogen atom, $\mu_{\rm H_2}$ is the mean 
molecular weight per hydrogen molecule ($\mu_{\rm H_2} = 2.8$ for gas composed 
of 71\% hydrogen, 27\% helium, and 2\% metals by mass 
\citep{kauffmann2008:mambo}), and $A_{\rm pixel}$ is the area of each pixel.  
The total mass of each outflow is then obtained by summing $M_{\rm v, pixel}$ 
over all velocity and spatial pixels encompassing the outflow.  
The velocities of integration are assumed to be symmetrical about the rest 
velocities and are chosen based on visual inspection of channel maps for 
each outflow.  To determine the lower bound of integration, which we define as 
$v_{\rm min}$, we select the lowest-velocity redshifted and blueshifted 
channels where the ambient cloud emission drops below 3$\sigma$ (measured at 
locations outside of the outflow lobes to avoid issues with separating 
ambient cloud and outflow emission).  Since we adopt symmetrical velocity 
limits for blueshifted and redshfited emission, the larger of these two 
(measured relative to rest) is 
then taken to be $v_{\rm min}$ and is listed in the second column of Table 
\ref{tab_outflow_dynamics}.  To determine the upper bound of integration, 
which we define as $v_{\rm max}$, we select the highest-velocity redshifted 
and blueshifted channels where outflowing gas is detected above 3$\sigma$.  
The larger of these two (again, measured relative to rest) 
is taken to be $v_{\rm max}$ and is listed in the 
third column of Table \ref{tab_outflow_dynamics}.  Average spectra for each 
outflow, with the ambient cloud velocity, $v_{\rm min}$, and $v_{\rm max}$ 
indicated, are shown in Appendix \ref{sec_appendix_spectra}.

Some ambient cloud emission 
is still apparent in Figures \ref{fig_redblue21} and \ref{fig_redblue32} 
since some of the channels above the lower bound contain ambient emission 
below 3$\sigma$ that integrates to levels above the 3$\sigma$ rms of the 
integrated maps.  None of this ambient emission is 
included in our calculations of outflow masses and dynamical properties since 
we first cut out all emission below 3$\sigma$ in each velocity channel before 
calculating these properties.  
The calculated masses (assuming a temperature of 50 K; see 
\S \ref{sec_correct_temperature}) are listed in the fourth column of Table 
\ref{tab_outflow_dynamics}.

The momentum and kinetic energy within each velocity channel in each pixel are 
calculated as $P_{\rm v, pixel} = M_{\rm v, pixel} \times v$ and 
$E_{\rm v, pixel} = \frac{1}{2} M_{\rm v, pixel} \times v^2$, respectively, where 
$v$ is the velocity of each channel with respect to the systemic velocity.  
The total momentum (\pflow) and kinetic energy (\eflow) of each outflow are 
then calculated by summing over the same velocity and spatial pixels as for the 
mass, and are listed in the fifth and sixth columns of Table 
\ref{tab_outflow_dynamics}.  Some authors instead define the total \pflow\ and 
\eflow\ as the total \mflow\ multiplied by ($v_{\rm char}$) or 
($\frac{1}{2} v_{\rm char}^2$), respectively, where $v_{\rm char}$ is the 
intensity (or mass) weighted outflow velocity 
\citep[e.g.,][]{andre1990:vla1623}.  Such a method is mathematically identical 
for \pflow\ but will underestimate the total \eflow\ since it will not fully 
account for the large fraction of total energy contained in the highest 
velocity gas.  Finally, the luminosity and force of each outflow 
are calculated as \lflow\ $=$ \eflow\ $/$ $\tau_{\rm d}$ and \fflow\ $=$ \pflow 
$/$ $\tau_{\rm d}$, respectively, where $\tau_{\rm d}$ is the dynamical 
time.  It is calculated as $\tau_{\rm d} = R_{\rm lobe} / v_{\rm max}$, with 
$R_{\rm lobe}$ and $v_{\rm max}$ (the average length of the red and blue lobes 
and the maximum velocity at which outflowing gas is detected above 3$\sigma$, 
respectively) listed in Tables \ref{tab_outflow_geometry} 
and \ref{tab_outflow_dynamics}.  The seventh, eighth, and ninth 
columns list $\tau_{\rm d}$, \lflow, and \fflow, respectively.  Note that 
the dynamical time of an outflow likely underestimates its true age due to 
rapid acceleration and decelaration in the outflow 
\citep{parker1991:outflows,masson1992:outflows,masson1993:outflows}.

Inspection of Figures \ref{fig_redblue21} and \ref{fig_redblue32} clearly 
show that some outflows extend beyond the mapped areas.  The values of 
\mflow, \pflow, and \eflow\ that we calculate for these outflows are thus 
lower limits and marked as such in Table \ref{tab_outflow_dynamics}.  
The calculated values of \lflow\ and \fflow, however, are reliable measures 
of the total outflow luminosities and driving forces as long as the energy 
and momentum injection rates are assumed to be constant over the lifetime 
of the detectable outflow.

\section{Correction Factors to Outflow Properties}\label{sec_correct}

In the above section, we calculated the masses and dynamical properties of 
the outflows studied here assuming optically thin, LTE emission at an 
excitation temperature of 50 K, and only integrating channels at velocities 
larger than those in which ambient cloud emission is detected.  Below we 
attempt to quantify the correction factors that must be applied to the 
values obtained under these simple assumptions.

\begin{figure}
\epsscale{1.2}
\plotone{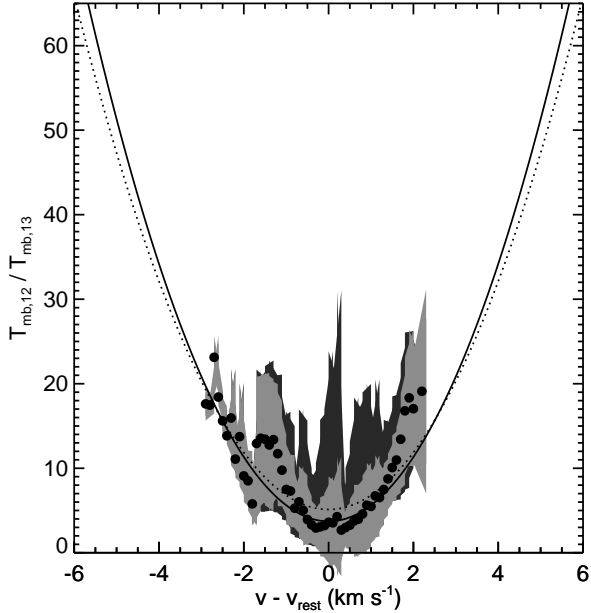}
\caption{\label{fig_opac21}$T_{\rm mb,12} / T_{\rm mb,13}$ 
as a function of velocity from 
rest for \cojtwo\ and \coojtwo.  Plotted are the mean ratio 
determined from all of the pointed observations (circles), the standard 
deviation at each velocity (light gray shading), the full extent 
at each velocity (dark gray shading), the best-fit 
second-order polynomial (solid black line), and the best-fit second-order 
polynomial after excluding all velocity channels with 
$|(v - v_{\rm rest})| < 1$ \kms\ (dotted line).}
\end{figure}

\subsection{Opacity}\label{sec_correct_opacity}

Numerous studies have established that the line-wings 
of outflows are typically optically thick in low-J transitions of 
\co\ \citep[e.g.,][]{goldsmith1984:outflows,cabrit1992:outflows,bally1999:outflows,arce2001:outflows,curtis2010:outflows}.  
Using our \coo\ data obtained as described above, we follow a standard method 
of correcting the outflow masses and dynamical properties 
\citep[e.g.,][]{goldsmith1984:outflows,curtis2010:outflows}.  Assuming that 
both \co\ and \coo\ are in LTE at the same excitation temperature, and 
further assuming identical beam-filling factors, the ratio of brightness 
temperatures between the two isotopologues is given as
\begin{equation}\label{eq_tau1}
\frac{T_{\rm mb,12}}{T_{\rm mb,13}} = \frac{1 - e^{-\tau_{12}}}{1 - e^{-\tau_{13}}} \qquad ,
\end{equation}
where $T_{\rm mb,12}$ and $T_{\rm mb,13}$ are the observed \co\ and \coo\ 
brightness temperatures, respectively, 
and $\tau_{12}$ and $\tau_{13}$ are the opacities of the \co\ and \coo\ 
transitions.  Assuming that the \coo\ is optically thin, Equation 
\ref{eq_tau1} can be rewritten as 
\begin{equation}\label{eq_tau2}
\frac{T_{\rm mb,12}}{T_{\rm mb,13}} = \frac{[^{12}\rm{CO}]}{[^{13}\rm{CO}]} \frac{1 - e^{-\tau_{12}}}{\tau_{12}}\qquad ,
\end{equation}
where $[^{12}\rm{CO}] / [^{13}\rm{CO}]$ is the 
abundance ratio, which is taken to be 62 \citep{langer1993:abundance}.  
Using this expression, $\tau_{12}$ can be determined numerically from the 
observed ratio $T_{\rm mb,12} / T_{\rm mb,13}$, and then the correction factor 
$\tau_{12} / ( 1 - e^{-\tau_{12}})$ can be applied to the \co\ data to correct the 
observed brightness temperatures to the values they would have in the optically 
thin limit.  As noted by \citet{toolsofradioastronomy}, 
Equation \ref{eq_tau2} overestimates the ratio of brightness temperatures by 
an amount that increases with $\tau_{12}$, due to the increasingly invalid 
assumption that the \coo\ is optically thin.  The most optically thick 
outflows in our sample have $\tau_{12} \sim 10-20$ (as derived below), which 
leads to overestimates in Equation \ref{eq_tau2} of 5\% -- 15\%.  These 
overestimates are small enough to have no significant effect on our results.

\subsubsection{$J=2-1$}\label{sec_opacity_coj2}

\begin{figure}
\epsscale{1.2}
\plotone{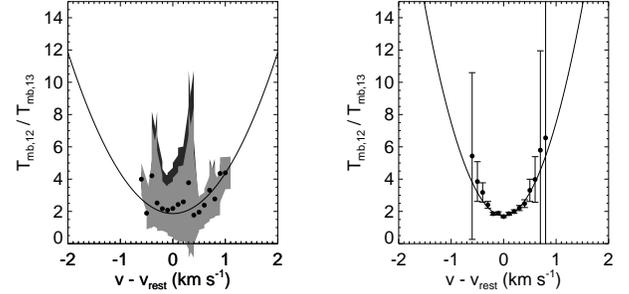}
\caption{\label{fig_opac32}$T_{\rm mb,12} / T_{\rm mb,13}$ 
as a function of velocity from 
rest for \cojthree\ and \coojthree.  {\it Left:}  Plotted are the mean ratio 
determined from all of the pointed observations (circles), the standard 
deviation at each velocity (light gray shading), the full extent 
at each velocity (dark gray shading), and the best-fit 
second-order polynomial (solid black line).  Only ratios used in the polynomial 
fit are plotted (see text for details).  {\it Right:}  Plotted is the 
ratio determined from the average \cojthree\ and \coojthree\ L673-7 spectra 
(circles with error bars), and the best-fit second-order polynomial.}
\end{figure}

As listed in Table \ref{tab_cso_13co}, we obtained pointed \coojtwo\ 
observations toward 17 positions in five different outflows.  To facilitate 
comparison between \co\ and \coo, we also obtained pointed \cojtwo\ 
observations toward these same positions on the same nights to remove 
uncertainties introduced by different telescope beams, efficiencies, and 
weather conditions.  For each of these 17 positions, we calculate 
$T_{\rm mb,12} / T_{\rm mb,13}$ as a function of velocity measured relative to 
rest (such that the systemic core velocity is equal to 0 \kms) for each 
velocity where both lines are detected at or above 3$\sigma$.

Since we only have select pointed \coo\ observations (obtaining full \coo\ maps 
of multiple outflows 
to the depths required to detect line-wings from outflows is prohibitively 
expensive in terms of telescope time), we average each of the 
$T_{\rm mb,12} / T_{\rm mb,13}$ 
from above to obtain a mean ratio in each velocity bin.  Following 
\citet{arce2001:outflows}, we then use linear least-squares to 
fit a second order polynomial to the mean 
ratio versus velocity, constrained to reach a minimum at rest (zero) 
velocity.  This allows us to correct for opacity even at velocities where the 
line-wings were not detected in \coo.  
For the fit we only consider velocities within $\pm$4 \kms\ 
from rest that have two or more measurements of the ratio.

\begin{deluxetable*}{lccccccccccccccc}
\tabletypesize{\scriptsize}
\tablewidth{0pt}
\tablecaption{\label{tab_outflow_corrections}Outflow Correction Factors}  
\tablehead{
\colhead{} & \multicolumn{5}{c}{Opacity} & \multicolumn{5}{c}{Low-Velocity} & \multicolumn{5}{c}{Sensitivity} \\
\colhead{Source} & \colhead{\mflow} & \colhead{\pflow} & \colhead{\eflow} & \colhead{\lflow} & \colhead{\fflow} & \colhead{\mflow} & \colhead{\pflow} & \colhead{\eflow} & \colhead{\lflow} & \colhead{\fflow} & \colhead{\mflow} & \colhead{\pflow} & \colhead{\eflow} & \colhead{\lflow} & \colhead{\fflow} 
}
\startdata
\multicolumn{16}{c}{\cojtwo} \\
\hline
IRAS 03235$+$3004                    & 4.2     & 4.1     & 3.9     & 3.8     & 4.2     & 6.7     & 4.4     & 2.9     & 2.8     & 4.3     & 1.6     & 1.6     & 1.6     & 1.8     & 1.8     \\
IRAS 03282$+$3035                    & 1.0     & 1.0     & 1.0     & 1.0     & 1.0     & 7.6     & 3.8     & 2.1     & 2.2     & 3.8     & 1.2     & 1.3     & 1.5     & 2.0     & 1.8     \\
HH211                                & 2.4     & 2.4     & 2.1     & 2.1     & 2.3     & 14.4    & 8.0     & 4.7     & 4.7     & 8.1     & 1.1     & 1.2     & 1.4     & 2.4     & 2.1     \\
L1709-SMM1\tablenotemark{a}          & 7.3     & 6.9     & 7.0     & 7.1     & 7.2     & \nodata & \nodata & \nodata & \nodata & \nodata & 1.3     & 1.4     & 1.4     & 1.4     & 1.4     \\
L1709-SMM2\tablenotemark{b}          & 3.8     & 3.5     & 3.2     & \nodata & \nodata & 1.6     & 1.3     & 1.2     & \nodata & \nodata & 1.2     & 1.2     & 1.2     & \nodata & \nodata \\
CB68\tablenotemark{c}                & 10.6    & 10.5    & 10.3    & 10.0    & 10.9    & \nodata & \nodata & \nodata & \nodata & \nodata & 2.4     & 2.3     & 2.2     & 2.9     & 2.8     \\
Aqu-MM2\tablenotemark{a}             & 1.6     & 1.4     & 1.4     & 1.4     & 1.5     & \nodata & \nodata & \nodata & \nodata & \nodata & 1.3     & 1.4     & 1.3     & 1.4     & 1.3     \\
Aqu-MM3                              & 1.7     & 1.6     & 1.6     & 1.6     & 1.6     & 2.6     & 1.8     & 1.3     & 1.3     & 1.7     & 1.1     & 1.2     & 1.1     & 1.5     & 1.5     \\
Aqu-MM5                              & 2.2     & 2.0     & 1.9     & 1.9     & 2.1     & 9.2     & 5.0     & 2.9     & 3.0     & 5.0     & 1.4     & 1.5     & 1.5     & 2.1     & 2.0     \\
SerpS-MM13\tablenotemark{a}          & 1.0     & 1.0     & 1.0     & 1.0     & 1.0     & \nodata & \nodata & \nodata & \nodata & \nodata & 1.2     & 1.2     & 1.3     & 1.5     & 1.3     \\
CrA-IRAS32\tablenotemark{a}          & 4.6     & 4.5     & 4.4     & 4.3     & 4.4     & \nodata & \nodata & \nodata & \nodata & \nodata & 1.5     & 1.6     & 1.6     & 2.0     & 1.8     \\
L673-7\tablenotemark{a}              & 2.3     & 2.2     & 2.0     & 1.9     & 2.1     & \nodata & \nodata & \nodata & \nodata & \nodata & 1.2     & 1.2     & 1.2     & 1.3     & 1.3     \\
B335\tablenotemark{c,d}              & 6.6     & 5.9     & 5.0     & 4.9     & 5.8     & \nodata & \nodata & \nodata & \nodata & \nodata & \nodata & \nodata & \nodata & \nodata & \nodata \\
L1152                                & 5.1     & 4.8     & 4.7     & 4.8     & 4.9     & 7.0     & 4.7     & 3.3     & 3.4     & 4.8     & 1.5     & 1.6     & 1.6     & 2.2     & 2.1     \\
L1157                                & 2.2     & 1.5     & 1.1     & 1.2     & 1.5     & 1.8     & 1.2     & 1.1     & 1.0     & 1.2     & 1.1     & 1.1     & 1.4     & 1.5     & 1.3     \\
L1165                                & 4.8     & 4.8     & 4.5     & 4.5     & 4.5     & 2.5     & 1.9     & 1.6     & 1.6     & 2.0     & 1.5     & 1.6     & 1.7     & 2.1     & 2.1     \\
L1251A-IRS3\tablenotemark{a}         & 3.1     & 2.8     & 2.6     & 2.6     & 2.8     & \nodata & \nodata & \nodata & \nodata & \nodata & 1.4     & 1.4     & 1.4     & 1.4     & 1.4     \\
\hline
\multicolumn{16}{c}{\cojthree} \\
\hline
IRAS 03235$+$3004\tablenotemark{a}   & 2.5     & 2.5     & 2.4     & 2.4     & 2.4     & \nodata & \nodata & \nodata & \nodata & \nodata & 1.0     & 1.0     & 1.2     & 1.6     & 1.4     \\
IRAS 03271$+$3013                    & 2.0     & 1.8     & 1.6     & 1.5     & 1.8     & 5.0     & 3.7     & 2.8     & 2.9     & 3.7     & 1.0     & 1.1     & 1.3     & 1.7     & 1.3     \\
IRAS 03282$+$3035                    & 1.0     & 1.0     & 1.0     & 1.0     & 1.0     & 5.5     & 3.1     & 2.0     & 1.9     & 3.1     & 1.3     & 1.4     & 1.6     & 2.2     & 2.0     \\
HH211                                & 2.1     & 2.2     & 2.0     & 2.1     & 2.2     & 12.9    & 9.6     & 7.7     & 7.4     & 9.5     & 2.2     & 2.6     & 3.4     & 5.7     & 4.5     \\
IRAS 04166$+$2706\tablenotemark{a,d} & 2.1     & 2.1     & 2.0     & 2.1     & 2.1     & \nodata & \nodata & \nodata & \nodata & \nodata & \nodata & \nodata & \nodata & \nodata & \nodata \\
IRAM 04191$+$1522\tablenotemark{d}   & 1.4     & 1.3     & 1.2     & 1.2     & 1.4     & 5.6     & 3.5     & 2.4     & 2.4     & 3.5     & \nodata & \nodata & \nodata & \nodata & \nodata \\
HH25\tablenotemark{a,d}              & 1.0     & 1.0     & 1.0     & 1.0     & 1.0     & \nodata & \nodata & \nodata & \nodata & \nodata & \nodata & \nodata & \nodata & \nodata & \nodata \\
HH26\tablenotemark{a,d}              & 1.0     & 1.0     & 1.0     & 1.0     & 1.0     & \nodata & \nodata & \nodata & \nodata & \nodata & \nodata & \nodata & \nodata & \nodata & \nodata \\
BHR86\tablenotemark{d}               & 1.8     & 1.6     & 1.5     & 1.5     & 1.6     & 6.0     & 4.1     & 3.1     & 3.0     & 4.0     & \nodata & \nodata & \nodata & \nodata & \nodata \\
IRAS 15398$-$3359\tablenotemark{a}   & 1.5     & 1.4     & 1.3     & 1.3     & 1.4     & \nodata & \nodata & \nodata & \nodata & \nodata & 1.3     & 1.4     & 1.5     & 1.7     & 1.6     \\
Lupus 3 MMS\tablenotemark{d}         & 2.0     & 1.8     & 1.8     & 1.7     & 1.8     & 9.2     & 6.0     & 4.1     & 4.2     & 6.0     & \nodata & \nodata & \nodata & \nodata & \nodata \\
L483\tablenotemark{d}                & 1.0     & 1.0     & 1.0     & 1.0     & 1.0     & 23.1    & 13.2    & 7.0     & 6.9     & 13.2    & \nodata & \nodata & \nodata & \nodata & \nodata \\
L673-7\tablenotemark{a}              & 2.1     & 2.0     & 1.9     & 1.9     & 2.0     & \nodata & \nodata & \nodata & \nodata & \nodata & 1.2     & 1.4     & 1.6     & 2.4     & 2.1     \\
L1157\tablenotemark{a}               & 2.1     & 1.5     & 1.3     & 1.3     & 1.6     & \nodata & \nodata & \nodata & \nodata & \nodata & 1.8     & 2.8     & 5.8     & 14.4    & 6.8     \\
L1228\tablenotemark{a,d}             & 1.3     & 1.2     & 1.1     & 1.1     & 1.2     & \nodata & \nodata & \nodata & \nodata & \nodata & \nodata & \nodata & \nodata & \nodata & \nodata \\
L1014\tablenotemark{a,d}             & 3.3     & 3.3     & 3.2     & 3.2     & 3.2     & \nodata & \nodata & \nodata & \nodata & \nodata & \nodata & \nodata & \nodata & \nodata & \nodata \\
L1165\tablenotemark{d}               & 2.2     & 2.0     & 1.8     & 1.8     & 2.0     & 10.8    & 7.0     & 5.1     & 5.1     & 7.2     & \nodata & \nodata & \nodata & \nodata & \nodata \\
\hline
\hline
Mean                                 & 2.8     & 2.6     & 2.5     & 2.4     & 2.6     & 7.7     & 4.8     & 3.3     & 3.4     & 5.1     & 1.4     & 1.5     & 1.7     & 2.6     & 2.1     \\
Median                               & 2.1     & 2.0     & 1.8     & 1.7     & 1.8     & 6.7     & 4.1     & 2.9     & 3.0     & 4.3     & 1.3     & 1.4     & 1.5     & 2.0     & 1.8     \\
Minimum                              & 1.0     & 1.0     & 1.0     & 1.0     & 1.0     & 1.6     & 1.2     & 1.0     & 1.0     & 1.2     & 1.0     & 1.0     & 1.1     & 1.3     & 1.3     \\
Maximum                              & 10.6    & 10.5    & 10.3    & 10.0    & 10.9    & 23.1    & 13.2    & 7.7     & 7.4     & 13.2    & 2.4     & 2.8     & 5.8     & 14.4    & 6.8     
\enddata
\tablenotetext{a}{No reliable Gaussian fit to the ambient cloud emission within 1 \kms\ of the rest velocity can be obtained.}
\tablenotetext{b}{Properties that require measurement of outflow lobe length ($\tau_{\rm d}$ and thus \lflow\ and \fflow) cannot be calculated due to the pole-on geometry of this outflow.}
\tablenotetext{c}{No low-velocity corrections are given because the minimum velocity over which the outflow emission is integrated is 1.0 \kms.}
\tablenotetext{d}{No corrections for sensitivity are given since the native resolution of the map is already 0.5 \kms\ or lower.}
\end{deluxetable*}

Figure \ref{fig_opac21} plots the mean $T_{\rm mb,12} / T_{\rm mb,13}$ as 
a function of velocity from rest for all points used in the fit, 
as well as the resulting second-order polynomial fit.  The fit is described 
by the equation 
\begin{equation}\label{eq_opac21}
T_{\rm mb,12} / T_{\rm mb,13} = (1.90 \pm 0.09) \, (\rm{v - v_{rest}})^2 + (3.72 \pm 0.48) \, ,
\end{equation} 
and has a reduced $\chi^2$ of 0.53.  Since it is possible that the lowest 
velocity channels are optically thick 
even in \coo, we also repeated the fit after 
excluding all channels where $|(v - v_{\rm rest})| < 1$ \kms.  The resulting fit 
is also plotted in Figure \ref{fig_opac21} and is within the uncertainties 
of the original fit and thus has no significant effect on our results.  
To correct all of our \cojtwo\ data for 
opacity, we take, at each velocity, the smaller of either the polynomial fit 
or 62 (the abundance ratio), use this value to calculate $\tau_{12}$ at 
each velocity numerically using Equation \ref{eq_tau2}, and then apply the 
velocity-dependent correction factor $\tau_{12} / ( 1 - e^{-\tau_{12}})$ to our 
data.

\subsubsection{$J=3-2$}\label{sec_opacity_coj3}

We obtained pointed \coojthree\ observations toward five positions in 
three different outflows, again 
with corresponding pointed \cojthree\ observations to facilitie comparison 
between the two isotopologues.  The decreased number of positions in the 
3--2 transition (five) compared to the 2--1 transition (17) was due to 
less available time in the required weather conditions.  As above, 
we average the results from each pointed observation to determine 
a mean $T_{\rm mb,12} / T_{\rm mb,13}$ for each velocity, and then fit a second-order 
polynomial, constrained to reach its minimum at rest (zero) velocity, to all 
velocities within $\pm$4 \kms\ from rest that have two or more individual 
measurements of $T_{\rm mb,12} / T_{\rm mb,13}$.  The resulting mean $T_{\rm mb,12} / T_{\rm mb,13}$ 
and best-fit polynomial are shown in the left panel of Figure 
\ref{fig_opac32}.  The fit is described by the equation
\begin{equation}\label{eq_opac32a}
I_{12} / I_{13} = (2.50 \pm 0.55) \, (\rm{v - v_{rest}})^2 + (1.87 \pm 0.27) \, ,
\end{equation} 
and has a reduced $\chi^2$ of 0.27.

\begin{figure*}
\epsscale{1.2}
\plotone{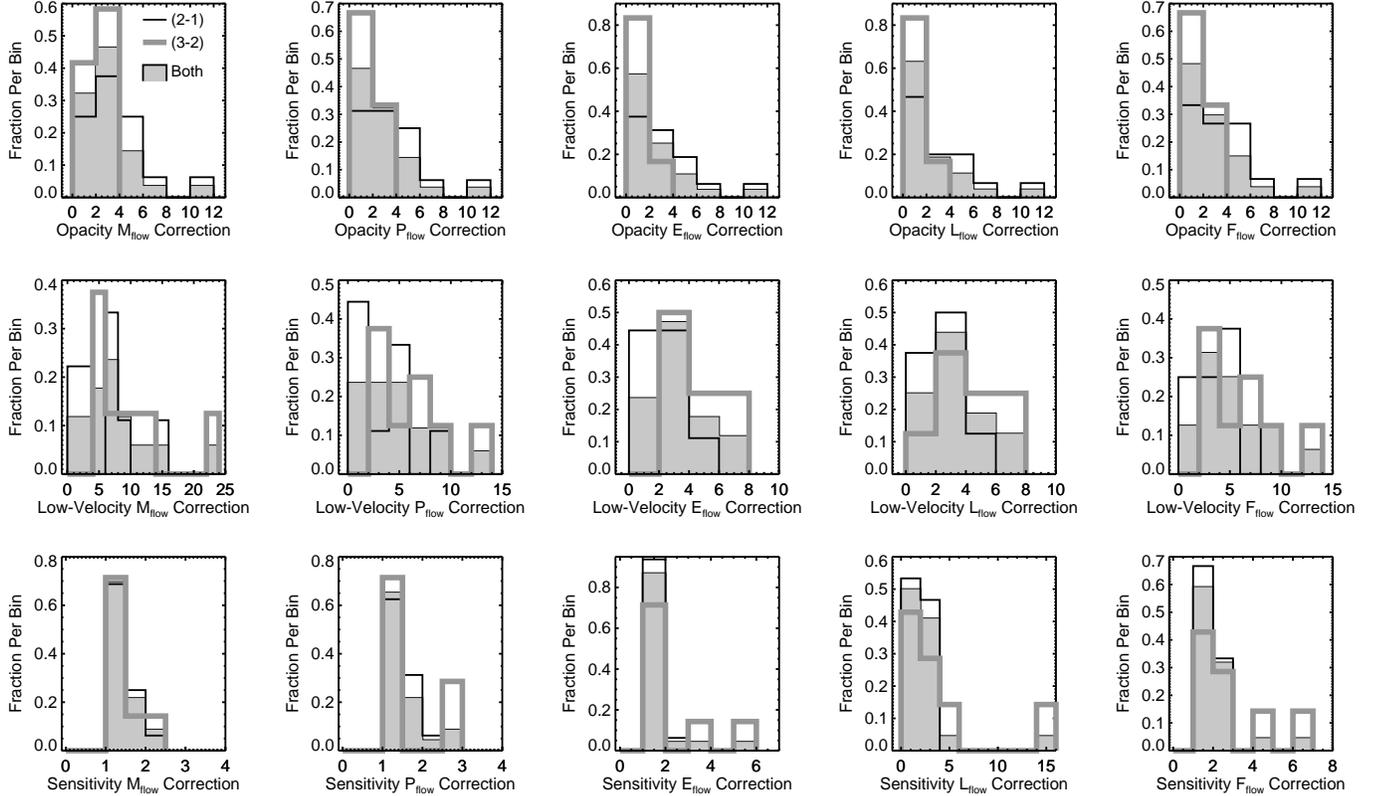}
\caption{\label{fig_hist_correct}Histograms showing the distributions of 
correction factors listed in Table \ref{tab_outflow_corrections}.  The top 
panels show the corrections for opacity, the middle panels show the corrections 
for low-velocity emission, and the bottom panels show the corrections for 
sensitivity.  From left-to-right, the panels plot the corrections for 
\mflow, \pflow, \eflow, \lflow, and \fflow.  The thin black histogram plots 
the corrections derived for the outflows mapped in \cojtwo, the thick gray 
histogram plots those derived for the outflows mapped in \cojthree, and the 
shaded histograms show the combined sample.}
\end{figure*} 

Given that only five pointings went into deriving this fit, we caution that it 
is extremely uncertain.  Indeed, inspection of Figure \ref{fig_opac32} 
shows that it is not even clear if a second-order polynomial is an 
approriate function to fit, and even if it is the fit is certainly not very 
well constrained.  As noted above, we also obtained a \coojthree\ map of the 
full extent of the L673-7 outflow.  This map is not sensitive enough for 
robust line-wing detection at each spatial position, thus instead we calculate 
the average \cojthree\ and \coojthree\ spectrum over all spatial pixels that 
encompass the outflow.  We calculate the ratio $T_{\rm mb,12} / T_{\rm mb,13}$ as a function 
of velocity from these average spectra for all velocities where both spectra 
are detected at or above 3$\sigma$, and again fit a second-order 
polynomial constrained to reach its minimum at rest (zero) velocity.  This 
fit is described by the equation 
\begin{equation}\label{eq_opac32b}
T_{\rm mb,12} / T_{\rm mb,13} = (5.77 \pm 0.68) \, (\rm{v - v_{rest}})^2 + (1.73 \pm 0.03) \, ,
\end{equation} 
and has a reduced $\chi^2$ of 0.48.  
The observed $T_{\rm mb,12} / T_{\rm mb,13}$ and 
best-fit polynomial are shown in the right panel of Figure \ref{fig_opac32}. 

We choose to use the fit to 
$T_{\rm mb,12} / T_{\rm mb,13}$ determined from L673-7 to 
determine opacity corrections for our \cojthree\ data because of the increased 
redundancy in using an entire outflow rather than only five pointings in three 
different outflows, and also because it is much clearer in this case that a 
second-order polynomial provides a good fit to the data.  The procedure for 
using the fit to derive velocity-dependent opacity corrections is the same as 
above for \cojtwo, except now using the L673-7 polynomial fit.  Since the 
L673-7 outflow is a relatively low-mass outflow compared to several others 
considered in this study and thus may be less optically thick, we caution that 
our results may underestimate the magnitude of the opacity corrections for some 
of the more massive outflows mapped in \cojthree.

\subsubsection{Opacity Corrections}\label{sec_correct_opacity_results}

Columns two through six of Table \ref{tab_outflow_corrections} list the 
resulting factors by which \mflow, \pflow, \eflow, \lflow, and \fflow\ 
increase when these opacity corrections are applied, and the top row of 
Figure \ref{fig_hist_correct} shows the distribution of opacity correction 
factors separately for the outflows mapped in each transition and combined.  
For the combined sample, we find that the outflow mass is increased by 
factors ranging from 1.0 to 10.6, with a mean (median) increase of 2.8 (2.1), 
and similar increases for the other properties.  
Using a similar procedure for outflows in Perseus mapped in \cojthree, 
\citet{curtis2010:outflows} found that their outflow masses increase by 
factors ranging from 1.8 to 14.3, with a median of 3.8.  Additionally, 
\citet{cabrit1992:outflows} found opacity corrections ranging from 1.0 to 
8.9, with a mean of 3.5, using a simpler method that applied one correction 
factor at all velocities.  In both cases our results are comparable.

\begin{figure}
\epsscale{1.16}
\plottwo{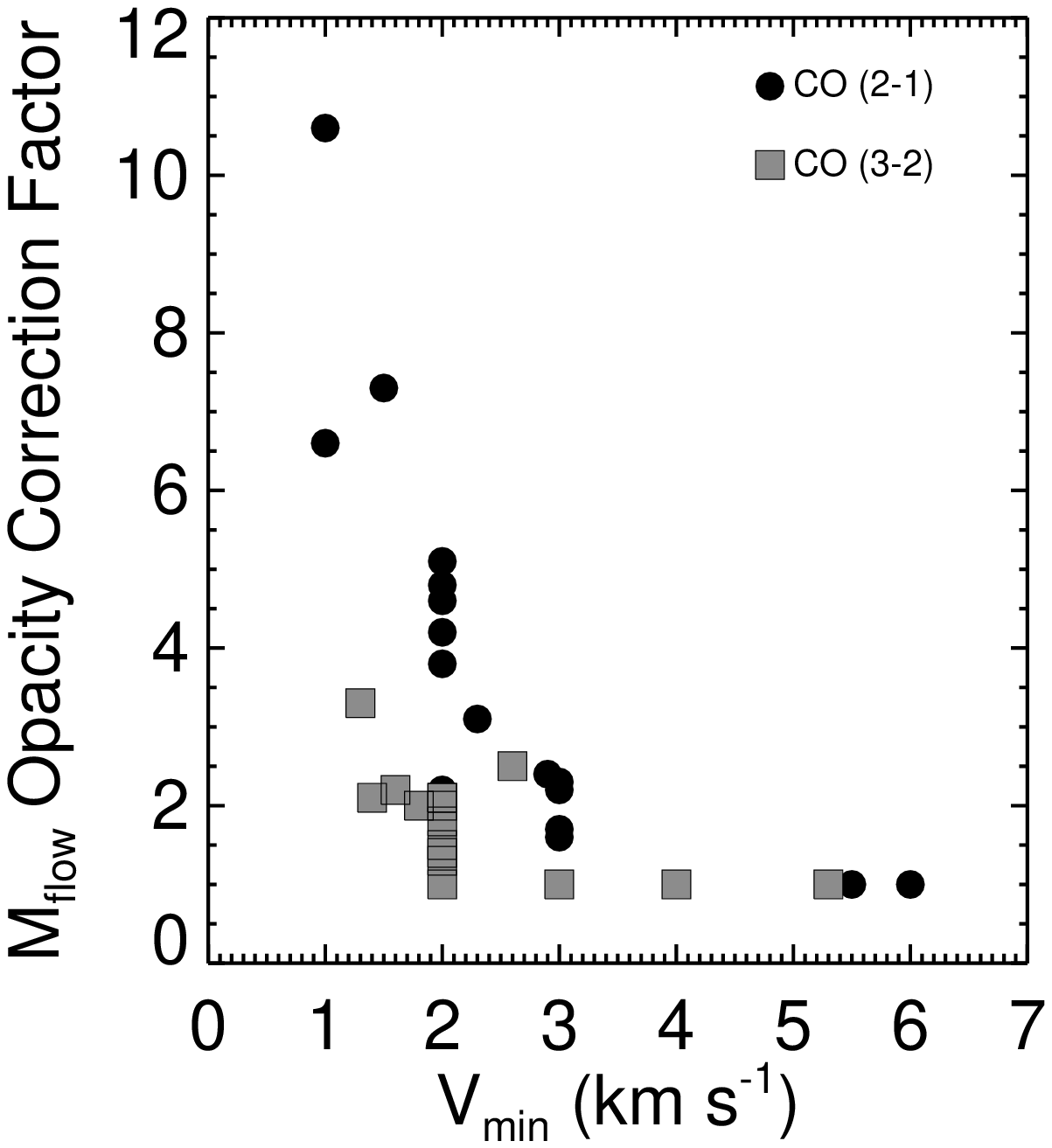}{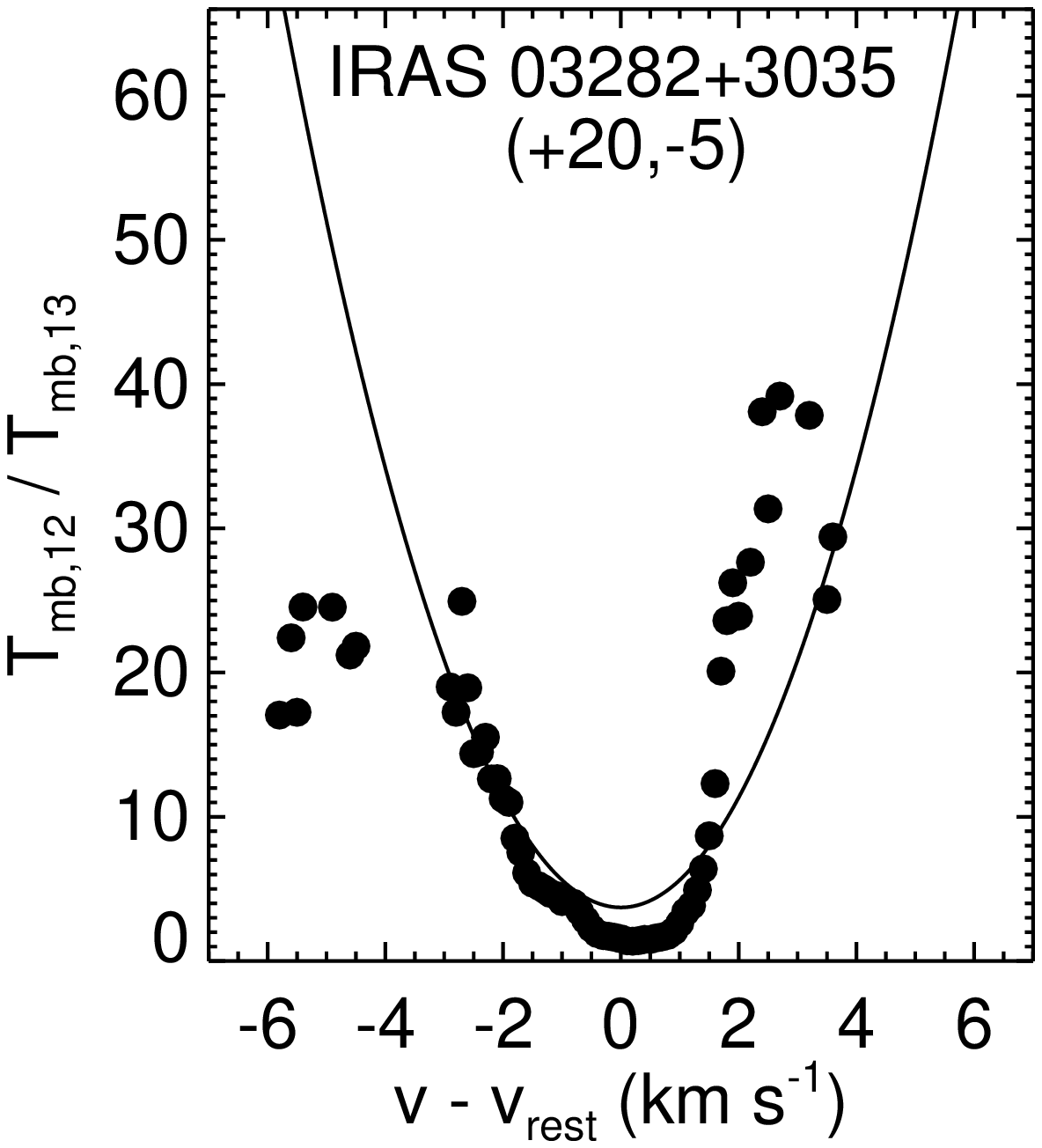}
\caption{\label{fig_opac}{\it Left:}  Total opacity correction factor for 
\mflow\ plotted vs.~the lower bound of the velocity range used to calculate 
the outflow properties, taken from Table \ref{tab_outflow_dynamics}.  Black 
circles show the corrections for outflows mapped in \cojtwo, and gray squares 
show the corrections for outflows mapped in \cojthree.  Similar trends are 
seen for the other outflow dynamical properties (\pflow, \eflow, \lflow, and 
\fflow).  
{\it Right:}  $T_{\rm mb,12} / T_{\rm mb,13}$ determined from observations of 
\cojtwo\ and \coojtwo\ toward one of the positions observed in the IRAS 
03282$+$3035 outflow (circles), with the position labeled in the top center 
of the panel.  The solid black line shows the best-fit second-order polynomial 
to the mean $T_{\rm mb,12} / T_{\rm mb,13}$ from all 17 pointed observations and 
is the same as that displayed in Figure \ref{fig_opac21}.}
\end{figure}

Both our results and previous studies 
\citep[e.g.,][]{cabrit1992:outflows,curtis2010:outflows} find a range in 
opacity correction factors of approximately one order of magnitude.  
Since the velocity-dependent opacity corrections are largest 
at the lowest velocities where the emission is the most optically thick, 
the magnitude of the total correction is 
expected to depend on the lower bound of the velocity range used to calculate 
the outflow properties.  As confirmed by the left panel of Figure 
\ref{fig_opac}, most of the range in total opacity correction factors is 
indeed explained by such a trend.  This trend likely explains why 
\citet{vandermarel2013:outflows} concluded that opacity corrections are 
less than a factor of two and can thus be neglected, since their minimum 
velocities were typically $\sim$3 \kms.

Finally, we end this section by noting that there is some limited evidence 
that our method underestimates the opacity correction factors.  
As seen in Figures \ref{fig_opac21} and \ref{fig_opac32}, the second-order 
polynomial fits to the observed $T_{\rm mb,12} / T_{\rm mb,13}$ reach the abundance ratio 
of 62, implying fully optically thin emission, for all velocities beyond 
$\sim$4--6 \kms\ from rest.  However, the right panel of Figure 
\ref{fig_opac} plots the observed $T_{\rm mb,12} / T_{\rm mb,13}$ for the 
\cojtwo\ and 
\coojtwo\ observations of a position in the IRAS 03282$+$3035 outflow.  This 
is one of the only set of pointed observations where \coo\ is detected at or 
above 3$\sigma$ beyond 4 \kms\ from rest, and in this case 
$T_{\rm mb,12} / T_{\rm mb,13}$ 
at these higher velocities is clearly below the fit, suggesting the emission 
is more optically thick at these velocities than predicted by the fit.  
\coo\ observations with higher sensitivity than those presented here are 
required to test the generality of this result, and such observations should 
be possible in the near future with the Atacama Large Millimeter Array (ALMA) 
and the Cerro Chajnantor Atacama Telescope (CCAT).

\subsection{Excitation Temperature}\label{sec_correct_temperature}

An unknown parameter in the calculation of \mflow\ (and all other dynamical 
properties that depend on mass) is $T_{\rm ex}$, the excitation temperature 
of the outflowing gas.  Most studies adopt values of $T_{\rm ex}$ in the range 
of 10 -- 50 K \citep[e.g.,][]{parker1991:outflows,hatchell2007:outflows,curtis2010:outflows,dunham2010:l6737}.  However, \citet{vankempen2009:hh46,vankempen2009:outflows} 
used multiple transitions of \co\ (up to (6--5)) to derive warmer 
temperatures, in the range of 50--200 K, for a sample of six outflows.  
Similarly high temperatures were found by \citet{yildiz2013:highj} with 
{\it Herschel} high-J \co\ observations up to \cojten.

\begin{figure}
\epsscale{1.2}
\plotone{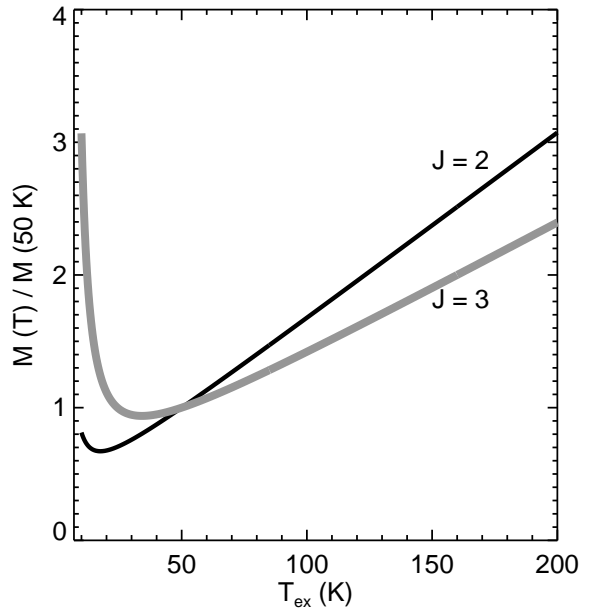}
\caption{\label{fig_masstex}Correction factors for outflow mass (and all other 
properties that depend on mass) for different assumed $T_{\rm ex}$, compared 
to the assumed value of 50 K.  Correction factors for both \cojtwo\ (black 
line) and \cojthree\ (gray line) are plotted.  See Appendix 
\ref{sec_appendix_equations} for details on the calculation.}
\end{figure}

\begin{figure*}
\epsscale{1.2}
\plotone{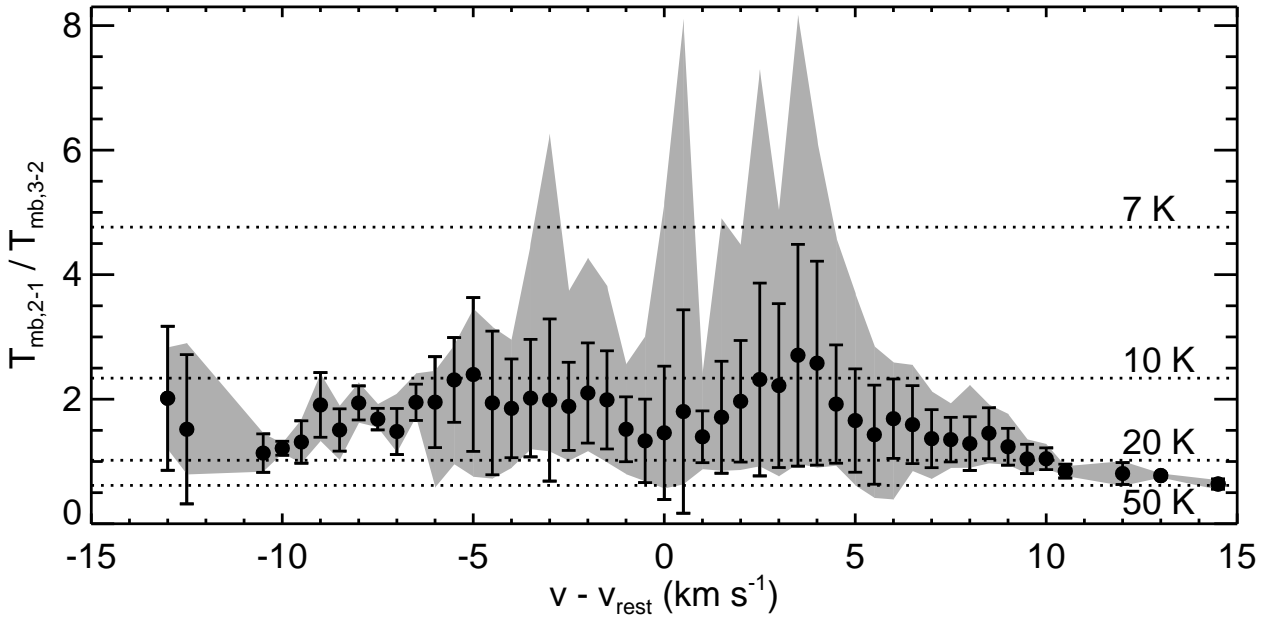}
\caption{\label{fig_ratiostex}The mean value of 
$T_{\rm mb,2-1} / T_{\rm mb,3-2}$ plotted 
versus velocity from rest for the six outflows mapped in both \cojtwo\ and 
\cojthree.  The averages are calculated over 0.5 \kms\ wide bins, with the 
standard deviation in each bin plotted as error bars and the total extent 
in each bin plotted as the gray shaded area.  The line ratios for LTE 
temperatures of 7, 10, 20, and 50 K are marked with dotted lines.}
\end{figure*}

Figure \ref{fig_masstex} shows the factors by which \mflow\ (and all other 
properties that depend on it) would change for $T_{\rm ex} = 10 - 200$ K, 
relative to the assumed value of 50 K (see Appendix 
\ref{sec_appendix_equations} for details on the calculation).  
These factors range from $0.7-3$, depending on transition and $T_{\rm ex}$.  
For both transitions, values above 50 K can only increase outflow properties, 
up to a factor of three compared to the assumption of 
$T_{\rm ex} = 50$ K.  Since we mapped six outflows 
(IRAS 03235$+$3004, IRAS 03282$+$3035, HH211, L673-7, L1157, and L1165) 
in both the (2--1) and (3--2) transitions of \co, here we use our data to 
study the excitation temperatures of these outflows.

For each of the six outflows, we corrected both transitions for opacity using 
our velocity-dependent corrections, re-gridded them onto the same velocity 
grid, convolved the \cojthree\ map with a Gaussian with a FWHM such that the 
output map matches the resolution of the \cojtwo\ map, aligned the 
convolved \cojthree\ and original \cojtwo\ maps onto the same spatial grid, 
calculated the mean spectra in each outflow lobe for each transition, and 
finally calculated $T_{\rm mb,2-1} / T_{\rm mb,3-2}$, 
the ratio of the mean spectra, for each lobe of each outflow.

Figure \ref{fig_ratiostex} displays the mean value of 
$T_{\rm mb,2-1} / T_{\rm mb,3-2}$ 
versus velocity from rest over all six outflows in 0.5 \kms\ bins and shows 
that $T_{\rm mb,2-1} / T_{\rm mb,3-2}$ ranges between $\sim 0.5-3$.  
Assuming LTE, nearly all of the velocities are consistent with $T_{\rm ex}$ 
in the range of 10 -- 20~K.  While there is very weak evidence for higher 
$T_{\rm ex}$ (up to 50 K) at the highest redshifted velocities, in general 
there is no clear trend in $T_{\rm mb,2-1} / T_{\rm mb,3-2}$ (and thus in 
implied $T_{\rm ex}$) with velocity.  In contrast, \citet{yildiz2013:highj} 
found clear evidence for increasing $T_{\rm ex}$ with velocity with higher-J 
{\it Herschel} observations of outflows.

\begin{figure}
\epsscale{1.2}
\plotone{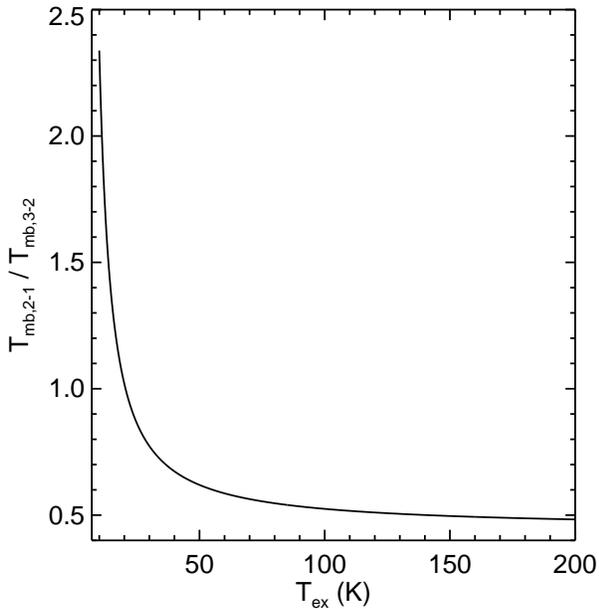}
\caption{\label{fig_lteratio}Expected $T_{\rm mb,2-1} / T_{\rm mb,3-2}$ 
plotted as a function of $T_{\rm ex}$, assuming LTE.  See Appendix 
\ref{sec_appendix_equations} for details on the calculation.}
\end{figure}

Our results seem to imply that the most appropriate assumptions for $T_{\rm ex}$ 
for the outflows studied here are those ranging from 10 to 20 K, which would 
lead to outflow properties that decrease by 20\% -- 30\% for those mapped in 
\cojtwo\ and increase by factors of 1 -- 3 for those mapped in \cojthree, 
compared to the values obtained by assuming $T_{\rm ex} = 50$ K.  
However, we caution that, by only considering \cojtwo\ and \cojthree, we are 
not sensitive to the presence of gas with $T_{\rm ex}$ above $\sim 50$ K, as 
clearly demonstrated by Figure \ref{fig_lteratio}, which shows that the 
line ratios change by only $\sim 0.1$ for $T_{\rm ex}$ between 50 and 200 K.
Higher-J transitions would be required to evaluate the 
existence of warmer gas.  To further reinforce this point, we calculated 
the ratio $T_{\rm mb,2-1} / T_{\rm mb,3-2}$ assuming an equal-mass mixture of 
warm (200 K) and cold (either 10 K or 50 K) gas is observed (see Appendix 
\ref{sec_appendix_equations} for details on the calculation, and note that, 
in the notation of Appendix \ref{sec_appendix_equations}, $A = 1$ for an 
equal-mass mixture of warm and cold gas).  If the resulting ratios 
were then assumed to arise from gas in LTE at a single temperature, the 
derived $T_{\rm ex}$ are 15.5 K for the mixture with cold gas at 10 K, and 
63 K for the mixture with cold gas at 50 K.  The warm, 200 K gas is almost 
completely invisible in the analysis of the ratio of 
$T_{\rm mb,2-1} / T_{\rm mb,3-2}$.

Since \citet{vankempen2009:hh46,vankempen2009:outflows} and 
\citet{yildiz2013:highj} found typical $T_{\rm ex}$ ranging from 50 to 200 K 
with higher-J transitions of \co, 
we adopt 50 K in this paper and note that our results may increase by 
up to factors of 3 if the temperatures are higher.  In reality, the gas 
in molecular outflows may not all be at the same excitation temperature; 
there may be variations both spatially and kinematically, and there may be 
very warm molecular gas in shocks 
\citep[e.g.,][]{green2013:digit,yildiz2013:highj,santangelo2013:herschel}.  
Indeed, \citet{downes2007:outflows} showed that 
the $T_{\rm ex}$ of their simulated outflows increased with increasing 
velocity, and \citet{yildiz2013:highj} found a similar trend in {\it Herschel} 
observations of low-mass protostars.  \citet{downes2007:outflows} 
cautioned that using a single temperature can lead to 
significant underestimates (by up to factors of $3-4$) in the outflow kinetic 
energy and mechanical luminosity, since both quantities depend on the square 
of velocity and thus give the most weight to the highest-velocity gas.  
Since our data do 
not show any clear trend between $T_{\rm ex}$ and velocity, and are generally 
insensitive to the presence of gas above 50 K anyway, we are unable to 
evaluate the effects of such an underestimate on our calculated outflow 
properties.

\begin{figure*}
\epsscale{1.2}
\plotone{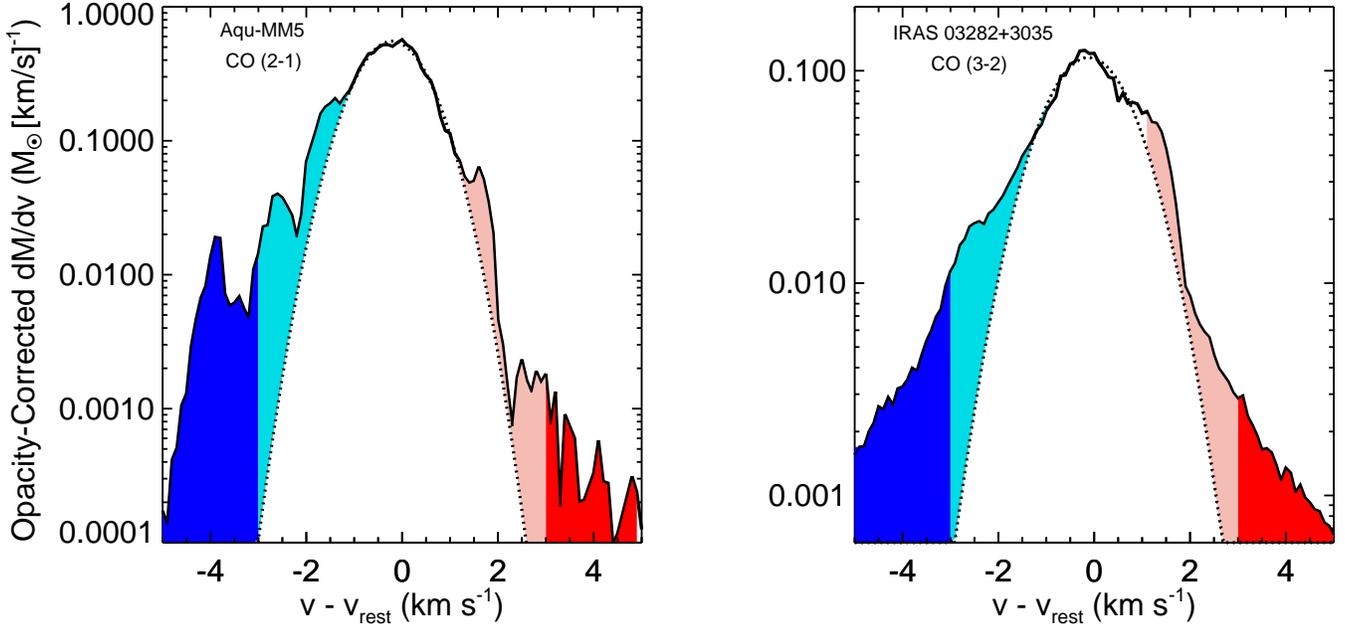}
\caption{\label{fig_lowvel_examples}Total mass spectra, $dM/dv$, for two 
outflows: Aqu-MM5 mapped in \cojtwo\ (left) and IRAS 03282$+$3035 mapped in 
\cojthree\ (right).  The mass spectra are calculated by summing the mass in 
each velocity channel (corrected for opacity using the 
velocity-dependent corrections derived in \S \ref{sec_correct_opacity}) over 
the total extent of the outflow, and are plotted as solid black lines.  The 
dotted lines show the Gaussian fits to all velocities within $\pm 1$ \kms, 
representing the ambient cloud emission.  The dark blue and red shaded areas 
show the total mass calculated by only integrating beyond $v_{\rm min}$ 
(note that a small amount of total mass is not displayed in the shaded regions 
since, for display purposes, both panels cut off at velocities smaller than $
v_{\rm max}$), and the light blue 
and red shaded areas show the extra mass added by integrating the difference 
between the solid and dotted curves between 1 \kms\ and $v_{\rm min}$.}
\end{figure*}

\subsection{Low-Velocity Outflow Emission}\label{sec_correct_lowvel}

To avoid erroneously including ambient cloud emission when calculating \mflow\ 
(and all other parameters that depend on \mflow), 
many studies take $v_{\rm min}$ to be the 
minimum velocity at which such emission is no longer detected, determined 
either by eye (e.g., this study), by comparing \co\ spectra on and off 
the outflow lobes \citep[e.g.,][]{maury2009:outflows}, or assumed to be a 
fixed value \citep[typically 2 \kms; e.g.,][]{hatchell2007:outflows,hatchell2009:outflows,curtis2010:outflows}.  
In this study $v_{\rm min}$ ranges from 1.0 -- 6.0 \kms, with a mean and median 
of 2.5 and 2.0, respectively.  However, since the typical escape velocities 
are much less than 1 -- 6 \kms\ (to give an example, the escape 
velocities from a central mass of 0.5 \msun\ at distances of 5000 -- 50000 AU 
range from 0.4 to 0.1 \kms), only integrating beyond a mean velocity of 2.5 
\kms\ clearly has the potential to miss some of the outflowing gas.  Combined 
with the fact that the mass spectra of molecular outflows steeply rise toward 
lower velocities \citep[e.g., Figure 7 of][]{arce2001:outflows}, it is 
apparent that our calculations likely miss a significant fraction of the total 
outflow mass 
\citep[see also][]{arce2001:outflows,downes2007:outflows,offner2011:outflows}.

To correct for this missing mass, early studies assumed that the intensity 
of the outflow emission is constant over low velocities dominated by ambient 
cloud emission and equal to the mean intensity just outside this velocity 
range \citep[e.g.,][]{bally1983:outflows,margulis1985:outflows}.  However, 
\citet{cabrit1990:outflows} showed that such corrections are arbitrary and 
often overestimate the total outflow mass.  In this study, we instead follow 
a procedure first outlined by 
\citet{arce2001:outflows} and recently adopted by \citet{offner2011:outflows} 
to analyze synthetic observations of simulated outflows.  First, for each 
outflow, we calculate the total mass spectrum, $dM/dv$, by summing 
the mass in each velocity channel (corrected for opacity using the 
velocity-dependent corrections derived in \S \ref{sec_correct_opacity}) over 
the total extent of the outflow.  This mass spectrum is composed of a central 
component arising from the ambient cloud that is approximately described as 
a Gaussian, and broad, high-velocity wings arising from the outflow.  We 
fit a Gaussian to the central component, only considering velocities within 
$\pm 1$ \kms\ from rest for the fit, subtract this Gaussian from the total 
mass spectrum, and then calculate the additional mass added to the outflow 
by integrating the difference for all velocities between 1 \kms\ and 
$v_{\rm min}$.  The extra momentum and 
kinetic energy added to the outflow are calculated in a similar manner, except 
by multiplying the mass in each velocity channel by the appropriate power of 
velocity.  Figure \ref{fig_lowvel_examples} shows two examples of this 
procedure, one for each of the two rotational transitions of \co\ considered 
in this paper.

\begin{figure}
\epsscale{1.2}
\plotone{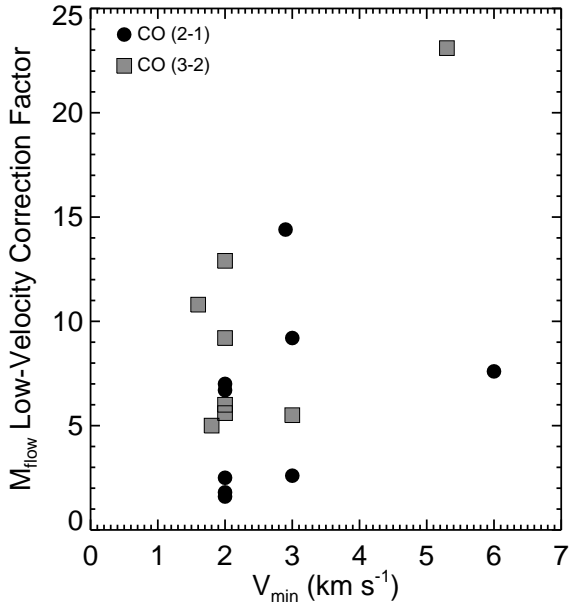}
\caption{\label{fig_lowvel}Total low-velocity correction factor for 
\mflow\ plotted vs.~$v_{\rm min}$, the lower bound of the velocity range 
used to calculate the outflow properties.  Black 
circles show the corrections for outflows mapped in \cojtwo, and gray squares 
show the corrections for outflows mapped in \cojthree.}
\end{figure}

Columns seven through 11 of Table \ref{tab_outflow_corrections} list the 
resulting factors by which \mflow, \pflow, \eflow, \lflow, and \fflow\ 
increase compared to the opacity-corrected values integrated between 
$v_{\rm min}$ and $v_{\rm max}$.  We do not 
list corrections when we can't obtain satisfactory Gaussian fits to the 
ambient cloud emission (usually due to offpositions contaminated with emission 
near the cloud rest velocities) or when $v_{\rm min} = 1$ \kms\ and 
corrections are thus uncessary.  The middle 
row of Figure \ref{fig_hist_correct} shows the distribution of correction 
factors separately for the outflows mapped in each transition and combined.  
For the combined sample, we find that the outflow mass is increased by 
factors ranging from 1.6 to 23.1, with a mean (median) of 7.7 (6.7).  The 
corrections are smaller for the other properties (increases by mean factors 
of 4.8, 3.3, 3.4, and 5.1 for \pflow, \eflow, \lflow, and \fflow, 
respectively), as expected since they depend on velocity to the first 
(\pflow, \fflow) or second (\eflow, \lflow) power and are less affected by 
emission at low velocities.  As demonstrated by Figure \ref{fig_lowvel}, there 
is no significant correlation between $v_{\rm min}$ and the size of the 
correction factors.  However, several outflows with $v_{\rm min} > 2.0$ \kms\ 
are not plotted here since satisfactory Gaussian fits could not be obtained 
due to contaminated off-positions, potentially masking the expected trend 
of increasing correction factors with increasing $v_{\rm min}$.

While our results indicate that significant fractions of the total mass, 
momentum, and energy of outflows can be missed by only integrating above 
a minimum velocity, we stress that the exact factors found here are highly 
uncertain and depend on the ambient cloud mass spectrum being well-fit by a 
simple Gaussian.  Nevertheless, our results are generally consistent with 
those of \citet{offner2011:outflows}, 
who applied the same procedure to their synthetic observations of simulated 
outflows and concluded that only integrating beyond 2 \kms\ from rest could 
lead to underestimates in \mflow\ by factors of 5--10.  However, their results 
were based on only comparing to the total ejected mass in the simulations, 
since they were unable to track the total entrained mass; the true 
underestimates may be even larger.

Finally, we note that both our results and those of 
\citet{offner2011:outflows} are unable to correct for the mass at the lowest 
velocities (in our case, within $\pm 1$ \kms\ from rest).  
Using a very different method based on comparing the spectra at each position 
in an outflow to a reference spectrum constructed from nearby, off-outflow 
positions, both \citet{maury2009:outflows} and \citet{vandermarel2013:outflows} 
did correct for missing mass all the way down to the ambient cloud velocity.  
\citet{maury2009:outflows} found that \mflow\ increases by factors ranging from 
3.9 to 42.1, with a mean (median) of 15.1 (12.7).  
These corrections, which they stress should be treated as upper limits, are 
approximately a factor of two larger than our mean and median corrections.  
On the other hand, \citet{vandermarel2013:outflows} found that \fflow\ 
only increases by factors that are generally less than $\sim 2$ (they do 
not discuss corrections for \mflow), lower than found either by us or by 
\citet{maury2009:outflows} and \citet{offner2011:outflows}.  At present we 
do not have a satisfactory explanation for this discrepancy and note this 
remains an open question subject to further study.  
While the exact corrections remain quite uncertain and dependent on the exact 
procedure used to develop them, our findings coupled with those of most other 
recent studies indicate that adopting minimum velocities 
for integrating outflow properties can lead to significant underestimates.

\subsection{Sensitivity}\label{sec_correct_velres}

With the very high spectral resolution of many of our maps, we 
can evaluate whether high-velocity outflow emission below the 
sensitivities of our observations affects our results.  While such emission 
is unlikely to significantly affect the total \mflow\ due to the steeply 
declining nature of outflow mass spectra (see \S \ref{sec_correct_lowvel} and 
Figure \ref{fig_lowvel_examples}), it may affect the total \pflow\ and 
\eflow, which are more heavily weighted toward the highest-velocity emission.  
To evaluate this effect, we smoothed each map with a native 
$\delta v \leq 0.2$ \kms\ down to $\delta v = 0.5$ \kms\ and recalculated 
the outflow properties, with 0.5 \kms\ chosen as the best compromise between 
increasing the sensitivity in high-velocity channels and retaining sufficient 
velocity resolution to fully resolve the kinematic structure of the outflows.  
Since $\delta v = 0.5$ \kms\ is too low of a velocity resolution to 
reliably fit to the ambient cloud emission at $\pm 1$ \kms, we only integrated 
for velocities above $v_{\rm min}$ and compared to the values obtained 
from the higher resolution maps over the same velocity range.

\begin{figure}
\epsscale{1.2}
\plotone{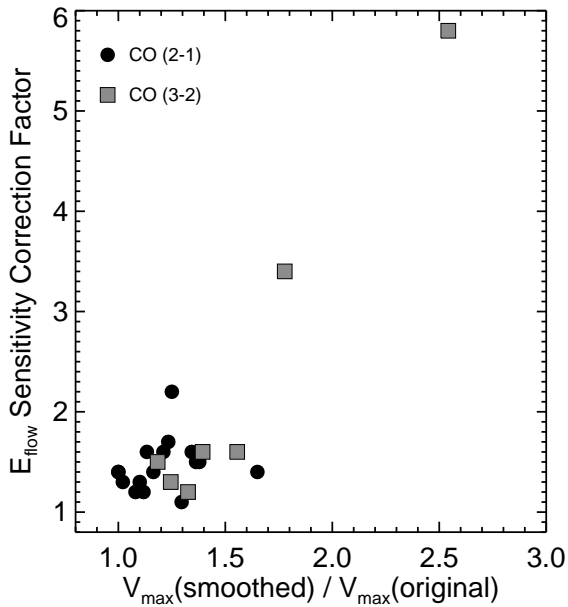}
\caption{\label{fig_velres}Total sensitivity correction factor for 
\eflow\ plotted versus the ratio of maximum velocities of detected emission 
in the maps smoothed to 0.5 \kms\ resolution to those in the original maps.  
Black circles show the corrections for outflows mapped in \cojtwo, and gray 
squares show the corrections for outflows mapped in \cojthree.}
\end{figure}

Columns 12 through 16 of Table \ref{tab_outflow_corrections} list the 
resulting factors by which \mflow, \pflow, \eflow, \lflow, and \fflow\ 
increase when these sensitivity corrections are made.  These factors are 
multiplicative with those listed in other columns and discussed in previous 
Sections.  The bottom row of Figure \ref{fig_hist_correct} shows the 
distribution of correction factors separately for the outflows mapped in each 
transition and combined.  For the combined sample, we find that the outflow 
mass is only increased by factors ranging from 1.0 to 2.4, with a mean 
(median) increase of 1.4 (1.3).  However, as expected, the corrections are 
larger for \pflow\ and \eflow\ (increases by mean factors of 1.5 and 1.7, 
respectively, and maximum increases by factors of 2.8 and 5.8, respectively), 
since both are weighted to higher-velocity emission.  
The corrections for \lflow\ and \fflow\ are even larger (increases by mean 
factors of 2.6 and 2.1), since their numerators increase while their 
denominators ($t_{\rm dyn}$) simultaneously decrease due to increases in 
$v_{\rm max}$.  Figure \ref{fig_velres} shows that larger correction 
factors are derived for larger increases in the maximum velocity at which 
outflow emission is detected between the original and smoothed maps.

These results emphasize that significant underestimates in the kinematic 
properties of outflows are possible when lacking sufficient sensitivity to 
detect the highest-velocity emisison.  Furthermore, they disagree with 
those of \citet{vandermarel2013:outflows}, who argued that the sensitivity 
and spectral binning of the observations do not significantly affect the 
calculated properties.  We note that 
qualitatively similar results to our own were found by 
\citet{arce2013:hh46}, who used ALMA observations of HH 46/47 to detect 
outflow emission at higher velocities than 
previously detected in single-dish observations with lower sensitivity, 
and calculated correction factors of about 1, 4, and 11 for 
\mflow, \pflow, and \eflow, respectively (Arce 2013, priv.~comm.).  
While it is impossible to quantify the magnitude of 
this effect for all cases, since it depends on the sensitivity of the 
observations, we note that the sensitivity of our observations are generally 
comparable to those of other large, single-dish surveys of molecular outflows 
\citep[e.g.,][]{bontemps1996:outflows,hatchell2007:outflows,hatchell2009:outflows,maury2009:outflows,curtis2010:outflows}.  Future studies should carefully 
evaluate the magnitude of this effect in their data.

Finally, we note that the extremely high velocity (EHV) components of 
molecular outflows that are common in outflows driven by massive protostars 
\citep[e.g.,][]{choi1993:ehv}, typically at velocities in excess of 
50 \kms\ from rest, are also sometimes found in outflows driven by low-mass 
protostars \citep[e.g.,][]{tafalla2004:iras04166}.  
As they are often both compact 
and weak, the beam dilution from our single dish observations with low spatial 
resolution render them undetectable in our data \citep[as confirmed by nondetections of EHV components for IRAS 03271$+$3013, IRAS 03282$+$3035, HH211, or IRAS 04166$+$2706, all of which are known to have such components;][]{bachiller1991:iras03282,gueth1999:hh211,tafalla2004:iras04166}.  While such components increase the total \mflow\ by negligible 
amounts, they can contain up to $2-4$ times as much momentum and energy as the 
lower-velocity outflow components \citep[e.g.,][]{tafalla2004:iras04166}.  
Sensitive interferometer observations with high spatial resolution are 
needed to search for EHV components missed by our maps.

\subsection{Other Possible Corrections}\label{sec_correct_other}

\begin{deluxetable}{lcccc}
\tabletypesize{\scriptsize}
\tablewidth{0pt}
\tablecaption{\label{tab_inclination}Inclination Corrections for Motions Along Jet Axis}
\tablehead{
\colhead{} & \colhead{Inclination} & \multicolumn{3}{c}{Corrections} \\
\colhead{Quantity} & \colhead{Dependence} & \colhead{$\langle i \rangle = 57.3$\degree} & \colhead{$i = 15$\degree} & \colhead{$i = 85$\degree}
}
\startdata
$R_{\rm lobe}$ & $1 / {\rm sin} \, i$ & 1.2 & 11.5 & 1.0 \\
$\tau_{\rm d}$ & ${\rm cos} \, i / {\rm sin} \, i$ & 0.6 & 11.4 & 0.09 \\
\mflow\ & \nodata & \nodata & \nodata & \nodata \\
\pflow\ & $1 / {\rm cos} \, i$ & 1.9 & 1.0 & 11.5 \\
\eflow\ & $1 / {\rm cos}^2 \, i$ & 3.4 & 1.10& 131.6 \\
\lflow\ & ${\rm sin} \, i / {\rm cos}^3 \, i$ & 5.3 & 0.09 & 1504.7 \\
\fflow\ & ${\rm sin} \, i / {\rm cos}^2 \, i$ & 2.9 & 0.09 & 131.1 
\enddata
\end{deluxetable}

Since we can only measure the radial component of the total velocity of 
outflowing gas and the projection of the outflow lobe size on the plane of the 
sky, corrections for source inclination, $i$, are necessary, where $i$ is the 
angle between the rotation/outflow axis and the observer ($i = 0$\degree\ 
corresponds to a pole-on system, and $i = 90$\degree\ corresponds to an 
edge-on system).  The second column of Table \ref{tab_inclination} lists 
the inclination dependence for each outflow property for outflows where all 
of the motion is along the jet axis.  Since we are unable to measure opening 
angles of the outflows mapped here, we are also unable to derive reliable 
inclination constraints (see \S \ref{sec_geometry}).  Thus, Table 
\ref{tab_inclination} lists the correction factors for a mean inclination angle 
$\langle i \rangle = 57.3$\degree\ (assuming all orientations are equally 
favorable) and for nearly pole-on (5\degree) and nearly edge-on (85\degree) 
inclinations.  For the mean inclination angle, \pflow, \eflow, \lflow, and 
\fflow\ increase, on average, by factors of 1.9, 3.4, 5.3, and 2.9, 
respectively.  The correction factors for \pflow\ and 
\eflow\ are always greater than or equal to 1.0 for all possible inclinations.  
For \lflow\ and \fflow, they are greater than or equal to 1.0 for 
$i \geq 38.2$\degree\ and $i \geq 24.4$\degree, respectively.  Since the 
probabilities of viewing sources at lower inclinations are only 21\% and 
18\%, respectively, these corrections are greater than or equal to 1.0 the 
majority of the time.

The corrections listed in Table \ref{tab_inclination} are only valid for 
outflows where all of the motion is along the jet axis.  Using simulations, 
\citet{downes2007:outflows} also investigated inclination corrections taking 
into account transverse motions due to sideways expansion.  They showed that 
the correction factor of $1 / {\rm cos} \, i$ for \pflow\ always overestimates 
the true momentum.  They found that, by coincidence, the uncorrected \pflow\ 
always agrees with the true value to within a factor of two since 
underestimates of the momentum along the jet axis are canceled by overestimates 
due to the erroneous inclusion of transverse momentum.  Similarly, they also 
showed that the correction factor of $1 / {\rm cos}^2 \, i$ for \eflow\ 
also overestimates the true energy.  Unlike for momentum, however, the 
uncorrected values of \eflow\ do still underestimate the total energy for 
many inclinations.  However, we note that these results only apply for 
outflows from Class 0 protostars that are driven solely by jets and which have 
not yet broken out of their parent clouds, so they may not apply to all of the 
outflows studied here.

\begin{deluxetable*}{lccccccccc}
\tabletypesize{\scriptsize}
\tablewidth{0pt}
\tablecaption{\label{tab_outflow_dynamics_corrected}Corrected Outflow Dynamical Properties}  
\tablehead{
\colhead{} & \colhead{$v_{\rm min}$\tablenotemark{a}} & \colhead{$v_{\rm max}$\tablenotemark{a}} & \colhead{\mflow} & \colhead{\pflow} & \colhead{\eflow}& \colhead{$\tau_{\rm d}$} & \colhead{\lflow} & \colhead{\fflow} & \\
\colhead{Source} & \colhead{(\kms)} & \colhead{(\kms)} & \colhead{(\msun)} & \colhead{(\msun\ \kms)} & \colhead{(ergs)} & \colhead{(yr)} & \colhead{(\lsun)} & \colhead{(\msun\ \kms\ yr$^{-1}$)}
}
\startdata
\multicolumn{9}{c}{\cojtwo} \\
\hline
IRAS 03235$+$3004\tablenotemark{b}           & 2.0 & 5.1  & $\geq$5.0\ee{-1} & $\geq$7.8\ee{-1} & $\geq$1.3\ee{43} & 4.2\ee{4} & 2.5\ee{-3} & 1.9\ee{-5} \\
IRAS 03282$+$3035           & 6.0 & 25.9 & 4.3\ee{-1} & 2.1\ee{0}  & 1.4\ee{44} & 1.1\ee{4} & 1.0\ee{-1} & 1.9\ee{-4} \\
HH211                       & 2.9 & 9.9  & 1.1\ee{-1} & 2.3\ee{-1} & 5.5\ee{42} & 7.3\ee{3} & 6.6\ee{-3}  & 3.4\ee{-5} \\
L1709-SMM1\tablenotemark{c} & 1.5 & 2.3  & $\geq$8.6\ee{-3} & $\geq$1.5\ee{-2} & $\geq$2.6\ee{41} & 3.3\ee{4} & $\geq$6.8\ee{-5} & $\geq$4.7\ee{-7} \\
L1709-SMM5\tablenotemark{b,d} & 2.0 & 5.7  & $\geq$5.7\ee{-2} & $\geq$1.3\ee{-1} & $\geq$3.2\ee{42} & \nodata   & \nodata    & \nodata    \\
CB68                        & 1.0 & 2.0  & 1.6\ee{-2} & 1.9\ee{-2} & 2.1\ee{41} & 5.4\ee{4} & 3.2\ee{-5} & 3.4\ee{-7} \\
Aqu-MM2\tablenotemark{c}    & 3.0 & 9.8  & $\geq$2.9\ee{-2} & $\geq$1.4\ee{-1} & $\geq$6.7\ee{42} & 1.1\ee{4} & $\geq$5.3\ee{-3} & $\geq$1.2\ee{-5} \\
Aqu-MM3                     & 3.0 & 9.2  & 1.6\ee{-1} & 5.1\ee{-1} & 1.5\ee{43} & 1.5\ee{4} & 8.7\ee{-3} & 3.0\ee{-5} \\
Aqu-MM5                     & 3.0 & 9.2  & 1.8\ee{-1} & 3.9\ee{-1} & 9.1\ee{42} & 2.5\ee{4} & 3.0\ee{-3} & 1.5\ee{-5} \\
SerpS-MM13\tablenotemark{b,c} & 5.5 & 14.3 & $\geq$8.8\ee{-2} & $\geq$6.4\ee{-1} & $\geq$5.2\ee{43} & 2.6\ee{4} & $\geq$1.7\ee{-2} & $\geq$2.3\ee{-5} \\
CrA-IRAS32\tablenotemark{c} & 2.0 & 4.6  & $\geq$1.1\ee{-2} & $\geq$2.7\ee{-2} & $\geq$6.7\ee{41} & 2.1\ee{4} & $\geq$2.6\ee{-4} & $\geq$1.2\ee{-6} \\
L673-7\tablenotemark{c}     & 3.0 & 8.1  & $\geq$5.5\ee{-2} & $\geq$2.1\ee{-1} & $\geq$7.7\ee{42} & 2.6\ee{4} & $\geq$2.3\ee{-3} & $\geq$7.4\ee{-6} \\
B335                        & 1.0 & 5.5  & 9.9\ee{-2} & 1.7\ee{-1} & 3.3\ee{42} & 4.2\ee{4} & 6.4\ee{-4} & 4.0\ee{-6} \\
L1152\tablenotemark{b}      & 2.0 & 4.7  & $\geq$7.5\ee{-1} & $\geq$1.2\ee{0}  & $\geq$1.9\ee{43} & 7.0\ee{4} & 2.4\ee{-3} & 1.7\ee{-5} \\
L1157                       & 2.0 & 25.6 & 6.1\ee{-1} & 3.1\ee{0}  & 1.5\ee{44} & 1.0\ee{4} & 1.1\ee{-1} & 1.8\ee{-4} \\
L1165                       & 2.0 & 3.7  & 1.6\ee{-1} & 3.1\ee{-1} & 6.2\ee{42} & 7.7\ee{4} & 6.7\ee{-4} & 4.2\ee{-6} \\
L1251A-IRS3\tablenotemark{c} & 2.3 & 5.3  & $\geq$1.1\ee{-1} & $\geq$3.4\ee{-1} & $\geq$1.1\ee{43} & 6.8\ee{4} & $\geq$1.3\ee{-3} & $\geq$5.1\ee{-6} \\
\hline
\multicolumn{9}{c}{\cojthree} \\
\hline
IRAS 03235$+$3004\tablenotemark{b,c}           & 2.6 & 5.7  & $\geq$1.1\ee{-3} & $\geq$3.5\ee{-3} & $\geq$1.3\ee{41} & 1.1\ee{4} & $\geq$9.6\ee{-5} & $\geq$3.1\ee{-7} \\
IRAS 03271$+$3013\tablenotemark{b}           & 1.8 & 6.1  & $\geq$1.8\ee{-2} & $\geq$3.6\ee{-2} & $\geq$8.2\ee{41} & 1.1\ee{4} & 6.4\ee{-4} & 3.2\ee{-6} \\
IRAS 03282$+$3035\tablenotemark{b}           & 3.0 & 13.8 & $\geq$6.6\ee{-2} & $\geq$1.9\ee{-1} & $\geq$7.0\ee{42} & 4.8\ee{3} & 1.1\ee{-2} & 4.0\ee{-5} \\
HH211                       & 2.0 & 2.7  & 5.6\ee{-2} & 1.2\ee{-1} & 2.6\ee{42} & 2.5\ee{4} & 1.5\ee{-3}  & 8.4d-6\ee{-3} \\
IRAS 04166$+$2706\tablenotemark{c}           & 2.0 & 2.5  & $\geq$6.5\ee{-4} & $\geq$1.5\ee{-3} & $\geq$3.4\ee{40} & 2.0\ee{4} & $\geq$9.6\ee{-6} & $\geq$5.3\ee{-8} \\
IRAM 04191$+$1522           & 2.0 & 7.7  & 4.2\ee{-2} & 1.4\ee{-1} & 2.0\ee{42} & 1.4\ee{4} & 1.2\ee{-3} & 6.4\ee{-6} \\
HH25\tablenotemark{c}                        & 4.0 & 10.5 & $\geq$1.8\ee{-2} & $\geq$8.5\ee{-2} & $\geq$4.5\ee{42} & 9.9\ee{3} & $\geq$3.7\ee{-3} & $\geq$8.5\ee{-6} \\
HH26\tablenotemark{c}                        & 4.0 & 24.5 & $\geq$2.7\ee{-1} & $\geq$2.0\ee{0}  & $\geq$1.8\ee{44} & 1.2\ee{4} & $\geq$1.3\ee{-1} & $\geq$1.6\ee{-4} \\
BHR86                       & 2.0 & 5.6  & 1.4\ee{-1} & 2.4\ee{-1} & 5.1\ee{42} & 3.4\ee{4} & 1.2\ee{-3} & 7.0\ee{-6} \\
IRAS 15398$-$3359\tablenotemark{c}           & 2.0 & 5.8  & $\geq$3.7\ee{-4} & $\geq$1.2\ee{-3} & $\geq$3.7\ee{40} & 2.5\ee{3} & $\geq$1.2\ee{-4} & $\geq$4.5\ee{-7} \\
Lupus 3 MMS                 & 2.0 & 4.0  & 3.1\ee{-2} & 4.8\ee{-2} & 8.1\ee{41} & 2.5\ee{4} & 2.7\ee{-4} & 1.9\ee{-6} \\
L483                        & 5.3 & 8.9  & 2.8\ee{-2} & 1.1\ee{-1} & 3.7\ee{42} & 1.2\ee{4} & 2.6\ee{-3} & 9.0\ee{-6} \\
L673-7\tablenotemark{c}                      & 2.0 & 4.6  & $\geq$1.0\ee{-2} & $\geq$2.6\ee{-2} & $\geq$7.0\ee{41} & 4.6\ee{4} & $\geq$1.2\ee{-4} & $\geq$5.5\ee{-7} \\
L1157\tablenotemark{c}                       & 1.4 & 21.1 & $\geq$1.5\ee{-1} & $\geq$5.5\ee{-1} & $\geq$3.8\ee{43} & 1.2\ee{4} & $\geq$2.6\ee{-2} & $\geq$4.7\ee{-5} \\
L1228\tablenotemark{c}                       & 2.0 & 12.0 & $\geq$8.7\ee{-2} & $\geq$3.2\ee{-1} & $\geq$1.4\ee{43} & 2.4\ee{4} & $\geq$4.8\ee{-3} & $\geq$1.3\ee{-5} \\
L1014\tablenotemark{c}                       & 1.3 & 3.0  & $\geq$3.1\ee{-4} & $\geq$5.6\ee{-4} & $\geq$9.9\ee{39} & 2.4\ee{4} & $\geq$3.5\ee{-6} & $\geq$2.3\ee{-8} \\
L1165                       & 1.6 & 4.0  & 5.7\ee{-2} & 8.3\ee{-2} & 1.4\ee{42} & 4.5\ee{4} & 2.6\ee{-4} & 1.9\ee{-6} 
\enddata
\tablenotetext{a}{$v_{\rm min}$ and $v_{\rm max}$ are measured relative to the ambient cloud velocity of each source.  They are the same for both blueshfited and redshifted emission since we adopt symmetrical velocity intervals (see text in \S \ref{sec_mass_dynamical} for details.)}
\tablenotetext{b}{The calculated values of \mflow, \pflow, and \eflow\ are lower limits only since the outflows extend beyond the mapped areas.}
\tablenotetext{c}{The calculated values of \mflow, \pflow, \eflow, \lflow, and \fflow\ are lower limits only since we are unable to obtain reliable Gaussian fits to the ambient cloud emission within 1 \kms\ from the rest velocity and thus unable to correct for low-velocity outflow emission.}
\tablenotetext{d}{Properties that require measurement of outflow lobe length ($\tau_{\rm d}$ and thus \lflow\ and \fflow) cannot be calculated due to the pole-on geometry of this outflow.}
\end{deluxetable*}

Ultimately, given our inability to determine inclinations for most sources 
and the uncertainties over the correct inclination factors to apply, we do 
not correct our outflow properties for inclination.  As a result, even our 
corrected outflow properties are strictly lower limits, since inclination 
corrections will generally only increase these properties based on the above 
arguments, especially for nearly edge-on systems.  

Additional correction factors must be applied if some of the outflowing gas 
is atomic.  \citet{downes2007:outflows} investigated this possibility with 
numerical simulations and found that the fraction of gas dissociated in 
strong shocks becomes progressively larger for gas outflowing at larger 
velocities.  They used these results to show that properties measured only from 
observations of molecular gas underestimated the true values by factors of 
1.6 for \mflow, $2-4$ for \pflow\ and \fflow, and $3-7$ for \eflow\ and 
\lflow.  We are unable to evaluate the effects of dissociation with our data.

Finally, the method of calculation itself can lead to significant differences 
in calculated outflow properties.  These effects were recently explored in 
detail by \citet{vandermarel2013:outflows}, who found a factor of six spread 
in \fflow\ depending on the exact method of calculation.  Our method of 
calculating \fflow\ is identical to their method M7, which they conclude is 
the least affected by uncertain observational parameters.

\section{Discussion}\label{sec_discussion}

\subsection{Correction Factors}\label{sec_discussion_corrections}

In the previous sections we have explored in detail the corrections to 
outflow properties that must be applied to correct for optical depth of the 
\co\ transitions, outflowing gas at low velocities that overlap with the 
velocities of the ambient cloud gas, and outflowing gas below the sensitivities 
of the individual maps.  These corrections are tabulated in Table 
\ref{tab_outflow_corrections}, and are multiplicative.  
Multiplying all three together, we find 
that the mean total correction factors are 13.1, 9.5, 7.6, 10.2, and 12.4 
for \mflow, \pflow, \eflow, \lflow, and \fflow, respectively.  In the most 
extreme cases, they extend up to factors of 59.6, 54.9, 52.4, 88.6, and 89.8 
for the five quantities, and possibly even higher since we are unable to 
determine corrections for low-velocity outflow emission for 15 of the 34 
outflows in our sample.  Outflow studies that fail to correct for one or 
more of these effects risk underestimating the masses and dynamical properties 
of the outflows by up to two orders of magnitude, and possibly even more in 
the most extreme cases.

Table \ref{tab_outflow_dynamics_corrected} presents the same properties for 
each outflow as Table \ref{tab_outflow_dynamics}, except now with the 
corrections listed in Table \ref{tab_outflow_corrections} applied.  
Even after applying these corrections, it is very likely that 
our final values of outflow masses and dynamical properties reported in 
Table \ref{tab_outflow_dynamics_corrected} are still underestimates.  The 
corrections for both opacity and outflowing gas at low velocities are 
conservative, as discussed in Sections \ref{sec_correct_opacity} and 
\ref{sec_correct_lowvel}, respectively, and the sensitivity corrections are 
limited by the fact that we can only smooth the original maps so far in 
velocity while still preserving the basic kinematic structure of the outflows.  
Furthermore, excitation temperatures greater than 50 K, such as those found 
by \citet{vankempen2009:hh46,vankempen2009:outflows}, non-uniform excitation 
temperatures with warmer gas at higher velocities, inclination effects, and 
dissociation of molecular gas in strong shocks will all work to further 
increase the properties of each outflow, as discussed in detail in previous 
Sections.  We thus caution that molecular outflows are almost certainly 
significantly more massive and energetic than found by most analyses of 
low-J rotational transitions of \co, including our own results 
presented in this study.  Exactly how much more massive and energetic is 
impossible to quantify in a general sense because it depends on the 
assumptions and methods of each particular study, as well as quantities 
(inclination, fraction of material that is dissociated, etc.) that are not 
always possible to measure.  Nevertheless, these results must be kept in mind 
when evaluating the masses and energtics of outflows and their impact on 
their environments.

\subsection{Comparing the Two Transitions of \co}\label{sec_discussion_transitions}

Six of the outflows in this study are mapped in both \cojtwo\ and \cojthree:  
IRAS 03235$+$3004, IRAS 03282$+$3035, HH211, L673-7, L1157, and L1165.  
Inspection of Tables \ref{tab_outflow_dynamics} and 
\ref{tab_outflow_dynamics_corrected} shows that the values of 
\mflow\ and all dynamical properties (\pflow, \eflow, \lflow, and \fflow) are 
systematically lower when calculated from the \cojthree\ transition.  
To further investigate this trend, we first remove any possible effects from 
different spatial and spectral resolutions and sensitivities by convolving 
the \cojthree\ maps (which always have higher spatial resolution) down to 
the resolution of the \cojtwo\ maps and then aligning both onto a common 
spatial and spectral pixel grid, masking out spatial pixels that are not 
covered by both transitions to ensure complete overlap.  We then calculate 
the rms noise in each map and clip out all emission below three times the 
larger of the noises in the two transitions, which effectively degrades the 
sensitivity of the deeper map to match that of the other transition.  
We then recalculate opacity-corrected outflow properties from each transition, 
integrating over a common velocity interval for each transition with 
$v_{\rm min}$ chosen to be large enough to eliminate all possible confusion 
with ambient cloud emission and $v_{\rm max}$ chosen to extend only to the 
smaller of the two $v_{\rm max}$ for each transition.  In this manner we ensure 
that we are only comparing emission from the outflows, over velocities where 
both outflows are detected.

\begin{deluxetable}{lcc}
\tabletypesize{\scriptsize}
\tablewidth{0pt}
\tablecaption{\label{tab_transitions}\cojtwo\ versus \cojthree\ as an Outflow Tracer}
\tablehead{
\colhead{Source} & \colhead{\mflow$^{21}$/\mflow$^{32}$} & \colhead{\fflow$^{21}$/\fflow$^{32}$} 
}
\startdata
IRAS 03235$+$3004 & 10.2 & 10.5 \\
IRAS 03282$+$3035 & 4.3  & 5.0  \\
HH211             & 1.6  & 1.4  \\
L673-7            & 11.4 & 7.0  \\
L1157             & 2.1  & 2.0  \\
L1165             & 20.0 & 13.5 
\enddata
\end{deluxetable}

\begin{figure}
\epsscale{1.2}
\plotone{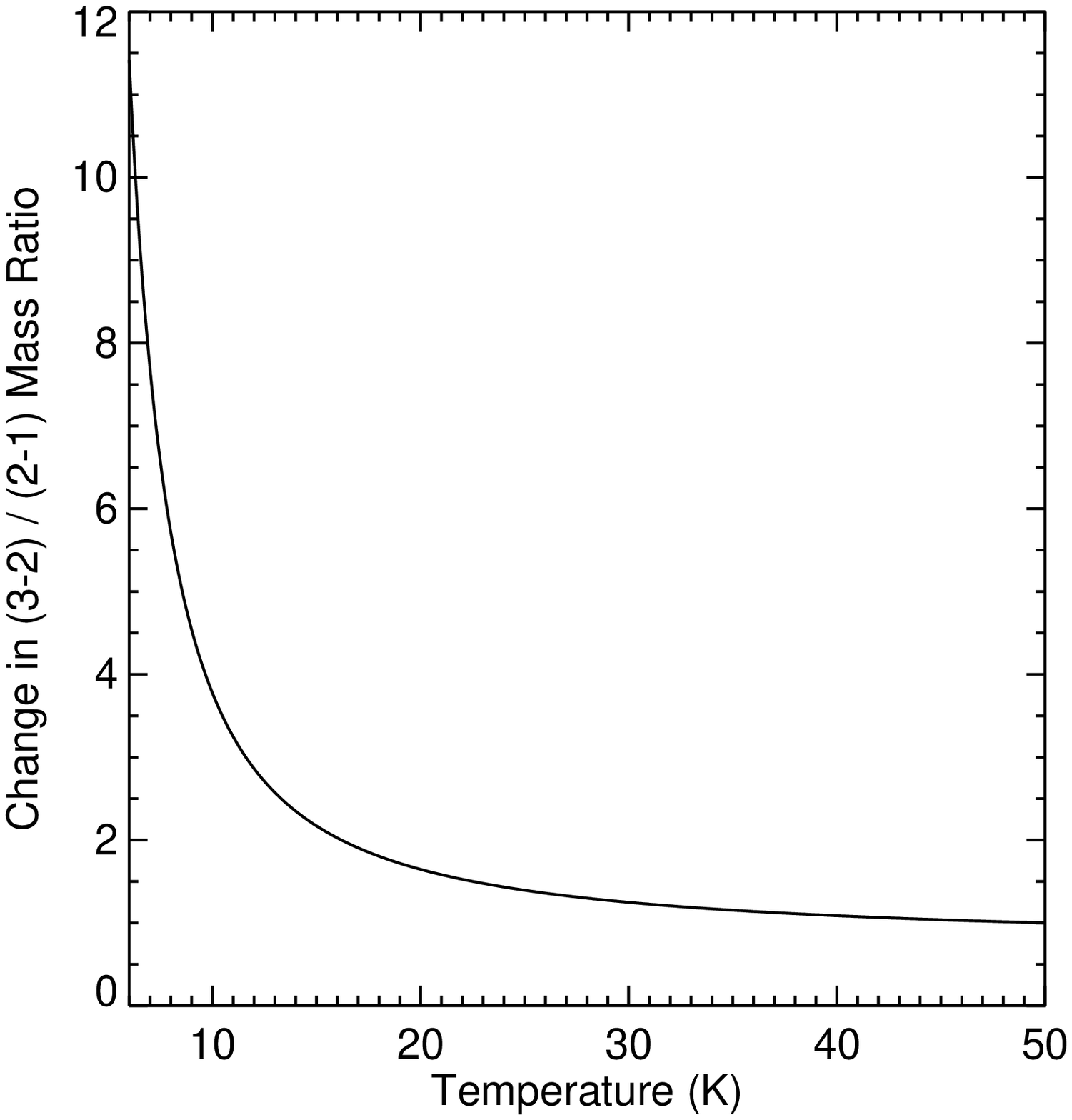}
\caption{\label{fig_transitions}The factor by which the ratio of \mflow\ 
calculated from \cojthree\ to that calculated from \cojtwo\ will change for 
different values of $T_{\rm ex}$, compared to the ratio calculated assuming 
$T_{\rm ex} = 50$ K.  See Appendix \ref{sec_appendix_equations} for 
details on the calculation.}
\end{figure}

The results of this process are listed in Table \ref{tab_transitions} for 
\mflow\ and \fflow; similar results are obtained for the other dynamical 
properties.  Both quantities are systematically higher when calculated from 
the \cojtwo\ maps than the \cojthree\ maps.  The mean increase in \mflow\ 
for these six outflows is 8.3, with individual values ranging from 1.6 to 20.  
As long as the gas has $T_{\rm ex} \ga 20$ K the intensity will be as bright 
or brighter in the (3--2) transition than the (2--1) transition, so missing 
(3--2) emission below the sensitivities of the maps is not a likely explanation 
since we only integrate emission above a common sensitivity.  Thus any emission 
bright enough to detect in \cojtwo\ should also be bright enough to detect 
in \cojthree, unless $T_{\rm ex}$ is significantly below 20 K (which we 
consider unlikely; see \S \ref{sec_correct_temperature}).  Another possible 
explanation for this discrepancy is that our assumed $T_{\rm ex}$ of 50 K 
is wrong, since the mass ratio between the two transitions will change for 
different temperatures.  However, as shown by Figure \ref{fig_transitions}, 
increasing the ratio of \mflow\ by a factor of 8.3 requires decreasing the 
assumed $T_{\rm ex}$ to below 10 K, and increasing this ratio by a factor of 
20 requires decreasing the assumed $T_{\rm ex}$ to $\sim$ 5 K (see Appendix 
\ref{sec_appendix_equations} for details on the calculation).  Such low 
temperatures are required because it is only at these very low temperatures 
that the relative population of the $J=3$ state compared to the $J=2$ state 
decreases substantially, but $T_{\rm ex} < 10$ K is highly unlikely for 
gas in molecular outflows (see \S \ref{sec_correct_temperature}).  A 
third possible explanation is that higher-J transitions of \co\ are dominated 
by emission of gas at higher temperatures, as suggested by recent 
{\it Herschel} detections of warm \co\ 
\citep[up to $\sim 1000$ K; e.g.,][]{green2013:digit,yildiz2013:highj,santangelo2013:herschel} associated with low-mass 
protostars.  However, as seen by Figure \ref{fig_masstex}, even increasing 
the assumed $T_{\rm ex}$ to 200 K for \cojthree\ fails to resolve the 
discrepancy for most sources, and such extreme temperature differences between 
the gas dominating the emission in two successive rotational levels with 
energies that are separated by less than 20 K are unlikely.  Non-LTE 
radiative transfer modeling would be required to fully explore these effects, 
but such work is beyond the scope of this paper.

An additional potential explanation recently put forth by 
\citet{ginsburg2011:outflows} is that the \cojthree\ line is subthermally 
excited and thus a poor tracer of total mass.  The critical density of the 
(3--2) transition is about 20 times higher than that of the (1--0) transition.  
\citet{ginsburg2011:outflows} used RADEX, a one-dimensional, non-LTE radiative 
transfer code \citep{vandertak2007:radex} to show that the total outflow mass 
calculated from \cojthree\ can be underestimated by up to 1--2 orders of 
magnitude for gas densities between 10$^{2}$ and 10$^{4}$ cm$^{-3}$.  Since 
the (2--1) transition may also be subthermally excited in some cases 
(its critical density is about six times higher than that for the ground-state 
transition) the difference between outflow masses calculated from the (3--2) 
and (2--1) transitions are likely equal to or less than the \cojthree\ 
underestimates calculated by \citet{ginsburg2011:outflows}, consistent 
with our results listed in Table \ref{tab_transitions}.
Since we have rejected sensitivity and the assumed $T_{\rm ex}$ 
as explanations for the discrepancy in outflow properties calculated from 
the (3--2) and (2--1) 
transitions, our results are consistent with the claim by 
\citet{ginsburg2011:outflows} that the \cojthree\ line is subthermally excited 
in molecular outflows and thus a poor tracer of total outflow mass.

While our results are based on only six outflows and require confirmation with 
a larger sample, they have important implications.  
Many recent studies focusing 
on topics including the evolution of outflows, the link between outflows and 
protostellar accretion, and the impact of outflow feedback on cluster-scale 
star formation have primarily used observations of \cojthree, both because 
it can be mapped with the same telescope at higher angular resolution than 
lower-J transitions and because it is easier to separate warm outflowing gas 
from cold ambient gas in higher-J transitions 
\citep[e.g.,][]{hatchell2007:outflows,hatchell2009:outflows,curtis2010:outflows,nakamura2011:serpsouth}.  If \citet{ginsburg2011:outflows} are correct and 
the \cojthree\ line is subthermally excited in outflows, as indeed suggested 
by our results, all of these studies have likely underestimated the total 
mass, momentum, and energy of the outflows in their samples, and conclusions 
based on these quantities will need to be revisited.  Confirming our results 
with a larger sample of outflows, and extending this comparison to the 
\cojone\ transition, are noted as critical directions for future work to 
pursue.

\section{Summary}\label{sec_summary}

In this paper we have presented the first results of a survey of 28 molecular 
outflows driven by low-mass protostars, all of which are sufficiently isolated 
spatially and/or kinematically to fully separate into individual outflows.  
Using a combination of new and archival data from several single-dish 
telescopes, 17 outflows were mapped in \cojtwo\ and 17 are mapped in \cojthree, 
with 6 mapped in both transitions.  We summarize our main results as follows:

\begin{enumerate}
\item For each outflow, we calculate and tabulate the mass (\mflow), momentum 
(\pflow), kinetic energy (\eflow), mechanical luminosity (\lflow), and force 
(\fflow) assuming optically thin emission in LTE at an excitation temperature, 
$T_{\rm ex}$, of 50 K.  We also tabulate the size of each outflow and its 
position angle on the sky.
\item For outflows mapped in both transitions of \co, line ratios suggest 
excitation temperatures ranging from 10 -- 20 K.  While there is very weak 
evidence for higher temperatures (up to 50 K) at the highest redshifted 
velocities, in general there is no clear trend with velocity.  
However, with only these two low-J rotational transitions of \co, 
we are insensitive to the presence of warmer gas.  We thus adopt 50 K as the 
most likely value based on results from other authors that find such 
temperatures.
\item All of the calculated outflow properties are significantly underestimated 
when calculated from the original data under the assumption of optically thin 
emission in LTE, with velocity ranges chosen to avoid contamination by 
ambient cloud emission.  Taken together, the effects of opacity, outflow 
emission at low velocities confused with ambient cloud emission, and emission 
below the sensitivities of the observations increase outflow masses and 
dynamical properties by an order of magnitude, on average, and factors of 
50--90 in the most extreme cases.
\item Different (and non-uniform) excitation temperatures, inclination effects, 
and dissociation of molecular gas will all work to further increase the masses 
and dynamical properties of outflows.  Molecular outflows are thus almost 
certainly more massive and energetic than commonly reported.
\item For outflows mapped in both transitions, the masses and dynamical 
properties are lower, on average, by about an 
order of magnitude when calculated 
from the \cojthree\ maps compared to the \cojtwo\ maps, even after accounting 
for different opacities, map sensitivities, and possible excitation temperature 
variations.  \citet{ginsburg2011:outflows} argued that the \cojthree\ line is 
subthermally excited in outflows, and our results support this finding.
\end{enumerate}

We have provided a systematic analysis of the uncertainties in and necessary 
corrections to typical calculations of outflow masses and 
dynamical properties.  Studies that neglect one or more of these effects 
will underestimate the properties of molecular outflows; not only  
does this indicate that outflows are more massive and energetic than commonly 
found, but it also suggests that outflows may have larger impacts on the 
turbulence and energetics of their environments than is often calculated 
based on studies of outflows in clustered regions 
\citep[e.g.,][]{arce2010:perseus,nakamura2011:serpsouth,plunkett2013:ngc1333}.  
In a forthcoming paper we will explore the effects of 
these corrections on our understanding of the evolution of outflows 
and the link between protostellar accretion and outflow activity.  Several 
avenues of future work remain necessary.   
First, larger, more sensitive maps of \coo\ are necessary in order to 
derive separate velocity-dependent opacity corrections for each outflow, and 
perhaps even for each position in each outflow, rather than the corrections 
averaged over many outflows that we derive here.  Additionally, higher-J 
transitions of \co\ are needed to fully evaluate the excitation temperatures 
of the outflows and identify any variations with position and/or velocity.  
Finally, a larger sample of outflows must be mapped in multiple transitions 
of \co\ in order to confirm our findings that the \cojthree\ line 
underestimates outflow masses and dynamical properties due to subthermal 
excitation, as recently argued by \citet{ginsburg2011:outflows}.

\acknowledgments
The authors express their gratitude to Stella Offner, Adele Plunkett, Xuepeng 
Chen, and Neal Evans for helpful discussions and/or for commenting on drafts 
of this manuscript.  We also thank Neal Evans for assistance with obtaining 
APEX data, and the anonymous referee for a set of comments that have improved 
the quality of this publication.  
We gratefully acknowledge the assistance provided by the staff of the 
CSO in obtaining some of the observations presented here.  
This work is based on data obtained with the following facilities:  
The Atacama Pathfinder Experiment (APEX), a collaboration between the 
Max-Planck-Institut fur Radioastronomie, the European Southern Observatory, 
and the Onsala Space Observatory; 
The Caltech Submillimeter Observatory (CSO), which is operated by the 
California Institute of Technology under cooperative agreement with the 
National Science Foundation (AST-0838261); 
The James Clark Maxwell Telescope (JCMT), which is operated by the Joint 
Astronomy Centre on behalf of the Science and Technology Facilities Council of 
the United Kingdom, the National Research Council of Canada, and (until 31 
March 2013) the Netherlands Organisation for Scientific Research.  
This publication makes use of data products 
from the Infrared Processing and Analysis Center/California 
Institute of Technology, funded by the National Aeronautics and Space 
Administration and the National Science Foundation.  These data were provided 
by the NASA/IPAC Infrared Science Archive, which is operated by the Jet 
Propulsion Laboratory, California Institute of Technology, under contract with 
NASA.  
This research has made use of NASA's Astrophysics Data System (ADS) 
Abstract Service, the IDL Astronomy Library hosted by the NASA Goddard Space 
Flight Center, the SIMBAD database operated at CDS, Strasbourg, France, and 
the facilities of the Canadian Astronomy Data Center (CADC) operated by the 
National Research Council of Canada with the support of the Canadian Space 
Agency. 
MMD acknowledges support as an SMA postdoctoral fellow.  MMD and HGA 
acknowledge support from the NSF through grant AST-0845619 to HGA.  
JEL was supported by the Basic Science Research Program through the National 
Research Foundation of Korea (NRF) funded by the Ministry of Education of the 
Korean government (grant No.~NRF-2012R1A1A2044689).  
The work of AMS was supported by the Deutsche Forschungsgemeinschaft priority 
program 1573 (“Physics of the Inter- stellar Medium”).

\bibliographystyle{apj.bst}
\bibliography{dunham_citations}


\appendix

\section{A.~~Summary of Individual Sources}\label{sec_appendix_sources}

In this section we provide a brief description of each source surveyed in this paper.

\subsection{IRAS 03235$+$3004}

IRAS 03235$+$3004 (hereafter IRAS03235) is a protostar embedded in the 
southwestern portion of the Perseus Molecular Cloud, on the western edge of 
L1455.  For this and all other sources in Perseus we adopt a distance of 
250 pc \citep[][and references therein]{enoch2006:bolocam}, consistent with 
the VLBI maser parallax distance of 235 $\pm$ 18 pc for NGC 1333 determined 
by \citet{hirota2008:parallax}.  IRAS03235 is embedded within 
a $0.5-2.5$ \msun\ core \citep{enoch2006:bolocam,enoch2009:protostars,hatchell2005:perseus,hatchell2007:perseusseds}
that exhibits spectroscopic signatures of infall \citep{gregersen2000:infall} 
and is located at a rest velocity of 5.1 \kms\ 
\citep{mardones1997:infall,kirk2007:perseus}.  
It is detected at 2 \um\ \citep{ladd1993:perseus} and is classified 
as a borderline Class 0/I object based on the full observed SED, 
including \emph{Spitzer Space Telescope} \citep{werner2004:spitzer} 
$3.6-70$ \um\ detections 
\citep{jorgensen2006:perseus,rebull2007:perseus}, 
with individual estimates of \tbol\ ranging between 68 
and 136 K \citep{mardones1997:infall,hatchell2007:perseusseds,enoch2009:protostars,evans2009:c2d}.  
IRAS03235 drives a bipolar molecular outflow detected by 
\citet{hatchell2009:outflows}.

\subsection{IRAS 03271$+$3013}

IRAS 03271$+$3013 (hereafter IRAS03271) is an embedded protostar located 
south of NGC1333 in the Perseus Molecular Cloud at a distance of 250 
pc (see above).  The surrounding core has a mass of $0.5-5$ \msun\ 
\citep{bachiller1991:iras03282,ladd1994:dcdc,hatchell2005:perseus,hatchell2007:perseusseds,enoch2006:bolocam,enoch2009:protostars}, 
is located at a rest velocity of 5.9 \kms\ 
\citep{bachiller1991:iras03282,ladd1994:dcdc,kirk2007:perseus,rosolowsky2008:nh3,hatchell2009:outflows,emprechtinger2009:n2hp}, 
and shows no 
evidence for strong deuteration or infall \citep{emprechtinger2009:n2hp}.  
IRAS03271 is detected in the near-infrared at 2 \um\ \citep{ladd1993:perseus} 
and in the mid-infrared at $3.6-70$ \um\ with \emph{Spitzer} 
\citep{jorgensen2006:perseus,rebull2007:perseus}
and is classified as a Class I protostar, with individual estimates of \tbol\ 
ranging between 97 and 133 K 
\citep{hatchell2007:perseusseds,enoch2009:protostars,evans2009:c2d}.  
It drives a bipolar molecular outflow 
\citep{bachiller1991:iras03282,hatchell2009:outflows} that extends to extremely 
high velocities (up to $\sim$40 \kms\ from rest), it is 
associated with a faint 2 \um\ reflection nebula that extends along the 
outflow axis \citep{connelley2007:nebulae}, and it has been identified as the 
driving source of three Herbig-Haro objects 
\citep[HH 368, 369, and 370;][]{wu2002:hh}.

\subsection{IRAS 03282$+$3035}

IRAS 03282$+$3035 (hereafter IRAS03282) is a well-studied, deeply embedded 
protostar located south of NGC1333 and west of B1 in the Perseus Molecular 
Cloud at a distance of 250 pc (see above).  The surrounding core has 
a mass of $0.8-6.3$ \msun\ \citep{bachiller1991:iras03282,barsony1998:perseus,shirley2000:scuba,motte2001:mm,hatchell2005:perseus,hatchell2007:perseusseds,enoch2006:bolocam,enoch2009:protostars}, 
is located at a rest velocity of $7.1$ \kms\ \citep{bachiller1991:iras03282,mardones1997:infall,gregersen2000:infall,hatchell2007:outflows}, 
shows strong deuteration \citep{roberts2002:deuteration,hatchell2003:deuteration,roberts2007:deuteration,emprechtinger2009:n2hp} 
and no significant evidence of infall or fast rotation 
\citep{mardones1997:infall,gregersen2000:infall,chen2007:ovro}.  
IRAS03282 is detected in the mid-infrared with \emph{Spitzer} 
\citep{jorgensen2006:perseus,rebull2007:perseus} and is classified as a Class 0 
protostar, with individual estimates of \tbol\ ranging between 23 and 60 K 
\citep{shirley2000:scuba,hatchell2007:perseusseds,enoch2009:protostars,evans2009:c2d}.

IRAS03282 drives a strong bipolar molecular outflow first detected by 
\citep{bachiller1991:iras03282}.  This outflow extends to extremely high 
velocities (greater than 50 \kms\ relative to the core rest velocity) 
and features a 
highly collimated, jet-like structure at the highest velocities comprised of a 
series of high-velocity ``bullets'' or clumps \citep{bachiller1991:iras03282}.  
The velocities and spacings of these clumps suggest an episodicity in the 
mass ejection (and thus likely in the underlying mass accretion) on timescales 
of $\sim 10^3$ yr \citep{bachiller1991:iras03282}.  This outflow is associated 
with near-infrared H$_2$ knots coincident with the high-velocity bullets, warm 
(T $> 50-100$ K) \ammonia, SiO emission, and enhanced CH$_3$OH abundance 
\citep{bally1993:iras03282,bachiller1993:nh3,bachiller1994:iras03282,bachiller1995:methanol}, 
and is consistent with models featuring time-dependent rather than 
steady-state jets \citep{bachiller1994:iras03282}.

\subsection{HH211}

HH211 was originally discovered as a jet detected in near-infrared continuum 
and narrow-band H$_2$ images \citep{mccaughrean1994:hh211}.
It is located in the southwestern 
portion of IC348 in the Perseus Molecular Cloud at a distance of 250  
pc (see above).  The near-infrared jet consists of multiple knots, and it 
coincides with a molecular outflow that extends up to $\sim$50 \kms\ from rest 
that is centered on a dense core detected in \ammonia\ and submillimeter and 
millimeter continuum observations \citep{bachiller1987:ic348,mccaughrean1994:hh211,gueth1999:hh211,tanner2011:hh211}.  
This core has a mass of $2.5 - 23$ \msun\ 
\citep{motte2001:mm,kirk2006:perseus,hatchell2005:perseus,hatchell2007:perseusseds,enoch2006:bolocam,enoch2009:protostars}, is located at a rest 
velocity of 9.1 \kms\ 
\citep{mardones1997:infall,gregersen2000:infall,hatchell2007:outflows},
 and shows moderately strong deuteration 
\citep{roberts2002:deuteration,hatchell2003:deuteration,roberts2007:deuteration,emprechtinger2009:n2hp}
but no conclusive evidence for infall 
\citep{mardones1997:infall,gregersen2000:infall}.  
Embedded within this core is a Class 0 protostar detected in the 
far-infrared with \emph{ISO} \citep{froebrich2003:farir} and in the 
mid-infrared with \emph{Spitzer}, but only at wavelengths longward of 24 \um\ 
\citep{rebull2007:perseus,evans2009:c2d}.  Individual estimates of \tbol\ for 
this protostar range from 21 to 31 K 
\citep{hatchell2007:perseusseds,enoch2009:protostars,evans2009:c2d}.

The molecular outflow driven by this source has been extensively studied over 
the past two decades.  At low velocities the outflow traces the shells of 
cleared cavities whereas at high velocities a highly collimated jet consisting 
of multiple knots is seen 
\citep{gueth1999:hh211,palau2006:hh211,lee2007:hh211,lee2009:hh211}.  
The outflow has also been detected in SiO emission ranging from the $J=1-0$ 
transition up to the $J=11-10$ transition, with line ratios suggesting very 
warm gas (T $> 300-500$ K) and a morphology of multiple clumps and knots 
matching that seen in CO and H$_2$ \citep{chandler2001:hh211,nisini2002:hh211,oconnell2005:hh211,palau2006:hh211,lee2007:hh211,lee2009:hh211}.  
\citet{lee2007:hh211} and \citet{lee2009:hh211} 
showed that the knots are moving at a transverse velocity of 170 \kms\ 
(measured from proper motion of the knots) and are spaced by $\sim$ 600 -- 
900 AU, implying a timescale of 17 -- 25 yr for the underlying episodicity 
in the mass ejection.  With knowledge of both the transverse velocity of the 
knots from proper motion and the radial velocity from their line observations, 
they concluded that the HH211 protostellar system has an inclination of 
85\degree\ and is thus nearly edge-on.

Finally, \citet{lee2009:hh211} resolved the millimeter continuum into two 
sources separated by $\sim$ 84 AU.  The stronger of the two sources is clearly 
the driving source of the outflow, and no second outflow driven by the 
companion is detected.  It is likely that the primary source dominates both 
the outflow emission and the observed SED, thus we assume for the purposes of 
this study that HH211 is a single object.

\subsection{IRAS 04166$+$2706}

IRAS 04166$+$2706 (hereafter IRAS04166) is a Class I protostar located in 
the Taurus Molecular Cloud at an assumed distance of 140 pc 
\citep{kenyon1994:taurus}
, with individual estimates of \tbol\ ranging from 75 K to 139 K 
\citep{chen1995:tbol,shirley2000:scuba,young2003:scuba}.  The surrounding 
dense gas is located at systemic velocity of 6.7 \kms\ and shows kinematic 
evidence of infall onto the Class I protostar 
\citep{mardones1997:infall,gregersen2000:infall}.  
A bipolar molecular outflow driven by IRAS04166 was 
first detected by \citet{bontemps1996:outflows} and later mapped in detail by 
\citet{tafalla2004:iras04166} and \citet{santiagogarcia2009:iras04166}.  
The latter two studies revealed a highly collimated outflow with an extremely 
high velocity component extending up to $\sim$50 \kms.

\subsection{IRAM 04191$+$1522}

IRAM 04191$+$1522 (hereafter IRAM04191) is a Class 0 protostar located in the 
southern part of the Taurus Molecular Cloud at an assumed distance of 140 pc 
\citep{kenyon1994:taurus}, 
consistent with the recent distance estimate to this source 
of 127 $\pm$ 25 pc by \citet{maheswar2011:distances}.  
It was originally detected in the far-infrared and submillimeter 
by \citet{andre1999:iram04191} and later in the mid-infrared with the 
\emph{Spitzer Space Telescope} by \citet{dunham2006:iram04191}, 
and features \lbol\ $\sim$ 0.1--0.15 
\lsun, \tbol\ $=27$ K, and \lbolsmm\ $= 5$ 
\citep{andre1999:iram04191,dunham2008:lowlum}.  
\citet{dunham2006:iram04191} used radiative transfer models to show that 
\lint\ $\sim$ 0.08 \lsun, where \lint\ is the internal luminosity and 
excludes any luminosity arising from external heating by the interstellar 
radiation field.

IRAM04191 drives a collimated, bipolar molecular outflow with well-separated 
red and blue lobes 
\citep{andre1999:iram04191,lee2002:outflows,lee2005:iram04191}.  It 
is embedded within a core of $\sim 1-3$ \msun\ that appears flattened along an 
axis perpendicular to the outflow and features extended subsonic infall, 
rotation, CO and \nthp\ depletion, and significant deuteration 
\citep{andre1999:iram04191,belloche2002:iram04191,belloche2004:iram04191,lee2005:iram04191}.  
The rest velocity of the core is taken to be 6.7 \kms\ 
\citep{lee2002:outflows,belloche2002:iram04191}.  \citet{lee2005:iram04191} 
speculated about the possible presence of an unseen binary 
companion based on the outflow morphology; this speculation was recently 
confirmed by \citet{chen2012:iram04191}, who detected a binary companion at 
1.3 mm in high angular resolution Submillimeter Array 
\citep[SMA;][]{ho2004:sma} observations.  As no 
evidence for this compansion is seen in the \emph{Spitzer} mid-infrared 
observations of IRAM04191, \citet{chen2012:iram04191} speculate that this 
companion does not contribute significantly to the bolometric luminosity of 
IRAM04191 or to the large-scale molecular outflow.  We thus assume for the 
purposes of this study that IRAM04191 is a single object.


\subsection{HH25 and HH26}

HH25 and HH26 are two Herbig Haro objects located in L1630 near the northern 
edge of the Orion Molecular Cloud Complex \citep{herbig1974:hh}, at an assumed 
distance of 430 pc \citep[e.g.,][]{antoniucci2008:hh}.  HH25 and HH26 are 
driven by embedded Class 0 and Class I sources, respectively (commonly known as 
HH25MMS and HH26IR), with HH25MMS located $\sim$1.5\am\ northeast of 
HH26IR.  The jets associated with these HH objects have been extensively 
mapped in various optical and near-infrared transitions 
\citep[e.g.,][]{davis1997:nearir,eisloffel1997:jets,schwartz1997:hh2526,chrysostomou2002:hh,caratti2006:jets}.  
An approximately east-west bipolar molecular outflow driven by HH26IR was 
discovered by \citet{snell1982:outflows}, and higher-resolution data presented 
by \citet{gibb1993:hh2526} revealed the presence 
of a second, nearly orthogonal bipolar molecular outflow driven by HH25MMS.  
Both sources are associated with dense cores detected as (sub)millimeter 
continuum sources 
\citep{lis1999:hh2526,johnstone2001:scuba,mitchell2001:scuba}, 
with total masses of $\sim$1.5 \msun\ (HH26IR) and $\sim$5 \msun\ 
(HH25MMS).  These cores are both located at a rest velocity of 10.1 \kms\ 
\citep{matthews1983:nh3,gibb1995:orion}.

\subsection{BHR86}

BHR86 is a cometary-shaped globule cataloged as the dark core BHR86, 
DC 303.8$-$14.2, and Sandqvist 160 in the surveys of 
\citet{bourke1995:catalog}, \citet{hartley1986:catalog}, and 
\citet{sandqvist1977:catalog}, respectively.  It is located 
in the northeast portion of the Chamaeleon II molecular cloud at a distance 
of 178 pc \citep{whittet1997:dcham}.  Measurements of the core systemic 
velocity range from 3.4 \kms\ to 4.3 \kms\ depending on the observed molecule 
and transition 
\citep{bourke1995:nh3,mardones1997:infall,lohr2007:astro}; here we adopt a 
rest velocity of 3.7 \kms\ based on the \ammonia\ observations presented by 
\citet{bourke1995:nh3}.

BHR86 harbors an embedded protostar first detected by the 
{\it Infrared Astronomical Satellite (IRAS)} 
\citep[IRAS 13036$-$7644;][]{gregorio1988:southiras}, and is associated with 
kinematic signatures of infall 
\citep{mardones1997:infall,lehtinen1997:bhr86}, 1.3 mm continuum emission 
\citep{henning1993:cham,henning1998:globules,launhardt2010:globules}, 3.6 and 
6 cm radio continuum emission \citep{lehtinen2003:bhr86},and a bipolar 
molecular outflow \citep{lehtinen1997:bhr86}.  The protostar and surrounding 
core were detected in \emph{ISO} far-infrared and \emph{Spitzer} mid-infrared 
observations by \citet{lehtinen2005:bhr86} and \citet{launhardt2010:globules},
respectively.  Both studies found that BHR86 harbors a Class 0 protostar 
with \tbol\ $\sim 60$ K, close to the Class 0/I boundary.  

\subsection{IRAS 15398$-$3359}

IRAS 15398$-$3359 (hereafter IRAS 15398) is a protostar embedded in the dense 
core B228 \citep{barnard1927:catalog} in the Lupus I Molecular Cloud at a 
distance of 150 pc \citep{comeron2008:lupus}.  It was originally identified as 
a protostar by \citet{heyer1989:b228}, who also discovered the Herbig-Haro 
object HH185 driven by this source.  With a measured \tbol\ of 48 K 
\citep{shirley2000:scuba}, IRAS15398 is a Class 0 protostar.  A bipolar 
molecular outflow driven by this source was discovered by 
\citet{tachihara1996:lupus} and later mapped in multiple transitions of CO by 
\citet{vankempen2009:highj}, who used the line 
ratios in the various transitions to determine a temperature of $\sim 100-200$ 
K for the outflowing gas.  The dense core B228 in which IRAS15398 is embedded 
has a mass of 0.3 -- 0.8 \msun\ 
\citep{reipurth1993:dust,shirley2000:scuba,shirley2002:scuba} and is located 
at a rest velocity of 5.1 \kms\ 
\citep{mardones1997:infall,hirota1998:hcn,vankempen2009:highj}.

Careful inspection of the \emph{Spitzer} source catalogs produced by the 
c2d \citep[Cores to Disks {\it Spitzer} Legacy Survey;][]{evans2003:c2d,evans2009:c2d} 
team\footnote{Available at http://irsa.ipac.caltech.edu/} at the position of 
IRAS15398 yield two mid-infrared sources located within 2\as\ of each other.  
One is the bright source detected at $3.6-70$ \um\ 
\citep[and in fact saturated at 3.6 and 4.5 \um;][]{chapman2007:lupus} that is 
associated with the \emph{IRAS} 
detection of the protostar.  The other is detected at $3.6-8$ \um\ with a 
rising SED consistent with being a young stellar object (YSO), 
but with a separation of 2\as\ it is 
not resolved into a separate source at 24 or 70 \um.  The nature of this source 
is unclear.  It could possibly be a binary companion, although no such 
companion is detected in the near-infrared in the multiplicity study conducted 
by \citet{connelley2008:nirmultiplicity} despite sufficient angular resolution 
in their observations to detect such a companion.  Furthermore, no such 
millimeter companion is detected in the multiplicity study conducted by 
\citet{chen2013:smamultiplicity} with the SMA, although their angular 
resolution is only marginally sufficient for such a purpose.  Even if it is a 
binary companion, it is fainter at 8 \um\ than IRAS15398 itself 
by a factor of 6 \citep{merin2008:lupus} and thus unlikely to dominate either 
the observed infrared and (sub)millimeter SED or the outflow.  For the 
purposes of this study we assume that IRAS15398 is a single object.

\subsection{Lupus 3 MMS}

Lupus 3 MMS is an embedded Class 0 protostar in the Lupus 3 Molecular Cloud 
at a distance of 200 pc \citep{comeron2008:lupus}.  It was first discovered by 
\citet{tachihara2007:lupus} in H$^{13}$CO$^+$ (1-0) and 1.2 mm continuum 
observations.  They also detected fan-shaped nebulosity and a jet-like 
feature extending to the southwest in near-infrared images and blueshifted 
\cojthree\ emission in pointed observations towards five positions southwest 
of the core, but did not fully map the region in CO.  Lupus 3 MMS was detected 
in the mid-infrared at $3.6-70$ \um\ with \emph{Spitzer} 
\citep{tachihara2007:lupus,chapman2007:lupus,merin2008:lupus} and exhibits a 
Class 0 SED with \tbol\ $= 39$ K 
\citep{tachihara2007:lupus,dunham2008:lowlum,evans2009:c2d}.  
The core systemic velocity is taken to be 4.8 \kms\ 
based on an unpublished \cooojtwo\ spectrum observed at the CSO.

\subsection{L1709-SMM1/5}

L1709-SMM1, more commonly known as Oph-IRS63 or IRAS 16285$-$2355, 
is a protostar located in the L1709 portion of the Ophiuchus molecular cloud 
at a distance of 125 pc \citep{degeus1989:scocenob} and a rest velocity of 
2.5 \kms\ \citep[e.g.,][]{visser2002:scuba}.  It drives a bipolar molecular 
outflow first detected by \citet{bontemps1996:outflows} and later mapped by 
\citet{visser2002:scuba}.  With individual estimates of \tbol\ ranging from 
270 to 363 K \citep[e.g.,][]{vankempen2009:oph,dunham2013:luminosities}, 
it is classified as a Class I protostar.  Its status as 
an embedded protostar, rather than a more evolved YSO with a disk observed at 
an edge-on inclination, was confirmed by dense gas observations presented by 
\citet{vankempen2009:oph}.

L1709-SMM5 is located about 2\am\ south of IRS63/SMM1.  It was first detected 
in submillimeter continuum images presented by \citet{visser2002:scuba}, 
signifying the presence of a dense core.  \citet{visser2002:scuba} 
also detected an outflow driven by 
this source, indicating the core harbors a protostar.  SMM5 was detected in 
\emph{Spitzer} infrared observations presented by 
\citet{jorgensen2008:scubaspitzer} and \citet{evans2009:c2d}, 
but very little is known about this source.  \citet{evans2009:c2d} 
calculated a \tbol\ of 700 K and classified it as a Class I 
source.  Given its proximity to IRS63/SMM1, we assume the same rest velocity of 
2.5 \kms.

\subsection{CB68}

CB68 is an isolated Bok Globule located north of the Ophiuchus molecular cloud 
cataloged as CB68 by \citet{clemens1988:catalog} and as the opacity class 5 
cloud L146 by \citet{lynds1962:catalog}.  The core has a total mass of 
0.1 -- 0.6 \msun\ 
\citep{launhardt1997:globules,huard1999:globules,young2006:scuba,launhardt2010:globules}, 
is located at a rest velocity of 5.2 \kms\ 
\citep[e.g.,][]{wang1995:collapse,codella1997:globules,launhardt1998:globules}, 
and is located at an adopted distance of 130 pc \citep{hatchell2012:ophnorth}.  
CB68 is associated with the Class 0 protostar IRAS 16544$-$1604 
\citep{clemens1988:catalog}, with individual estimates of \tbol\ ranging from 
50 to 74 K \citep{mardones1997:infall,launhardt2010:globules}.  A bipolar 
molecular outflow driven by this protostar was first detected and mapped by 
\citet{vallee2000:cb68}.

\subsection{L483}

L483 is a Lynds Opacity Class 6 cloud \citep{lynds1962:catalog} located at a 
rest velocity of 5.4 \kms\ 
\citep[e.g.,][]{dieter1973:clouds,fuller1993:cores,benson1998:dcdc} 
associated with the protostar IRAS 18148$-$0440 \citep{parker1988:iras}.  With 
individual estimates of \tbol\ ranging from 48 K to 60 K 
\citep{gregersen1997:infall,mardones1997:infall,shirley2000:scuba,visser2002:scuba,jorgensen2009:prosac}, this protostar is classified as Class 0, although 
\citet{tafalla2000:l483} argued 
that it is in transition between Class 0 and Class I based on various indirect 
evolutionary indicators.  L483 drives a bipolar molecular outflow first 
detected and mapped by \citet{parker1991:outflows}.  Furthermore, it is 
associated with a H$_2$O maser \citep{xiang1992:h2omasers} and a bipolar 
infrared nebula that is aligned with the molecular outflow and shows 
morphological and photometric variability on timescales of months 
\citep{fuller1995:l483,connelley2009:l483}.  L483 is a relatively bright 
protostar, with \lbol\ $\sim$ 10 \lsun\ 
\citep[e.g.,][]{parker1988:iras,shirley2000:scuba}, has a surrounding core 
mass of $1-2$ \msun\ \citep[e.g.,][]{shirley2000:scuba,shirley2002:scuba,jorgensen2002:envelopes,visser2002:scuba}, 
and is associated with spectroscopic evidence of infall motions 
\citep{myers1995:infall}, although the latter point is somewhat 
controversial given that some molecular line observations show evidence of 
expansion rather than infall motions \citep{park2000:l483}.  L483 is located at 
an assumed distance of 200 pc 
\citep[e.g.,][]{parker1988:iras,hilton1995:distances,shirley2000:scuba}

\subsection{Aqu-MM2/3/5}

Aqu-MM2, Aqu-MM3, and Aqu-MM5 are three protostars in the Aquila rift, located 
about 0.5\degree\ northwest of the protostellar cluster Serpens South that 
was recently discovered by \citep{gutermuth2008:serpsouth}.  All three 
protostars were discovered by \citep{maury2011:aquila}, who presented 
\emph{Herschel} far-infrared and MAMBO millimeter continuum emission maps of 
parts of Aquila, and were each found to drive bipolar molecular outflows by 
\citet{nakamura2011:serpsouth}.  \citet{maury2011:aquila} 
calculated \tbol\ of 26, 46, and 188 K for Aqu-MM2, 3, and 5, respectively, 
while \citet{dunham2013:luminosities} calculated \tbol\ of 80 and 
250 K for Aqu-MM3 and 5, thus classifying Aqu-MM2 as Class 0, Aqu-MM3 as either 
Class 0 or early Class I, and Aqu-MM5 as Class I.  All three protostars are 
located at an assumed distance of 260 pc, although we acknowledge that the true 
distance may be as large as 430 pc \citep[see discussions in][]{gutermuth2008:serpsouth,maury2011:aquila,dunham2013:luminosities}.

The exact rest velocity of each protostar is not well known.  Serpens South 
is located at a rest velocity of $\sim 8$ \kms, although there is some 
variation on the order of $\sim 1$ \kms\ within the cluster due to velocity 
gradients \citep[e.g.,][]{nakamura2011:serpsouth,kirk2013:serpsouth}.  
Given the proximity of these protostars to Serpens South and the extent of the 
velocity gradients within the cluster, we expect the rest velocities of these 
protostars to be within 1--2 \kms\ of 8 \kms.  Indeed, inspection of our 
\cojtwo\ maps show widespread emission from the ambient cloud between about 5.5 
and 12.5 \kms.  We thus estimate the rest velocity as 9 \kms\ but acknowledge 
it is uncertain by a few \kms.  Future dense gas tracers are required to 
better determine the rest velocities of these cores.

\subsection{SerpS-MM13}

SerpS-MM13 is a protostar in the southern part of the Serpens South cluster 
\citep{gutermuth2008:serpsouth} that was detected by \emph{IRAS} as IRAS 
18274$-$0212.  It is located at an assumed distance of 260 pc and core rest 
velocity of 8 \kms\ (see above).  \citet{maury2011:aquila} 
measured a \tbol\ of 131 K, classifying it as a Class I protostar, and 
\citet{nakamura2011:serpsouth} detected and mapped a bipolar molecular 
outflow driven by this source.  \citet{connelley2008:nirmultiplicity} 
included it in their near-infrared survey of protostellar multiplicity and did 
not detect any evidence for multiplicity in this source.

\subsection{CrA-IRAS32}

CrA-IRAS32 is a protostar in the Corona Australis star-forming region, located 
at a distance of 130 pc \citep{neuhauser2008:cra}.  First discovered by 
\citet{wilking1992:cra} as IRAS 18595$-$3712, it is located approximately 
15\am\ southeast of the well-studied Coronet cluster and is located at a rest 
velocity of 5.6 \kms\ \citep[e.g.,][]{vankempen2009:highj}.  With individual 
estimates of \tbol\ ranging from 61 -- 148 K 
\citep{chen1997:unknown,dunham2013:luminosities}, CrA-IRAS32 is either a late 
Class 0 or early Class I protostar.  It drives a bipolar molecular  
outflow partially mapped with both single-dish \citep{vankempen2009:highj} and 
interferometer CO observations \citep{peterson2011:cra}.  Cra-IRAS32 has been 
the subject of extensive studies at both infrared 
\citep[e.g.,][]{olofsson1999:cra,connelley2007:nebulae,haas2008:cra,seale2008:outflows,peterson2011:cra} 
and (sub)millimeter wavelengths 
\citep{chini2003:cra,nutter2005:cra,vankempen2009:highj,peterson2011:cra}, 
all of which have revealed an embedded protostar that is driving an outflow and 
is associated with both infrared nebulosity indicative of an outflow cavity and 
strong dust continuum and gas molecular line emission.  
\citet{peterson2011:cra} additionally used SMA dust continuum observations to 
infer the presence of a disk with a mass of 0.024 \msun.

\subsection{L673-7}

L673-7 is part of the Lynds Opacity Class 6 \citep{lynds1962:catalog} 
cloud complex L673.  L673-7 was first identified as a distinct core by 
\citet{lee1999:catalog}, who 
concluded it was a starless core based on no detection of an associated 
embedded YSO in \emph{IRAS} data.  Three related studies searching for infall 
motions towards starless cores using different dense gas tracers included 
L673-7 \citep{lee1999:infall,lee2004:infall,sohn2007:infall};
 none found any evidence for infall motions in this dense core.  
\citet{dunham2010:l6737} detected a protostar with a very low luminosity 
(\lint\ $\sim$ 0.04 \lsun) embedded in the L673-7 core in \emph{Spitzer} 
observations, and a molecular outflow driven by this protostar in \cojtwo\ 
observations taken at the CSO.  The outflow detection confirms the tentative 
CO line wings noted earlier by \citep{park2004:co}.  \citet{dunham2010:l6737} 
calculated \tbol\ $=16$ K from the observed SED, classifying 
L673-7 as a Class 0 protostar.  They assumed a distance of 300 pc based on 
earlier work by \citep{herbig1983:hh}, but acknowledged that this distance is 
highly uncertain \citep[see][for a full discussion]{dunham2010:l6737}.  
Recently, \citet{maheswar2011:distances} derived a distance of 240 pc based on 
an analysis of $A_V$ vs.~distance for stars detected by 2MASS, and we adopt 
this distance in this study.  The systemic velocity of L673-7 is 7.1 \kms\ 
\citep{lee2004:infall,sohn2007:infall}.

\subsection{B335}

B335 is an isolated Bok Globule \citep{bok1947:globules} catalogued as the dark 
core B335, CB199, and L663 (Opacity Class 6) in the surveys of 
\citet{barnard1927:catalog}, \citet{clemens1988:catalog}, and 
\citet{lynds1962:catalog}, respectively.  
It is associated with  a bipolar molecular 
outflow, first discovered by \citet{frerking1982:outflows}, a compact infrared 
source, first detected by \citet{keene1983:b335}, and kinematic 
evidence of infall consistent with inside-out collapse as predicted by 
\citet{shu1977:sis} \citep{zhou1993:b335,choi1995:b335,mardones1997:infall}.  
The infrared source, a Class 0 protostar with \tbol\ = 28 K 
\citep{shirley2000:scuba} is detected and cataloged by \emph{IRAS} as IRAS 
19345$+$0727.  The core is located at a rest velocity of 8.3 \kms\ 
\citep{zhou1993:b335,mardones1997:infall,evans2005:b335}.  While most studies 
assume a distance of 250 pc following \citet{tomita1979:globules}, 
\citet{stutz2008:b335} revise this distance to 60 -- 200 pc, 
and \citet{olofsson2009:b335} revise it to 90 -- 120 pc, both based on 
analyses of extinction versus distance for stars close in projection to the 
core.  We follow \citet{stutz2008:b335} and adopt a distance of 150 pc.

B335 was the first Bok Globule recognized as a site of low-mass star formation, 
and as a result has been the focus of an extensive list of studies over the 
past three decades \citep[see][and references therein]{stutz2008:b335}.  It 
is one of the most well-studied Class 0 protostars, with detailed observations 
across the wavelength spectrum and numerous dedicated modeling efforts.  As 
such, it has played a central role in developing the current understanding of 
low-mass star formation.

\subsection{L1152}

L1152 is a Lynds Opacity Class 5 cloud located in Cepheus 
\citep{lynds1962:catalog}, at an assumed distance of 325 pc 
\citep[see][and references therein for a detailed discussion of the distances to various portions of Cepheus]{kirk2009:cepheus}.  It harbors the 
protostar IRAS 20353$+$6742 first detected by \citet{beichman1986:iras}  This 
protostar was originally classified as Class I 
\citep[e.g.,][]{bontemps1996:outflows,mardones1997:infall}, but more recent 
studies yield bolometric temperatures ranging from 17 to 33 K 
\citep{kirk2009:cepheus,tobin2011:kinematics} 
and classify it as a Class 0 protostar.  A 
bipolar molecular outflow driven by the IRAS source was first detected and 
partially mapped by \citet{bontemps1996:outflows}.  
A second, more evolved young star also driving an outflow is 
located several arcminutes to the northeast; the redshifted emission seen 
in the northeast of our \cojtwo\ map that appears unrelated to the outflow 
may be related to this second outflow and the associated HH376A.  The core 
is located at a rest velocity of 2.5 \kms\ 
\citep[e.g.,][]{benson1989:dcdc,mardones1997:infall,tobin2011:kinematics}.

\subsection{L1157}

L1157 is a Lynds Opacity Class 5 cloud located in Cepheus at a rest velocity of 
2.6 \kms\ \citep[e.g.,][]{gregersen1997:infall}.  While individual  
distance estimates to this object range from 250 -- 450 pc, we follow 
\citet{kirk2009:cepheus} and adopt a distance of 300 pc.  L1157 harbors the 
Class 0 protostar IRAS 20386$+$6751, with individual estimates of \tbol\ 
ranging from 29 -- 44 K 
\citep{gregersen1997:infall,shirley2000:scuba,tobin2011:kinematics}.
A collimated, bipolar molecular outflow driven by this protostar was first 
detected and mapped by \citet{umemoto1992:l1157}, who noted evidence of both 
temperature and molecular abundance enhancement in the blue lobe of the 
outflow and argued in favor of shock heating.  \citet{mikami1992:l1157} 
detected SiO emission toward the blue lobe, citing this as further evidence of 
shocks.  Follow-up studies at higher resolution confirmed these results and 
showed strong evidence for both episodicity and precession in the L1157 outflow 
\citep[e.g.,][]{zhang1995:l1157,tafalla1995:l1157,gueth1997:l1157}.  
\citet{bachiller1997:l1157} conducted a large line survey and found 
several additional examples of rare molecules with greatly enhanced abundances 
in the outflow, leading \citet{bachiller2001:l1157} to identify this as the 
prototype of ``chemically active outflows.''  Numerous spectral line surveys 
have targeted L1157 in the past decade and confirmed the chemical complexity of 
its outflow 
\citep[e.g.,][]{arce2008:l1157,codella2010:l1157,yamaguchi2012:l1157}.  
\citet{tobin2013:cepheus} recently confirmed that it shows no signs of 
multiplicity down to size scales of 100 AU using data from the Very Large 
Array (VLA).

\subsection{L1228}

L1228 is a Lynds Opacity Class 1 cloud \citep{lynds1962:catalog} 
associated with the Class I source IRAS 20582$+$7724 
\citep{haikala1989:l1228}, with individual 
measurements of \tbol\ between 79 -- 388 K 
\citep{arce2006:outflows,kauffmann2008:mambo,kirk2009:cepheus}.  A bipolar 
molecular outflow is driven by the \emph{IRAS} source at a position angle 
of 79\degree\ and with an opening angle of 95\degree\ 
\citep{winnewisser1988:l1228,haikala1989:l1228,tafalla1997:l1228,arce2004:l1228,arce2006:outflows}.  
This outflow is associated with the HH objects HH199 and HH200 
\citep{bally1995:l1228,devine2009:l1228}, 
and is both eroding and dispersing the dense core \citep{arce2004:l1228} 
and destroying large dust grains \citep{chapman2009:cores}.  
The dense core has a mass of $\sim$ 1 \msun\ 
\citep{young2006:scuba,kauffmann2008:mambo} and is located at a rest velocity 
of $-8.0$ \kms\ \citep{anglada1997:nh3,larionov1999:catalog,arce2004:l1228}.  
Distance estimates range from 150 to 300 pc 
\citep{benson1989:dcdc,anglada1997:nh3,kun1998:cepheus,kirk2009:cepheus}; 
we adopt a distance of 200 pc following \citet{kun1998:cepheus} and 
\citet{kirk2009:cepheus}.

\citet{bally1995:l1228} argued that the associated HH object HH200 is actually 
driven by a T Tauri star 1.5\am\ northwest of the Class I protostar, and 
that the observed molecular outflow contains overlapping emission from 
the main outflow and a weak secondary outflow 
also driven by the T Tauri star.  This interpretation is also favored by 
\citet{devine2009:l1228}.  However, while the case for HH200 being driven by 
this secondary source appears robust, \citet{tafalla1997:l1228} presented 
higher-spatial resolution CO maps and argued that there is no evidence for a 
second component driven by the T Tauri star in the molecular outflow.  Our JCMT 
data presented here, with even higher spatial resolution, agrees with 
\citet{tafalla1997:l1228}.  \citet{arce2004:l1228} detected a secondary 
dust core with a very low mass (0.006 \msun) located 5\as\ northwest of the 
primary source in OVRO 2.7 mm continuum observations.  The very low mass 
of this core coupled with no detections in the near- or mid-infrared 
\citep{connelley2008:nirmultiplicity,kirk2009:cepheus} 
argue that, if real, it is at a very early evolutionary stage and not a 
significant contributor to either the observed SED or molecular outflow.  We 
thus assume that L1228 is a single object for the purposes of this study.

\subsection{L1014}

L1014 is a Lynds Opacity Class 6 cloud \citep{lynds1962:catalog} 
originally believed to be starless based on no associated \emph{IRAS} source 
and no detected molecular outflow in single-dish observations 
\citep{visser2001:scuba,visser2002:scuba,crapsi2005:starless}.
\citet{young2004:l1014} detected a protostar 
with \lint\ $\sim$ 0.09 \lsun\ in \emph{Spitzer} observations, leading to 
the discovery of a new class of very low luminosity objects 
\citep[VeLLOs;][]{difrancesco2007:ppv,dunham2008:lowlum}, most in cores 
previously classified as starless.  This detection was quickly followed by 
that of a weak, compact outflow in SMA observations \citep{bourke2005:l1014} 
and extended near-infrared nebulosity aligning with the outflow morphology 
\citep{huard2006:l1014}.  The detection of a protostar was somewhat of a 
surprise given that \citet{crapsi2005:starless} characterized L1014 as only a 
moderately evolved dense core based on the detection of moderate depletion 
and deuteration and no clear kinematic signatures of infall.

The measured \tbol\ of L1014 is $50-66$ K 
\citep{young2004:l1014,dunham2008:lowlum}, classifying it as a Class 0 
protostar.  Individual estimates of the mass of the L1014 core range from 0.7 
to 3.6 \msun\ \citep{visser2001:scuba,visser2002:scuba,young2004:l1014,huard2006:l1014,kauffmann2008:mambo}, 
and the rest velocity is 4.2 \kms\ \citep{crapsi2005:starless}.  
\citet{young2004:l1014} 
assumed a distance of 200 pc but noted the true distance was only strongly 
constrained to be less than 1 kpc based on a lack of foreground stars.  
\citet{morita2006:l1014} argued for a larger distance of $400-900$ pc based on 
comparing the positions of three nearby T Tauri stars in color-magnitude 
space with stellar evolutionary models, although it is unclear if these three 
sources are truly associated with L1014, and if their results would change 
by comparing to stellar evolutionary models that include the effects of the 
early accretion history \citep[e.g.,][]{baraffe2009:episodic,hosokawa2011:episodic,baraffe2012:episodic}.  
\citet{maheswar2011:distances} derived a distance of 258 $\pm$ 50 pc based 
on an analysis of $A_V$ vs.~distance for stars detected by 2MASS, and we adopt 
this distance in this study.

\subsection{L1165}

L1165 is a Lynds Opacity Class 6 cloud \citep{lynds1962:catalog} associated 
with the bright Class I source IRAS 22051$+$5848 \citep{parker1988:iras}, 
suggested to be a candidate FU Orionis object based on the detection of CO 
bands in absorption \citep{reipurth1997:hh}.  \citet{parker1991:outflows} 
discovered a molecular outflow driven by this \emph{IRAS} source.  
\citet{reipurth1997:gianthh} detected an HH object (HH 354) 11\am\ to the 
northeast of the \emph{IRAS} source, along the same axis connecting IRAS 
22051$+$5848 with the small reflection nebula GY 22 
\citep{gyul1982:hh}.  They also noted a major cavity in the large-scale 
cloud structure along this same axis and suggested that the molecular outflow, 
HH object, reflection nebula, and cloud cavity are all part of one 
parsec-scale outflow system.  The L1165 core is at a rest velocity of 
$-$1.6 \kms\ based on the \ammonia\ observations presented by 
\citet{sepulveda2011:nh3}.

The distance to L1165 is not well characterized, with two different distance 
assumptions dominating the literature.  \citet{reipurth1997:hh}, 
\citet{reipurth1997:gianthh}, and \citet{sepulveda2011:nh3} all adopt a 
distance of 750 pc based on the assumption that L1165 is associated with the 
IC 1396 region.  On the other hand, \citet{dobashi1994:cepheus}, 
\citet{visser2002:scuba}, and \citet{tobin2010:protostars} 
all adopt a distance of 300 pc based on physical and 
kinematic association with a source of known distance.  While we adopt the 
closer distance of 300 pc in this study, we find no convincing evidence to 
prefer one distance over the other and acknowledge this distance is quite 
uncertain.

Very recently, \citet{tobin2013:cepheus} detected a companion source in 
VLA and CARMA data with a projected separation of $\sim 100$ AU.  From their 
CARMA data they found a 1 mm flux ratio of 5 between the primary and secondary, 
implying that most of the system mass is in the primary.  We thus assume that 
the primary source dominates both the outflow and continuum spectral energy 
distribution of L1165.  However, \citet{tobin2013:cepheus} 
showed that each lobe may originate from different sources in the 
system.  As their results are tentative, additional observations with 
very high spatial resolution are required to test these assumptions.  

\subsection{L1251A-IRS3}

L1251A is one of five cores located within the Lynds Opacity Class 5 
cloud L1251 \citep{lynds1962:catalog}.  It was first revealed as a separate 
core in molecular emission line maps presented by \citet{sato1994:l1251}.  
\citet{kun1993:l1251} derived a distance estimate of 
300 pc and noted that it is associated with three \emph{IRAS} sources with 
colors consistent with being YSOs.  Higher spatial resolution infrared images 
with \emph{Spitzer} revealed four YSOs within L1251A, two of which 
(denoted L1251A-IRS3 and L1251A-IRS4) are associated with (sub)millimeter dust 
continuum emission tracing dense cores, mid-infrared jets, and molecular 
outflows \citep{lee2010:l1251a}.  The molecular outflow driven by L1251A-IRS3 
is the dominant outflow in the region, and while the SRAO \cojtwo\ map 
presented here includes and detects both outflows, we only focus on the 
L1251A-IRS3 outflow given the relatively low spatial resolution and 
sensitivity of these data.  L1251A-IRS3 is a low luminosity Class 0 
protostar \citep[\lbol\ $= 0.8$ \lsun, \tbol\ $= 24$ K;][]{lee2010:l1251a}, 
with a surrounding core mass of 12 \msun\ located at a rest velocity of $-$3.9 
\kms\ \citep{sato1994:l1251,barranco1998:nh3,lee1999:catalog,lee2010:l1251a}.

\section{B.~~Average Spectra of Each Outflow}\label{sec_appendix_spectra}

\begin{figure*}
\epsscale{1.2}
\plotone{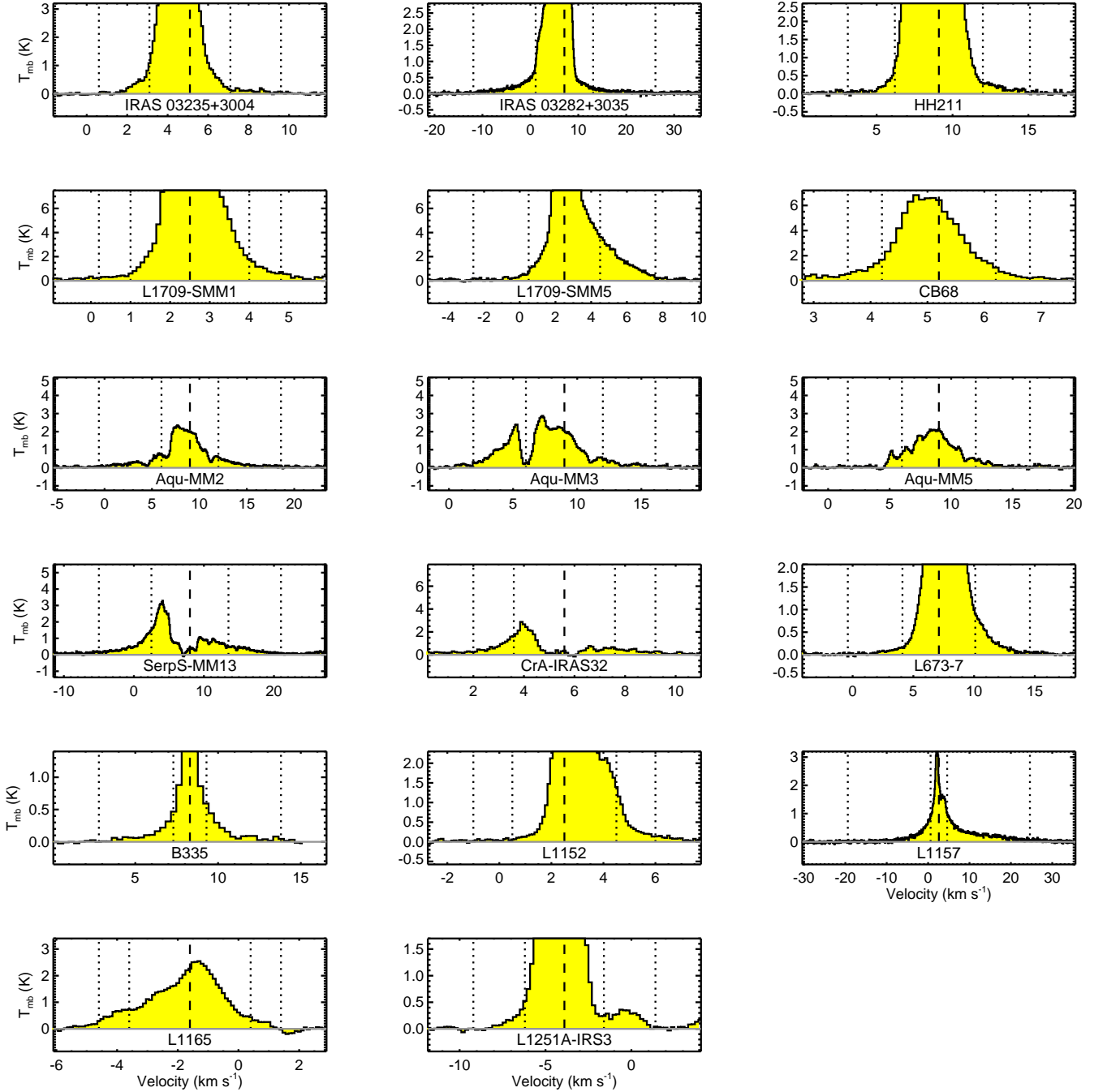}
\caption{\label{fig_spectra21}Average spectra for each of the outflows mapped 
in \cojtwo, averaged over all spatial pixels encompassed by the outflow lobes.  
The dashed vertical line in each panel marks the ambient 
cloud velocity.  The dotted vertical lines in each panel mark $v_{\rm min}$ and 
$v_{\rm max}$, which are chosen to be symmetrical about the cloud velocity 
(see \S \ref{sec_mass_dynamical} for details).}
\end{figure*}

\begin{figure*}
\epsscale{1.2}
\plotone{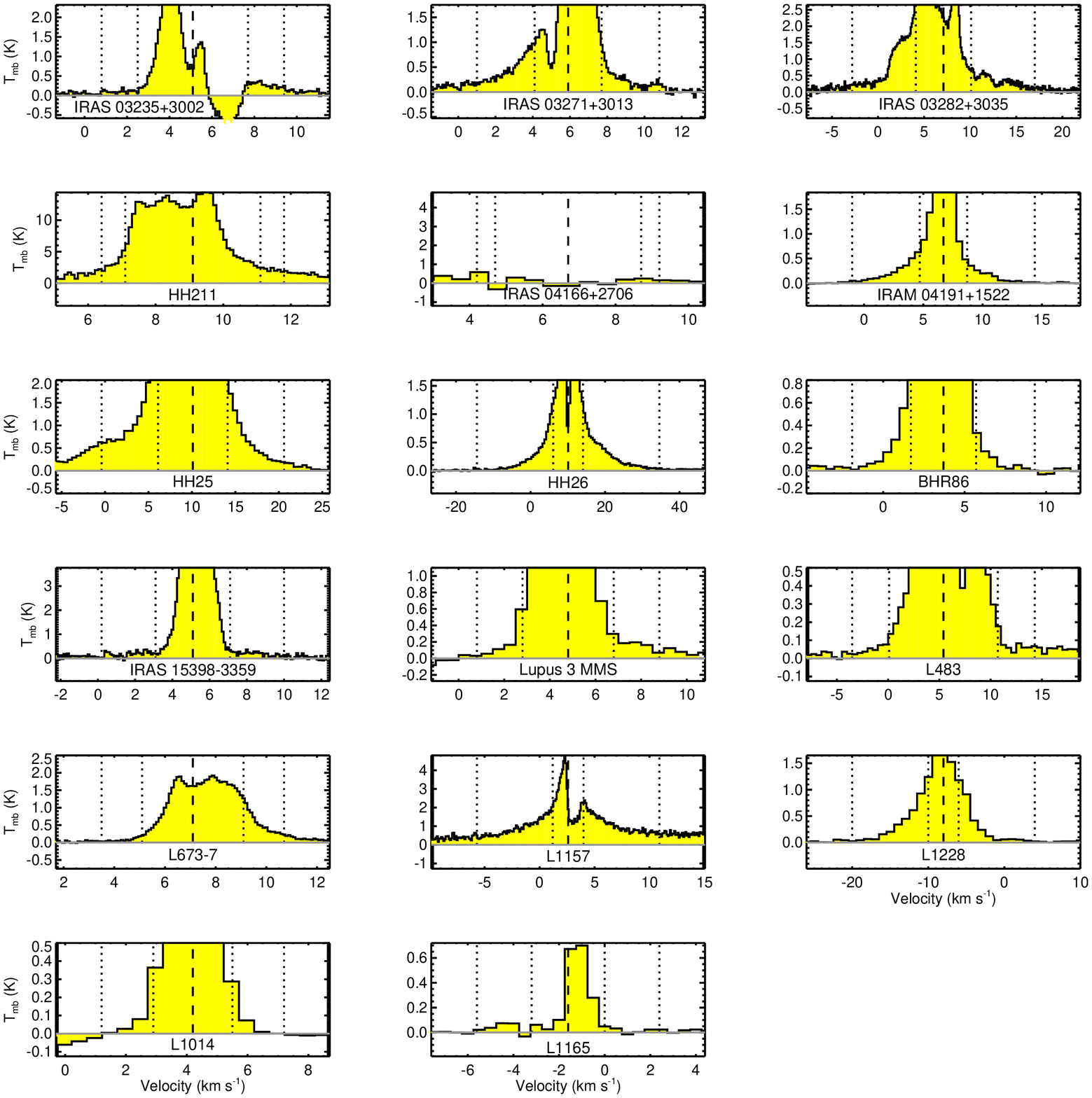}
\caption{\label{fig_spectra32}Same as Figure \ref{fig_spectra32}, except for 
the outflows mapped in \cojthree.}
\end{figure*}

Figures \ref{fig_spectra21} and \ref{fig_spectra32} display the average 
spectra toward each outflow mapped in \cojtwo\ and \cojthree, respectively.  
The spectra are averaged over all spatial pixels encompassed by the outflow 
lobes (the same pixels over which the outflow masses and dynamical properties 
are integrated).  The dashed vertical line in each panel marks the ambient 
cloud velocity.  The dotted vertical lines in each panel mark $v_{\rm min}$ and 
$v_{\rm max}$, which are chosen to be symmetrical about the cloud velocity 
(see \S \ref{sec_mass_dynamical} for details).  We emphasize here that, 
since these are average spectra over the full outflow lobes, their velocity 
structures may not always match the marked $v_{\rm min}$ and $v_{\rm max}$, 
which are chosen to eliminate ambient cloud emission at all spatial 
positions and also include all of the highest-velocity emission from outflowing 
gas.  Some spectra clearly show absorption features near the cloud velocities 
due to contaminated off positions; in many cases these features prevent us 
from correcting for the outflow emission at low velocities, as indicated 
in Tables \ref{tab_outflow_corrections} and 
\ref{tab_outflow_dynamics_corrected}.  The y-axis range of each panel was 
adjusted to emphasize the linewings from the outflows and thus sometimes 
results in the ambient cloud emission near the rest velocity being cut off.

\section{C.~~Optically Thin Emission from \co\ in LTE}\label{sec_appendix_equations}

\subsection{C.1~~Calculating Column Densities}

To derive the total column density from \co\ observations, we first start 
with the definition of integrated intensity in main-beam temperature units, 
$\int T_{\rm mb} \, d\nu$, where $T_{\rm mb}$ is the brightness temperature of 
the emission:

\begin{equation}\label{eq_appendix_tmb1}
\int T_{\rm mb} \, d\nu = \int \frac{c^2}{2 \, k \, \nu^2} \, B_{\nu}(T) \, (1 - e^{-\tau_{\nu}}) \, d\nu \approx \int \frac{h \, \nu}{k} \, \frac{1}{e^{\frac{h \nu}{k T}} - 1} \, \tau_{\nu} \, d\nu \,
\end{equation}
where $B_{\nu}(T)$ is the Planck function at temperature $T$, $\tau_{\nu}$ 
is the optical depth of the transition, and the right-most expression assumes 
that the emission is optically thin ($\tau_{\nu} << 1$).  
The optical depth of a transition from 
lower state J to upper state J+1 can be expressed in terms of the Einstein B 
coefficients for absorption ($B_{\rm J,J+1}$) and stimulated emission 
($B_{\rm J+1,J}$), 

\begin{equation}\label{eq_appendix_tau1}
\tau_{\rm J,J+1} = \frac{h \, \nu}{4 \, \pi} \, (N_{\rm J}(\nu) \, B_{\rm J,J+1} - N_{\rm J+1}(\nu) \, B_{\rm J+1,J}) \, ,
\end{equation}
where $N_{\rm J}(\nu)$ and $N_{\rm J+1}(\nu)$ are the column densities in the 
lower and upper states, respectively.  Defining $g_{\rm J}$ and $g_{\rm J+1}$ as 
the degeneracies of the lower and upper states, respectively, and using the 
relation between the Einstein B coefficients, 

\begin{equation}
g_{\rm J+1} \, B_{\rm J+1,J} = g_{\rm J} \, B_{\rm J,J+1} \, ,
\end{equation}
and the Boltzmann equation to relate $N_{\rm J}(\nu)$ and $N_{\rm J+1}(\nu)$ in 
LTE, 

\begin{equation}
\frac{N_{\rm J+1}(\nu)}{N_{\rm J}(\nu)} = \frac{g_{\rm J+1}}{g_{\rm J}} \, e^{-\frac{h \, \nu}{k \, T}} \, ,
\end{equation}
Equation \ref{eq_appendix_tau1} can be expressed as 

\begin{equation}\label{eq_appendix_tau2}
\tau_{\rm J,J+1} = \frac{h \, \nu}{4 \, \pi} \, N_{\rm J}(\nu) \, B_{\rm J,J+1} \, (1 - e^{-\frac{h \, \nu}{k \, T}}) \, .
\end{equation}
By using the following LTE relation between column density in a state J, 
$N_{\rm J}$, and total column density, $N$, where $Q(T)$ is the partition 
function ($Q(T)=\sum_{J=0}^{\infty} \, g_J \, e^{-E_J/kT}$), 

\begin{equation}
N_{\rm J}(\nu) = \frac{g_{\rm J}}{Q(T)} \, e^{-\frac{E_{\rm J}}{k \, T}} \, N(\nu) \, ,
\end{equation}
the optical depth of the transtion can be expressed in terms of the total 
column density of the molecule:

\begin{equation}\label{eq_appendix_tau3}
\tau_{\rm J,J+1} = \frac{h \, \nu}{4 \, \pi} \, B_{\rm J,J+1} \, (1 - e^{-\frac{h \, \nu}{k \, T}}) \, \frac{g_{\rm J}}{Q(T)} \, e^{-\frac{E_{\rm J}}{k \, T}} \, N(\nu) \, .
\end{equation}

If we assume that the interval of integration in Equation 
\ref{eq_appendix_tmb1} is much smaller than the frequency ($d\,\nu << \nu$), 
such that quantities that depend on $\nu$ can be treated as constants of 
integration, and use the fact that the energy released by a transition from 
state J+1 to state J is $E_{\rm J+1,J} = h \nu = E_{\rm J+1} - E_{\rm J}$, 
Equation \ref{eq_appendix_tau3} can be substituted into Equation 
\ref{eq_appendix_tmb1} to yield the expression

\begin{equation}\label{eq_appendix_tmb2}
\int T_{\rm mb} \, d\nu = \frac{B_{\rm J,J+1} \, h^2 \, \nu^2}{4\, \pi \, k} \, \frac{g_{\rm J}}{Q(T)} \, e^{-\frac{E_{\rm J+1}}{k \, T}} \, \int N(\nu) \, d\nu \, .
\end{equation}
Finally, recognizing that $\int N(\nu) \, d\nu$ is the total column density 
of the molecule, $N$, defining $X_{\rm CO}$ as the abundance of \co\ relative to 
H$_2$, changing the variable of integration from frequency 
to velocity, and using the relationship for $B_{\rm J,J+1}$ in terms of the dipole 
moment of the molecule $\mu$, $B_{\rm J,J+1} = \frac{32 \, \pi^4}{3\, h^2 \, c} \, \mu^2 \, \frac{({\rm J}+1)}{(2{\rm J}+1)}$, the total column density of H$_2$ can be calculated as 

\begin{equation}\label{eq_appendix_column}
N_{\rm H_2} = X_{\rm CO} \frac{3 \, k}{8 \, \pi^3 \, \nu \, \mu^2} \frac{(2{\rm J}+1)}{({\rm J}+1)} \frac{Q(T)}{g_{\rm J}} \, e^{\frac{E_{\rm J+1}}{k T}} \int T_{\rm mb} \, dV \, .
\end{equation}
In the rest of this Appendix we will express this as 
$N_{\rm H_2} = f({\rm J},T,X_{\rm CO}) \, I$, where $I$ is the integrated 
intensity ($I = \int T_{\rm mb} \, dV$) measured in K \cms, and 

\begin{equation}
f({\rm J},T,X_{\rm CO}) = X_{\rm CO} \frac{3 \, k}{8 \, \pi^3 \, \nu \, \mu^2} \frac{(2{\rm J}+1)}{({\rm J}+1)} \frac{Q(T)}{g_{\rm J}} \, e^{\frac{E_{\rm J+1}}{k T}} \, .
\end{equation}

\subsection{C.2~~Line Intensity, Column Density, and Mass Ratios For Different 
Excitation Temperatures}

The ratio of integrated intensities in two different transitions into lower 
states ${\rm J_1}$ and ${\rm J_2}$, $I_{\rm J_1} / I_{\rm J_2}$, can be expressed 
as $f({\rm J_2},T,X_{\rm CO}) / f({\rm J_1},T,X_{\rm CO})$.  
The correction factor to total 
outflow mass as a function of temperature, compared to the values obtained 
assuming $T = 50$ K, is calculated as the ratio of column densities 
calculated assuming a temperature $T$ 
to those calculated assuming a temperature 
of 50 K, $N_T / N_{50 K} = f({\rm J},T,X_{\rm CO}) / f({\rm J},50 K,X_{\rm CO})$.  
The factor by which the ratio of outflow mass calculated from two different 
transitions into lower states ${\rm J_1}$ and ${\rm J_2}$ will change assuming 
temperatures other than 50 K is then simply the ratio of the above expression 
calculated for both ${\rm J_1}$ and ${\rm J_2}$.

Finally, we consider the case where we have a mixture of gas at two 
temperatures, $T_1$ and $T_2$, with a ratio of mass (or column density) 
at the two temperatures $A = M_{T1} / M_{T2} = N_{T1} / N_{T2}$.  
Assuming the emission is optically thin and there is 
no self-absorption, the total measured integrated intensity is the sum of 
the intensity of gas at each temperature,

\begin{equation}\label{eq_appendix_itotal}
I_{\rm total} = I_{T_1} + I_{T_2}  = I_{T_1} \, \left( 1 + \frac{f({\rm J},T_1,X_{\rm CO})}{A \, f({\rm J},T_2,X_{\rm CO})} \right) \, .
\end{equation}
The ratio of total measured integrated intensity in two different transitions 
with lower states ${\rm J_1}$ and ${\rm J_2}$ is then simply the ratio of 
Equation \ref{eq_appendix_itotal},

\begin{equation}
I_{\rm total,{\rm J_1}} / I_{\rm total,{\rm J_2}} = \frac{I_{T_1,{\rm J_1}}}{I_{T_1,{\rm J_2}}} \, \frac{\left( 1 + \frac{f({\rm J_1},T_1,X_{\rm CO})}{A \, f({\rm J_1},T_2,X_{\rm CO})} \right)}{\left( 1 + \frac{f({\rm J_2},T_1,X_{\rm CO})}{A \, f({\rm J_2},T_2,X_{\rm CO})} \right)} = \frac{f({\rm J_2},T_1,X_{\rm CO})}{f({\rm J_1},T_1,X_{\rm CO})} \, \frac{\left( 1 + \frac{f({\rm J_1},T_1,X_{\rm CO})}{A \, f({\rm J_1},T_2,X_{\rm CO})} \right)}{\left( 1 + \frac{f({\rm J_2},T_1,X_{\rm CO})}{A \, f({\rm J_2},T_2,X_{\rm CO})} \right)}\, .
\end{equation}
This ratio can then be used to calculate the temperature of isothermal gas 
in LTE that would give rise to the same line ratio, using the above expression 
for the ratio of integrated intensities in two different transitions.

\end{document}